\documentclass[modern]{aastex701}
\usepackage{graphicx}

\graphicspath{{}}
\usepackage{amsmath}
\usepackage{comment}

\begin{document}

\title{\textcolor{black}{Exploring the Interior Structure and Mode of Tidal Heating in Enceladus}}

\author[orcid=0000-0002-2306-2576,sname=Bagheri]{Amirhossein Bagheri}
\altaffiliation{Corresponding Author}
\affiliation{California Institute of Technology}
\email[show]{abagheri@caltech.edu}  

\author[orcid=0000-0003-1412-6395,gname=Mark, sname=Simons]{Mark Simons} 
\affiliation{California Institute of Technology}
\email{simons@caltech.edu}

\author[orcid=0000-0001-9896-4585, gname=Ryan,sname=Park]{Ryan S. Park}
\affiliation{Jet Propulsion Laboratory, California Institute of Technology}
\email{ryan.s.park@jpl.nasa.gov}

\author[orcid=0000-0001-7857-8513, gname=Alexander, sname=Berne]{Alexander Berne}
\affiliation{California Institute of Technology}
\email{aberne@caltech.edu}

\author[0000-0001-5617-207X,gname=Douglas, sname=Hemingway]{Douglas Hemingway}
\affiliation{Jackson School of Geosciences, University of Texas at Austin}
\email{Douglas.Hemingway@utexas.edu}

\author[0000-0002-4611-3209,gname=Mohit, sname=Melwani Daswani]{Mohit Melwani Daswani}
\affiliation{Jet Propulsion Laboratory, California Institute of Technology}
\email{mohit.melwani.daswani@jpl.nasa.gov}

\author[0000-0002-4242-3293, gname=Steven, sname=Vance]{Steven D Vance}
\affiliation{Jet Propulsion Laboratory, California Institute of Technology}
\email{steven.d.vance@jpl.nasa.gov}




\begin{abstract}

Enceladus is among the most intriguing bodies in the solar system due to its astrobiological potential. Determining the extent and duration of habitability (i.e., \textit{sustained habitability}) requires characterizing the interior properties and the level and distribution of tidal heating in Enceladus. \textcolor{black}{Inferring the intensity of geophysical activity in the core has direct implications for the potential hydrothermal activity and supply of chemical species important for habitability to the ocean.} We build a statistical framework to constrain the interior using estimates of libration, shape, heat flux, gravity, and total mass. We use this framework to examine the extent that geodetic measurements can improve our understanding of the interior structure, with an emphasis on partitioning of dissipation between the shell and the core. We quantify plausible ranges of gravitational ($k_2$) and displacement ($h_2$, $l_2$) tidal Love numbers consistent with existing observations. We demonstrate that measuring $k_2$ alone can \textcolor{black}{only} constrain the total tidally dissipated energy, but not its radial distribution. However, measuring the amplitude and phase of $h_2$ or $l_2$ facilitates determining the extent of tidal dissipation in the shell and the core. We provide the precisions required for measuring $k_2$, $h_2$, and $l_2$ that enable distinguishing between the main tidal heating scenarios, i.e., in the shell versus the core. We also explore the effect of the structural heterogeneities of the shell on the tidal response. Lastly, we evaluate the efficacy of future geodetic measurements to constrain key interior properties essential to understand the present-day (\textit{instantaneous}) and long-term \textcolor{black}{(\textit{sustained)}} habitability at Enceladus.
\end{abstract}

\keywords{Tides, Tidal heating, Love numbers, Libration, Gravity, Icy moons}


\section{Introduction and Background}\label{sec:intro}

Flybys of Saturn's moon Enceladus by the \textit{Cassini} spacecraft revealed multiple intriguing features suggesting an active surface and interior: jets that continuously erupt from a region centered on the satellite's south pole \citep{porco_etal06} and feed Saturn's diffuse E-ring with water-ice crystals \citep{kempf_etal10}; a very bright surface \citep{hendrix_etal10}; a high surface heat flux \citep{spencer_etal06, howett_etal11}; and, importantly, evidence for a subsurface ocean \citep{porco_etal06, iess_etal14, thomas_etal16, park_etal24}. This subsurface water body likely meets the criteria for \textit{instantaneous} habitability \citep{cockell_etal24}: An extensive liquid water reservoir containing salts and complex organic compounds and the expected chemical disequilibrium required for supporting metabolism. These conditions can occur through water-rock interactions, presumably in the inferred hydrothermal systems at or below the seafloor \citep{postberg_etal09, howett_etal11, hsu_etal15,cable_etal21,cockell_etal24}. These lines of evidence for a potentially habitable environment position Enceladus as a compelling target for future exploration by orbiters and landers \citep{PlanetaryDecadal2023}. Investigating the extent to which these conditions might have been sustained at Enceladus requires understanding the geophysical processes in the interior over geological timescales (see \citet{cockell_etal24} and \citet{PlanetaryDecadal2023} for discussions of \textit{sustained or dynamical habitability}).

Analysis of \textit{Cassini}'s imaging, thermal mapping, and radiometric tracking data explored the global shape, gravity field, libration (i.e., diurnal wobble of the ice shell), and the surface topography of Enceladus, providing insights into the interior properties. The large amplitude of forced physical libration suggests that the ice crust is mechanically decoupled from the rocky core, likely by a global ocean \citep{nadezhdina_etal16, thomas_etal16,vanhoolst_etal16,park_etal24}. The presence of a global ocean is also supported by the analysis of the low-order gravity field \citep{McKinnon15, cadek_etal16}.

The total rate of heat loss at Enceladus remains poorly constrained \citep{howett_etal25}. Estimates of the ice shell thickness and surface temperatures suggest a total heat loss of approximately 25-40~GW \citep{hemingwayMittal19, nimmo_etal23}, while the measured heat output from the South Polar Terrain (SPT) alone is estimated to be 4-19~GW \citep{howett_etal11}.
With an estimated radiogenic heating rate of \textcolor{black}{less than 0.3~GW \citep{vance_etal2007,robertsNimmo08, nimmo_etal18}} and minimal additional heating due to libration \citep{Shao_Nimmo22}, the most likely heat source is therefore dissipation due to the eccentricity tides exerted by Saturn \citep[e.g.,][]{nimmo_etal23}.
Initial analyses suggested that given the \textcolor{black}{orbital configuration of } Enceladus and Saturn, tidal dissipation is insufficient to explain the estimated heat loss \citep{meyerWisdom07}. However, later studies suggested that Saturn is supplying tidal energy and angular momentum to its satellites at a higher rate than previous estimates \citep{lainey_etal17}, \textcolor{black}{suggesting a} resonance locking between the moons and the planet's internal oscillation modes \citep{fuller_etal16,lainey_etal17, nimmo_etal18, lainey_etal20, jacobson22}.
Alternatively, there may exist transient periods of tidal heating. In this case, the present-day ocean would be a manifestation of a cycle of ocean formation and freezing as a result of a dynamically unstable orbit \citep{meyerwisdom08, goldreich_etal25}, or instead as a result of changes in the mean motion resonance in the case of an ongoing feedback between the moon's thermal and orbital evolution. \citep{ojakangasstevenson86,meyerwisdom08episodic, renaud_etal21, goldreich_etal25}. A transient ocean could also be the result of recent catastrophic events that affect the Saturnian system \citep[e.g.,][]{Asphaug_Reufer13, wisdom_etal22, teodoro_etal23}.

It remains unknown whether \textcolor{black}{tidal dissipation occurs mainly} in Enceladus's core or ice shell. Distinguishing between the existing end-member scenarios is vital for assessing Enceladus's sustained habitability \citep[e.g.,][]{PlanetaryDecadal2023, cockell_etal24}.
A tidally active core can drive hydrothermal activity at the bottom of the ocean and the continuous contact between water and fresh rock, potentially supporting the chemical reactions necessary for life to develop and evolve over geological timescales \citep{vance_etal2007,hsu_etal15, choblet_etal17}.
A comprehensive description of the global heat budget would make it possible to distinguish between a primordial and a present-day hydrothermal source of organics, volatiles, and nanosilica materials in the plumes.
Studies have considered viscous dissipation in the ice shell as the mechanism \textcolor{black}{to maintain the global ocean \citep[e.g.][]{robertsNimmo08, shoji_etal13, efroimsky18, kangFlierl2020}. These models require a mean viscosity of $\eta_{shell}\approx10^{13-15}$~Pa.s which is significantly lower than the estimated ice viscosity at the temperatures expected in the outer shell of Enceladus \citep[e.g.][]{GoldsbyKohlstedt01}}. Alternative studies propose dissipation in water-filled fractures \citep{kiterubin16} or in a highly porous and dissipative rocky core with a very low effective viscosity \citep{roberts15,choblet_etal17, souvcek_etal19, liao_etal20, rovira_etal22, ayguncadek2023}. \textcolor{black}{Dissipation due to resonance effects in the ocean has also been proposed \citep{tyler20}, but was found incompatible with the estimated shell and ocean thicknesses \citep{hayMatsuyama17, matsuyama_etal18}.}

Considering the ambiguities in understanding the interior, geophysical investigations of Enceladus must account for the possibility of heating in the deep interior. \textcolor{black}{
\citet{ermakov_etal21} and \citet{Genova_etal24} explore models of the interior by future measurements the potential tidal Love number} $k_2$ along with libration \textcolor{black}{and} gravity-topography admittance to constrain the interior structure (i.e., thicknesses and densities of layers) at Enceladus. However, the ranges of interior parameters considered in these studies \textcolor{black}{assume a nearly rigid core and thus inherently produce negligible tidal dissipation in the core. Therefore, these studies do not address the challenge of discriminating} between the different scenarios of tidal dissipation. As we will show, a tidally active core will affect the time-dependent gravity field of the body and produce a gravitational signal that resembles that of a tidally dissipative ice shell, resulting in a non-unique interpretation of the amplitude and phase of $k_2$. In addition, gravity and topography measurements have revealed structural heterogeneities in the Enceladus ice shell \citep{Yin2015,park_etal24, Schenk_McKinnon24}. 
\textcolor{black}{Structural heterogeneities in the shell can potentially affect its tidal response compared to a spherically symmetric body \citep[e.g.,][]{zhong_etal12, qin_etal14, wahr_etal14, lau_etal15, dmitrovskii_etal22, berne_etal23}} by driving deformation at higher harmonic degrees (e.g., $l = 3,4$) \textcolor{black}{through} mode coupling. Such a coupling can produce variations in degree-2 Love numbers evaluated at different spherical harmonic orders, i.e., $k_{2m}$ and $h_{2m}$ \citep[e.g.,][]{bagheri_etal21, pou_etal22, berne_etal23}. 
\textcolor{black}{
The next sections of the paper are organized as follows. In Section~\ref{sec:methods}, we define the geophysical observables used in this study—tidal response, layer-resolved tidal heating, physical libration, and gravity coefficients—and detail their computation. In Section~\ref{sec:shellthicknesslibration}, we analyze the sensitivity of physical libration to shell properties and its correlation with the tidal response. In Section~\ref{sec:heatgenerationclasses}, we explore tidal-dissipation regimes in the shell and core that are consistent with the inferred heat-production rate, and we delineate the associated ranges of tidal Love numbers and rheological parameters. In Section~\ref{sec:roadmap}, we present a Bayesian MCMC framework that integrates all available observations to infer plausible interior models and to estimate ranges for as-yet unmeasured tidal Love numbers. We show the degeneracy of using $k_2$ alone to identify where tidal dissipation occurs at Enceladus, assess whether joint observations of $k_2$ and $h_2$ can distinguish dissipation regimes, and derive the minimum measurement precision required. Using precisions that can be expected from future missions, we perform inversions of geodetic data (static gravity, complex tidal Love numbers, and libration) to evaluate their ability to constrain interior structure. Finally, in Section~\ref{sec:concludiscussions}, we summarize the main conclusions and discuss the limitations of our analysis.}

\section{Methods}\label{sec:methods}

We describe the geophysical models needed to compute the geodetic quantities in our forward calculation. \textcolor{black}{We describe the tidal response in the form of the complex Love numbers and maximum surface deformations and the viscoelastic models used in computing the tidal response, followed by our derivation for the tidal heating in the shell and in the core. We also describe modeling the physical libration, and calculation of the static gravity coefficients. Detailed derivations are provided in the appendices.}

\subsection{Tidal Response}\label{sec:tidesandviscoelasticity}
The eccentric orbit of Enceladus around Saturn drives periodic gravitational tidal forcing of the moon. The  forcing  potential, truncated at spherical harmonic degree 2 and to first order in eccentricity, can be written as \citep{murraydemott99}:
\begin{equation}
    U_2 (r,\theta, \phi, t) = r^2 \omega^2 e \big[-\frac{3}{2}P_{2}^{0}(\lambda) \cos \omega t +  P_{2}^{2}(\lambda)(\frac{3}{4} \cos \omega t \cos 2\phi +\sin \omega t \sin 2 \phi) \big],
\end{equation}
where $r$ is the radius, $\theta$ is the co-latitude, $\phi$ is the longitude, $t$ is time, $\omega$ is the frequency of the forcing tide equal to the orbital frequency 53.07 $\mu \rm Rad.s^{-1}$, $e$ is the orbital eccentricity equal to 0.0047, $P_{2}^0$ and $P_{2}^2$ are the un-normalized associated Legendre polynomials, and $\lambda = \cos \theta$. For a spherically symmetric body, the tidal deformations in radial and tangential directions are described by the \textcolor{black}{degree 2} displacement Love numbers $h_2$ and $l_2$ \citep{melchior83}:
\begin{align}
 &   u_r(r,\theta, \phi, t) = \frac{ h_2(r)}{g(r)} U_2 (r,\theta, \phi, t),~ \notag \\
 &   u_{\theta}(r,\theta, \phi, t) = \frac{ l_2(r)}{g(r)} \frac{\partial  U_2 (r,\theta, \phi, t)}{\partial \theta},~ \notag \\
 &   u_{\phi}(r,\theta, \phi, t) = \frac{1}{\sin \theta} \frac{ l_2(r)}{g(r)} \frac{\partial U_2(r,\theta, \phi, t) }{\partial \phi},
\end{align} 
where $g(r)$ is the gravitational acceleration at radius $r$. The resulting perturbation in the gravitational potential field of Enceladus ($W_2$) is calculated through gravitational tidal Love number $k_2$ as:
\begin{equation}
   W_2(r,\theta, \phi, t) = k_2(r) U_2 (r,\theta, \phi, t).
\end{equation}
$k_2$, $h_2$, and $l_2$ are sensitive to the global scale properties of the body and primarily depend on its density and rheological structure. The maximum displacements in the radial and horizontal directions are expressed as \citep{park_etal20}:
\begin{align}\label{eq:deltaD}
 &   \max (\Delta R) = 12 \sqrt{2/7} \frac{|h_2| R_E^2 \omega^2 e}{g} , \notag \\
 &   \max( \Delta E) = \frac{12 |l_2| R_E^2 \omega^2 e}{g},  \notag \\
 &   \max (\Delta N) = \frac{9|l_2| R_E^2 \omega^2 e}{g},
\end{align}
where $R_E$ is the radius of Enceladus, and $\Delta R$,  $\Delta E$, and  $\Delta N$ are the radial, easterly, and northerly displacements on the surface, respectively. 

\textcolor{black}{In general, for a body with lateral variations in structure, the spherical harmonic expansion of the tidal response depends on both the degree (here $l=2$) and the order $m$. Thus,
whereas for a spherically symmetric body, $k_{22} = k_{20}= k_2$ and $h_{22} = h_{20}=h_2$, for a non-spherically symmetric body these quantities can be different \textcolor{black}{from each other}. In this case, $k_{2m} = W_{2m}(r,\theta,\phi,t)/U_{2m}(r,\theta,\phi,t)$, and $h_{2m} = g H_{2m}(r,\theta,\phi,t)/U_{2m}(r,\theta,\phi,t)$, where $W_{2m}$ and $H_{2m}$ are the spherical harmonic expansion terms of the potential and surface displacement (assuming l=2 forcing only).}

Because of the viscoelastic behavior of materials, the tidal Love numbers are complex values, implying that there is a phase lag between the forcing potential and the tidal response. \textcolor{black}{The total heat generated by tides can be directly computed from the complex $k_2$ \citep{pealeCassen78, segatz_etal88}}:
\begin{equation}\label{tidalheatingEquation}
    \dot E = \frac{21}{2} \frac{n^5 R_E^5}{G} e^2 \frac{|k_2|}{Q},
\end{equation}
   where $\dot E$ is the total power generated by tides, $n$ is the orbital mean motion,  $G$ is the gravitational constant, and $Q$ is the tidal quality factor given by \citep[][]{goldreichsari66}:
\begin{equation}
        Q^{-1} = \sin (|\epsilon_{k_2}|).
\end{equation}
\textcolor{black}{Here, $\epsilon_{k_2}$ is the phase lag of the perturbation in the gravitational potential.} 
In most cases, the tidal quality factor $Q$ is simply defined as $\tan(\epsilon_{k_2})$. However, this simplification is only correct when the phase lag is small such that $\tan(\epsilon_{k_2})\approx \sin(\epsilon_{k_2})$. For bodies like Enceladus  that may experience intense tidal heating and  high tidal phase lag, this approximation is no longer valid \citep{efroimsky12tidal}. 
Here, we explicitly consider phase lags, as opposed to the quality factor $Q$.

\noindent

\subsection{Viscoelastic Model}
To predict the tidal response of Enceladus, we must consider rheological models that \textcolor{black}{quantify} shear dissipation in rock and icy regions. 
When both viscous and elastic mechanisms are active, the viscoelastic response depends on the shear modulus $\mu$, frequency of excitation $\omega$, and the effective viscosity $\eta$. The Maxwell viscoelastic model provides a reasonable description of a material's behavior at timescales that are significantly shorter or longer than the Maxwell (relaxation) time of the body defined as $\tau_M = \eta/\mu$.
$\tau_M$ represents the approximate timescale at which the behavior transitions from being dominated by the elastic regime to the viscous regime. 
However, a Maxwell model may not be appropriate for describing tidal dissipation in planetary bodies that have tidal periods ranging from hours to a few days, which may be near the relaxation time of the body. Applicability of Maxwell's model for capturing tidal dissipation has been questioned for multiple planetary bodies \citep{bills_etal05, castillorogez_etal11, bagheri_etal19, petricca_etal24}.
Several models based on micro-scale physical processes and laboratory measurements have been proposed to mimic the behavior of a material in time scales between the elastic and viscous regimes \citep[e.g.,][]{castillorogez_etal11, bagheri_etal19, bagheri_etal21, petricca_etal24, tobie_etal25}.
In particular, the Andrade model has been \textcolor{black}{favored} to model tidal dissipation in celestial bodies taking into account the effect of anelastic behavior in the transition from dominantly elastic to dominantly viscous regimes \citep{castillorogez_etal11, RenaudHenning18, padovan_etal14, bagheri_etal19, bagheri_etal22}. 
A detailed discussion of viscoelastic models can be found in \cite{JacksonFaul10, FaulJAckson15, bagheri_etal19}. 
We consider both Maxwell and Andrade viscoelastic models because of the potential proximity of the relaxation time of the ice shell to Enceladus's orbital period \textcolor{black}{(see Appendix~\ref{sec:viscoelasticmodels})}, but most of our results assume the Andrade rheology.

\textcolor{black}{
\subsection{Partitioning of tidal heating between the shell and the core}\label{sec:beutheestimate}}

The total tidal energy dissipated in the body can be computed by measuring the amplitude and phase lag of $k_2$, independent of the choice of rheological model (see equation~\eqref{tidalheatingEquation}). 
We can compute the tidal heating in a specific region of the satellite, such as the core or the shell, by integrating the radial shear dissipation in that part of the body. 
\textcolor{black}{Thus, if we compute the contribution of each of the layers to $|k_2|/Q$ or identically to $\Im(k_2)$ of the body, that is, evaluating at each interface, we can compute the associated tidal heating for the part of the body that is interior to the interface. Here, we use equations (33), (35), and (37) in \citet{tobie_etal05} to derive the contribution of the core and the shell to the global tidal heating:}


\begin{equation}\label{corediss}
    \dot E_{core} = -\frac{21 \pi}{2}  \omega^5 e^2 R_E^4 \frac{R_{core}}{\pi G}\Im{(\hat k_2^{core}}).
\end{equation}
Here, $\hat k_2^{core}$ is the piecewise gravitational Love number associated with the core.  For a homogeneous core, $\hat k_2^{core}$ can be computed from the deformation at the top of the core \textcolor{black}{($\Delta R_{core}$)}:
\begin{equation}\label{corelove}
  \Im{(\hat k_2^{core}}) =  \frac{4\pi G\rho_{core}R_{core}^4}{5R_E^3} \Im{(\Delta R_{core})^4}.
\end{equation}
\textcolor{black}{Details of this derivation are presented in Appendix \ref{sec:heatpartioningderivation}. 
The tidal heating in the ice shell is then calculated by subtracting the contribution of the core from the total energy dissipated:
\begin{equation}\label{shelldiss}
    \dot E_{shell} = \dot E - \dot E_{core}.
\end{equation}
Tidal dissipation in the ocean is assumed negligible \citep{hayMatsuyama17, matsuyama_etal18}}.
Equations~\eqref{corediss} and~\eqref{shelldiss} provide a method to calculate the partitioning of the global tidal heating for a given interior structure model. 
Thus, these equations require \textit{a priori} assumptions or independent constraints on the rheological and structural properties to compute the shear dissipation in the core. 
Here, we derive analytical expressions for the partitioning of tidal heating on the basis of the observable quantities, i.e., the tidal Love numbers and the density structure of the body. 
We extend the formulation provided by \cite{beuthe19} which uses membrane mechanics to obtain the contribution of each layer to the imaginary part of the tidal Love number. The shell's contribution can be written as \citep{beuthe19}:

\begin{equation}\label{beuthegeneral}
    \Im{(\hat k_2^{shell})} = \left| \frac{k_2+1}{k_2^0 + 1} \right|^2\Im{(k_2^0)} - \xi_2 |h_2|^2 \Im{(\Lambda_T)},
\end{equation}
where, following the terminology used in \citet{beuthe19}, $\xi_2 = 3\rho_{ocean}/5\rho_E$, $\rho_E$ is the density of the whole body, $k_2$ and $h_2$ are the tidal Love numbers of the whole body, $k_2^0$ is the fluid-crust gravitational Love number (i.e., assuming the crust is fluid), and $\Lambda_T$ is the shell spring constant introduced by \citet{beuthe19}. \textcolor{black}{In Appendix \ref{sec:heatpartioningderivation}, we demonstrate that the tidal dissipation occurring in the shell can be expressed as:}
\begin{equation}\label{finalheatinshellwithlove}
 \dot E_{\text{shell}} = \frac{105}{10} \frac{(\omega R)^5}{G} \xi_2 \,|h_2|^2  \bigg[ \Im \bigg( {\frac{1}{h_2} \big( k_2+1 + 5 \frac{\delta \rho}{\rho} \epsilon \big)} \bigg)\bigg] \,  e^2.
\end{equation}
Here, $\delta \rho$ is the density difference between the ocean and the crust and $\epsilon$ is the ratio of the thickness of the crust to Enceladus's radius. Equation~\eqref{finalheatinshellwithlove} expresses the tidal heating in the shell using measurements of the \textcolor{black}{amplitudes and phases} of the Love numbers ($k_2$, $h_2$, $\epsilon_{k_2}$, $\epsilon_{h_2}$), and information on the density structure. 
The densities of the layers can be constrained by the static gravity field and libration, and are easier to constrain compared to the rheology of the layers required in Equation~\eqref{beuthegeneral}. \textcolor{black}{Based on Equation~\eqref{finalheatinshellwithlove}, the uncertainty in the magnitude of tidal heating occurring in the shell can be computed from the uncertainties on the density structure and the complex Love numbers.}
The tidal heating in the core can be computed by subtracting the tidal dissipation in the shell from the total heat. This derivation is general and can be used for other icy satellites for which the necessary geodetic observations are or will become available.

\subsection{Physical Libration}\label{sec:introlibrationgravity}

Owing to Enceladus's moderate orbital eccentricity and slightly elongated shape, it is subject to periodic torques that force harmonic oscillations in its orientation, called physical librations. 
This libration is superimposed on the overall synchronous orbit of the moon around Saturn. 
The amplitude of physical libration depends mainly on the moments of inertia of the body, and whether the outer surface layer, i.e., the ice shell, is coupled to the deep interior \textcolor{black}{\citep[e.g.,][]{thomas_etal16}, and to a lesser} extent to the rigidity of the shell.
The estimated libration of Enceladus based on the analysis of \textit{Cassini} observations is too large to be explained by an ice shell that is mechanically coupled to the core, implying a global ocean \citep{thomas_etal16}.
Recent work by \cite{park_etal24} suggests a slightly lower amplitude of libration, although still higher than expected if the ice shell and the core are coupled.
\textcolor{black}{We note that presence of a frictionless interface, such as found in soft-bedded glacier systems on Earth \citep{johnstonMontesi17} might also decouple the shell from the core. While the scenario with no global ocean cannot be ruled out by the existing constraints, here we assume the presence of a global subsurface ocean.}

Tides can influence the amplitude of libration by various mechanisms: modifying the gravitational torque exerted by the planet on the satellite; gravitational coupling of the libration of the ice shell to the solid interior through the torque between the periodic tidal bulge of one solid layer and the permanent shape of another; and the slight periodic modification of the polar moment of inertia by the zonal tides.
The amplitude of libration of the outer ice shell is given by \citep{vanhoolst_etal13}:
\begin{equation}\label{eq:libamplitude}
\gamma = \frac{ 1 }{C_c C_s} \frac{4e [ K_3 K_5 + K_{2}K_{6} - n^2C_c K_3]}{(n^2-\sigma_1^2)(n^2-\sigma_2^2)},
\end{equation}
where $\gamma$ is the amplitudes of libration of the shell, $K_1$, $K_2$, $K_3$, $K_4$, $K_5$ and $K_6$ describe the torques between the planet, the shell, and the core, $C_c$ and $C_s$ are the polar moments of inertia of the core and the shell, $\mathcal{M}$ is the mean anomaly, and 

\begin{equation}\label{eq:naturalomega}
\sigma_{1,2}^2 = \frac{K_1C_c + K_5C_s \pm \left[4(K_2K_4 - K_1K_5)C_cC_s + (K_1C_c + K_5C_s)^2\right]^{1/2}}{2C_cC_s}.
\end{equation}
$\sigma_{1,2}$ represents the natural frequencies of free libration of the body \citep{vanhoolst_etal13}. 
To compute the flattening of each of the layers, as required for the calculation of the torques between layers in equations (\ref{eq:libamplitude}, \ref{eq:naturalomega}), we use the approach of \citet{tricarico14} that provides hydrostatic equilibrium figures for moons in eccentric orbits, along with the global shape from the best-fitting ellipsoid model suggested by \citet{park_etal24}. 

\subsection{Static Gravity Field}\label{sec:introgravity}
Current gravity data for Enceladus is limited to that obtained from three \textit{Cassini} fly-bys. Therefore, we are limited to gravity field coefficients of spherical harmonic degrees $l=2,3$.
To further constrain the interior structure, we use the moment of inertia and the $l = 3$ zonal potential coefficients ($J_3$). \textcolor{black}{Direct measurement of the moment of inertia of Enceladus is only possible by measuring the body's precession. Alternatively, \cite{hemingway_etal18} propose to derive the moment of inertia from long-wavelength $l=2$ gravity-topography admittance measurements ($Z_{22}$ and $Z_{20}$). See \citet{McKinnon15}, \citet{hemingway_etal18}, and \citet{hemingwayMittal19} for details.} \textcolor{black}{This approach requires underlying assumptions about the non-hydrostatic component of the gravity field and the compensation mechanism, as discussed in \citet{hemingwayMittal19}.}
Here, we only consider the moment of inertia when we explore the potential impact of future observations.
To compute the zonal non-hydrostatic degree-3 static gravity coefficient ($J_3$), we follow an approach similar to that described in \citet{Genova_etal24}. \textcolor{black}{The gravity-topography admittance is computed as \citep{turcotte_etal81, hemingway_Matsuyama17}}:
\begin{equation}
    Z_l = 4 \pi G  \bigg(\frac{l+1}{2l+1}\bigg)  \rho_{\rm shell} \Bigg[ 1 - \bigg(1-\frac{D_{shell}}{R_E} \bigg)^{l+2} \bigg(\frac{g_t}{g_b} \bigg) C_l^0  \Bigg].
\end{equation}
Here, $D_{shell}$ is the mean shell thickness, $C_l^0$ is a compensation factor computed based on the effective elastic thickness $D_e$ \citep{turcotte_etal81}, and 
$g_t$ and $g_b$ denote the gravitational acceleration at the top and bottom of the shell:
\begin{equation*}
    g_t = \frac{GM}{R_E^2},\\
    \end{equation*}
    \begin{equation}
    g_b = \frac{GM}{(R_E - D_{\rm shell})^2} - \frac{4}{3}\pi G \rho_{\rm shell} \frac{R_E^3 - (R_E-D_{\rm shell})^3}{(R_E-D_{\rm shell})^2}.
\end{equation}
$J_3$ is computed as:
\begin{equation}
J_3 = \frac{Z_3 R_E^2H_{30}^{n\mathrm{hyd}}}{4GM}.
\end{equation}
$H_{30}^{nhyd}$ is the non-hydrostatic topography coefficient. \textcolor{black}{We compute $H_{30}^{nhyd}$  by using the formulation of \citet{tricarico14} to calculate the hydrostatic shape coefficients, and subtracting them from the observed shape coefficients of \citet{park_etal24}.}  This formulation implies an unconstrained degree of compensation.
Following \citet{Genova_etal24}, we assume Airy compensation, implying $C_l^0 = 1$. This assumption does not affect the uncertainties we derive for the observations and model parameters.

\begin{figure}[ht] 
\begin{center}
\vspace{5mm}
\includegraphics[width=.7\textwidth]{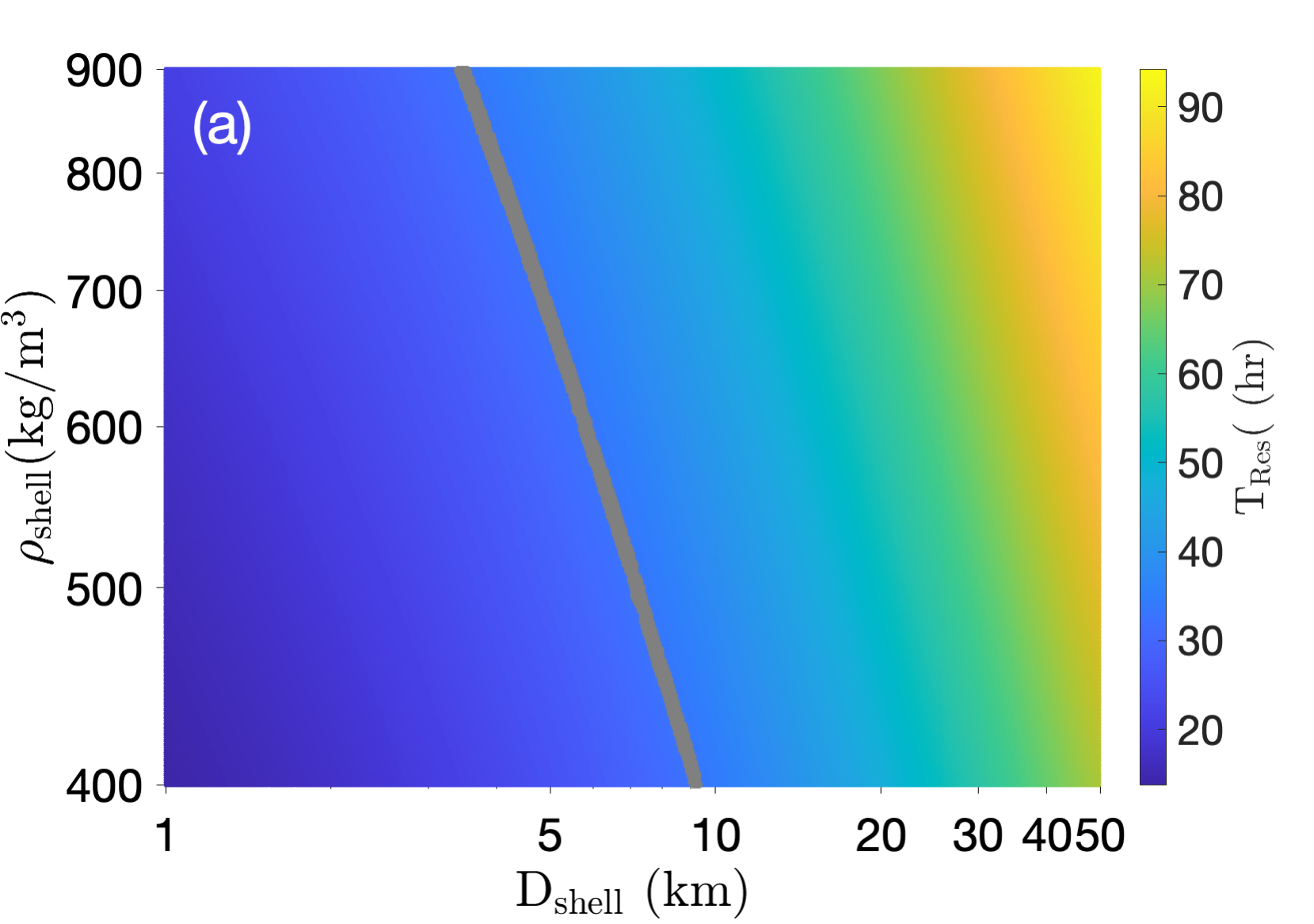}
\includegraphics[width=.7\textwidth]{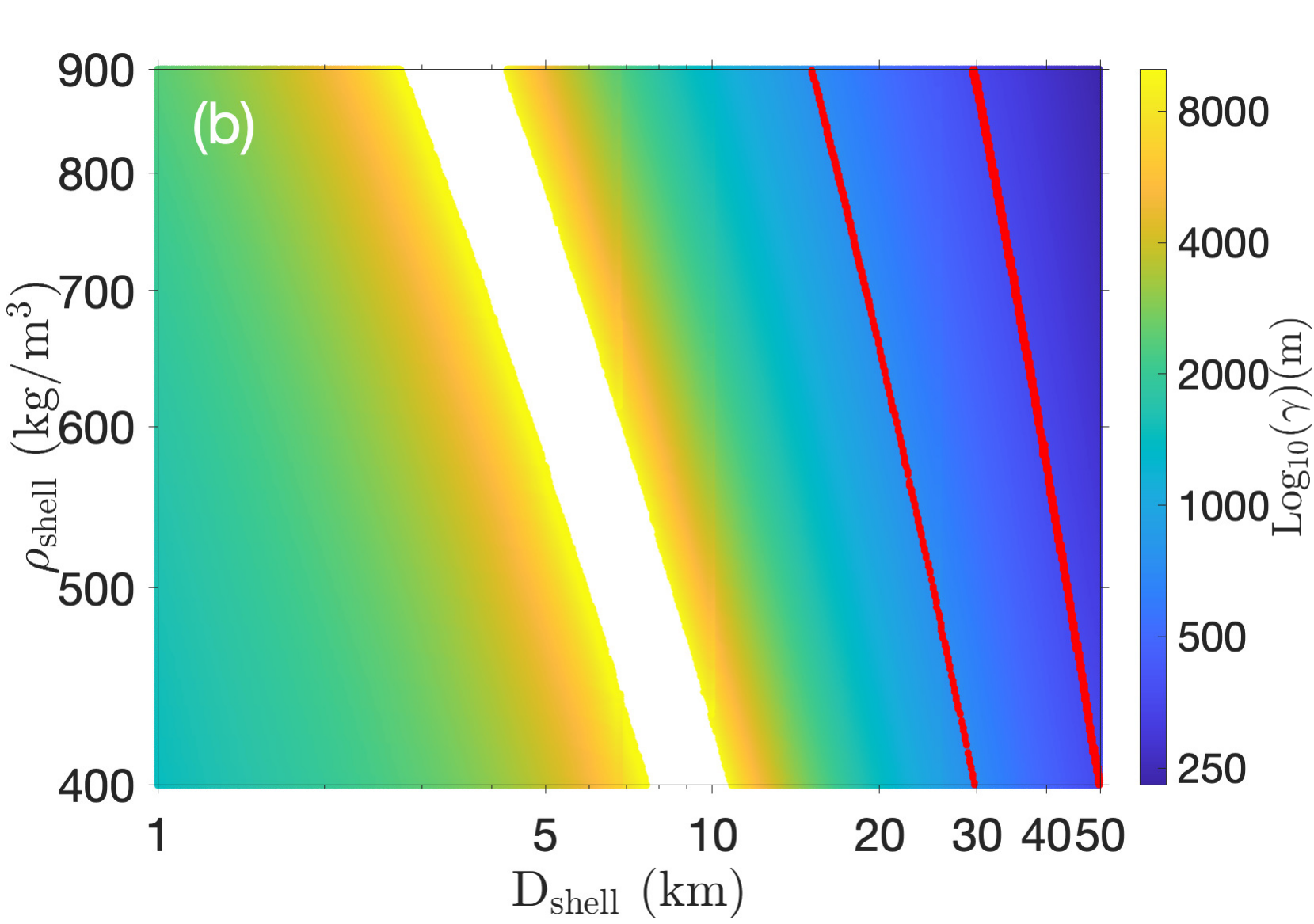}
\caption{Variations of (a) Period of natural (free) libration and (b) amplitude of forced libration of Enceladus as a function of shell thickness and density. The orbital period of Enceladus is shown by gray line in (a). The end-member values of amplitude of libration inferred by \citet{thomas_etal16, nadezhdina_etal16, park_etal24} (370~m$<\gamma<750$~m) are shown by red lines in (b). The results are presented for $\rho_{\rm ocean}=1000~\rm kg/m^3$, $R_{\rm core}=197$~km, $\rho_{\rm shell}=900~\rm~kg/m^3$, $\mu_{\rm shell}=3$~GPa, $\kappa_{\rm shell}=10$~GPa, $\eta_{shell} = 6\times10^{13}$~Pa.s, $\rho_{\rm core}=2320~\rm~kg/m^3$,  $\mu_{\rm core}=40$~GPa, $\kappa_{\rm core}=100$~GPa, $\eta_{core} = 10^{19}$~Pa.s. The horizontal and vertical axes and the libration amplitude color scale are logarithmic.} 
\label{figure:librationnatural}
\end{center}
\end{figure}

\textcolor{black}{
\section{Sensitivity of Physical libration to shell properties}\label{sec:shellthicknesslibration}}
\textcolor{black}{Enceladus's measured physical libration amplitude is one of the key observations to constrain its interior. In general, the librational response of an icy ocean world is \textcolor{black}{mostly dependent on} the inertial, elastic forces, and, in the sufficiently large moons, on the gravitational coupling between the interior layers and shell. Enceladus's ice shell thickness is not negligible compared to the radius}. Therefore, it is expected that libration amplitude
mainly depends on the moment of inertia of the shell, determined by the shell thickness and density \citep[e.g.,][]{hemingwayNimmo_24}. Rigidity of the shell can have a small effect. Using the formulation in Section~\ref{sec:introlibrationgravity}, we compute the libration amplitude and the natural libration frequency as a function of shell density and thickness -- the two parameters that determine the shell's moment of inertia (Figure~\ref{figure:librationnatural}).
We use a range for the amplitude of the libration of 360~m$<\gamma<$750~m, which spans the estimates of \citet{nadezhdina_etal16}, \citet{thomas_etal16}, and  \citet{park_etal24}.  
\textcolor{black}{To avoid unnecessary complexities, throughout this paper, we assume that all layers, i.e., shell, ocean, and core have uniform physical properties. We do not account for any temperature or pressure dependence of the layer properties.}

The sensitivity of the libration to the shell thickness is not uniform for different ranges of the ice shell properties~(Figure~\ref{figure:librationnatural}). \textcolor{black}{Variations of the libration amplitude with the shell properties near the natural libration frequency are very steep, resulting in a high sensitivity. The natural free libration period of the shell depends on the physical and mechanical properties of the shell (Figure~\ref{figure:librationnatural} (a), and equation~\ref{eq:naturalomega}). If the shell properties result in a natural period close to the orbital period of Enceladus, the resonant response will imply dramatically high amplitudes of libration (Panel (b) Figure~\ref{figure:librationnatural}). As a result of this behavior, if the mean thickness of the ice shell is less than 10~km, assuming shell densities in the range 400--900~kg/m$^3$}, high sensitivity to the libration amplitude is observed. The sensitivity of libration to the shell thickness significantly decreases when the orbital period is not near the natural libration period.
All currently available estimates of the physical libration of Enceladus are lower than those, suggesting a nearly resonant behavior that rules out a mean ice shell thickness of 5--10~km. 
\textcolor{black}{The amplitude of libration shows small sensitivity to the elastic response. We also expect low sensitivity to the shell viscosity (Appendix~\ref{sec:librationrigidity}). If the shell viscosity is near $\sim$12.5$\times$10$^{13}$~Pa.s, around the lower bound for the nominal relaxation viscosity of the ice, the effect of the viscoelastic deformation  of the shell is approximately 10\%. Only in an extreme case where the ice shell has a significantly low viscosity, i.e., dominated by a viscous behavior, the viscoelastic response of the shell can significantly affect the libration amplitude.
As expected, libration is largely insensitive to the response of the core because of the body's small size and therefore weak gravitational effect}. 


\section{End-member scenarios for tidal dissipation in Enceladus}\label{sec:heatgenerationclasses}

\begin{figure}
\centering
\hspace{8mm}\includegraphics[width=.51\textwidth]{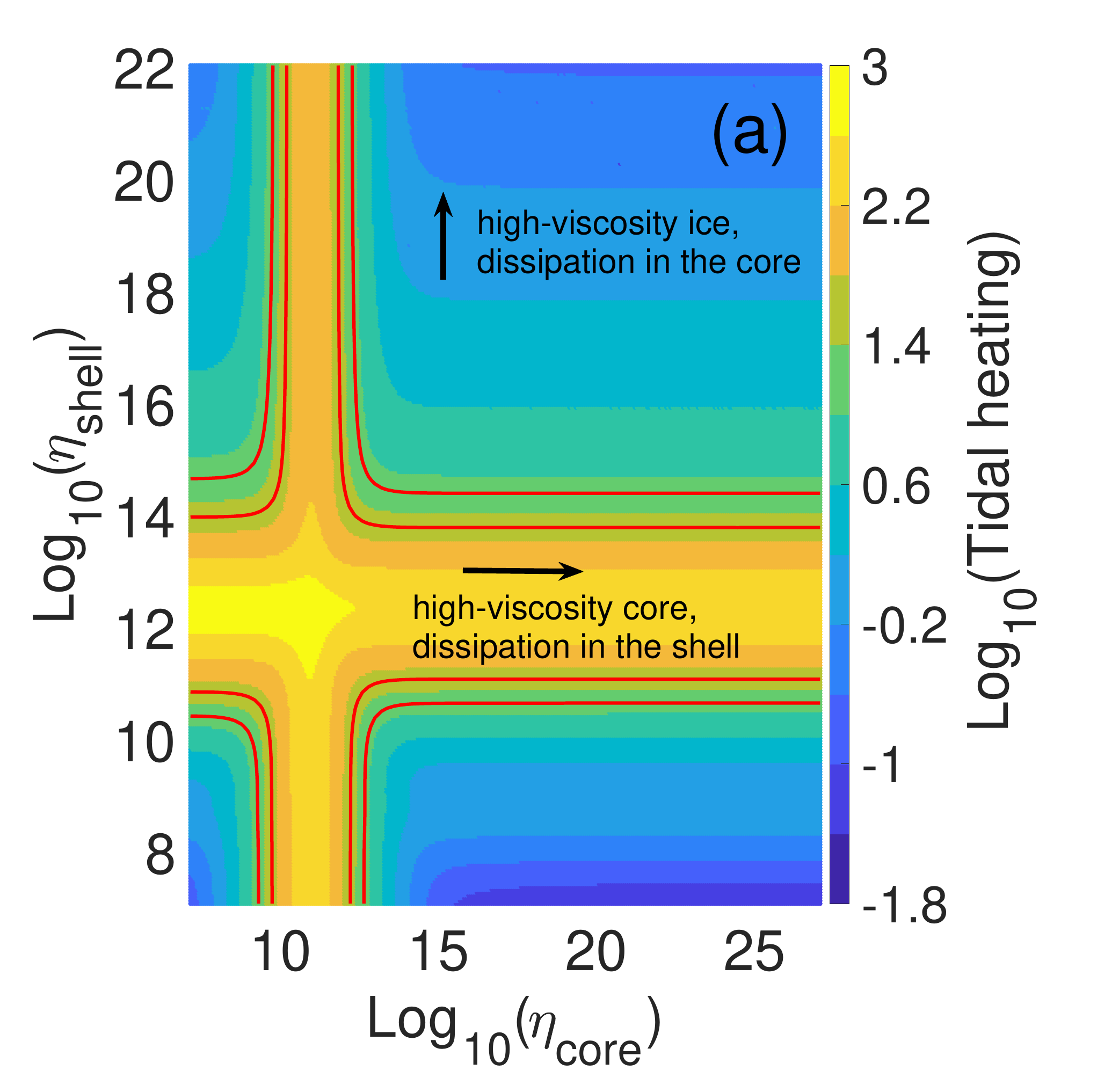}\
\includegraphics[width=.41\textwidth]{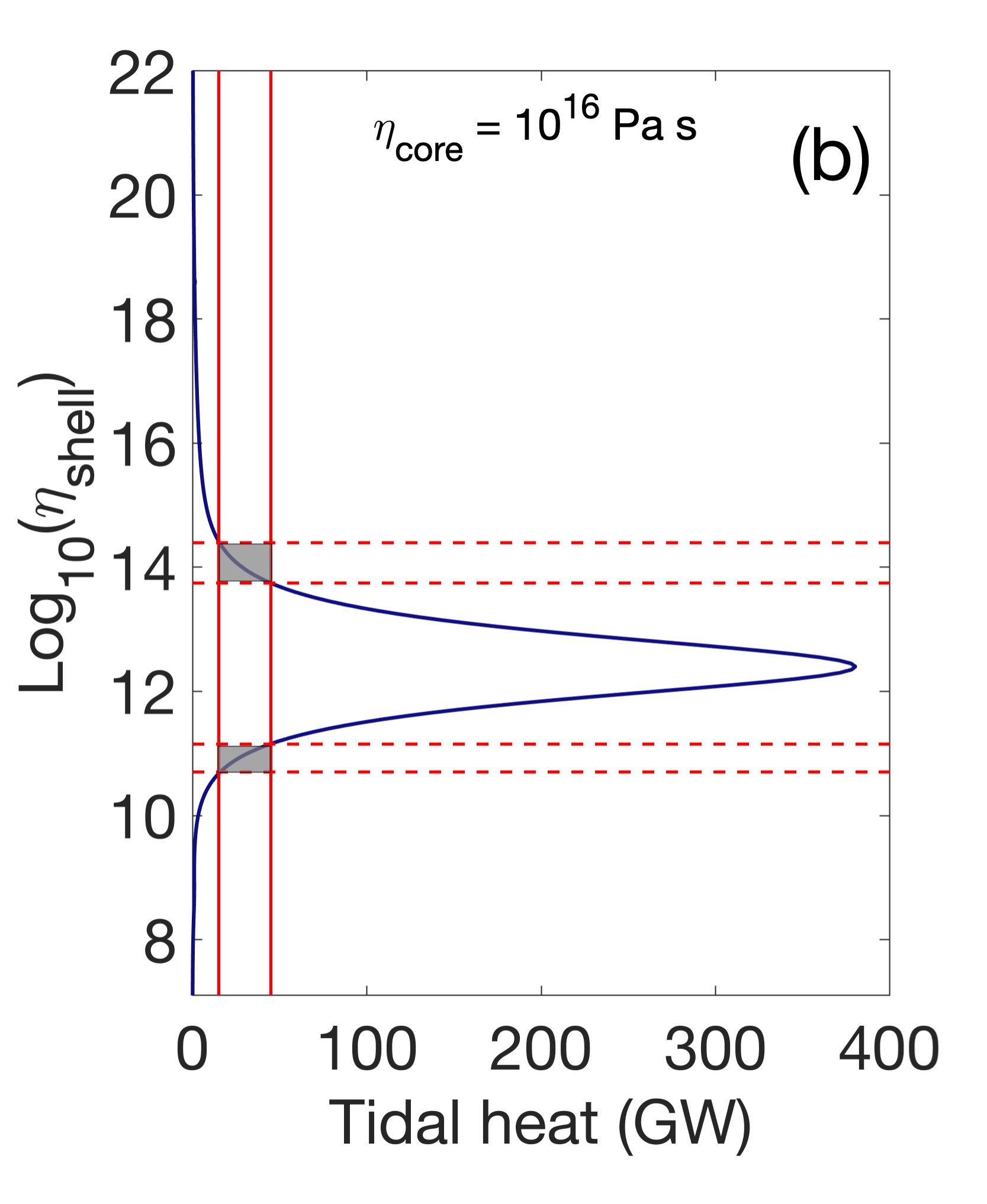}\\
\vspace{6mm}
\hspace{-6.8cm}\includegraphics[width=.4\textwidth]{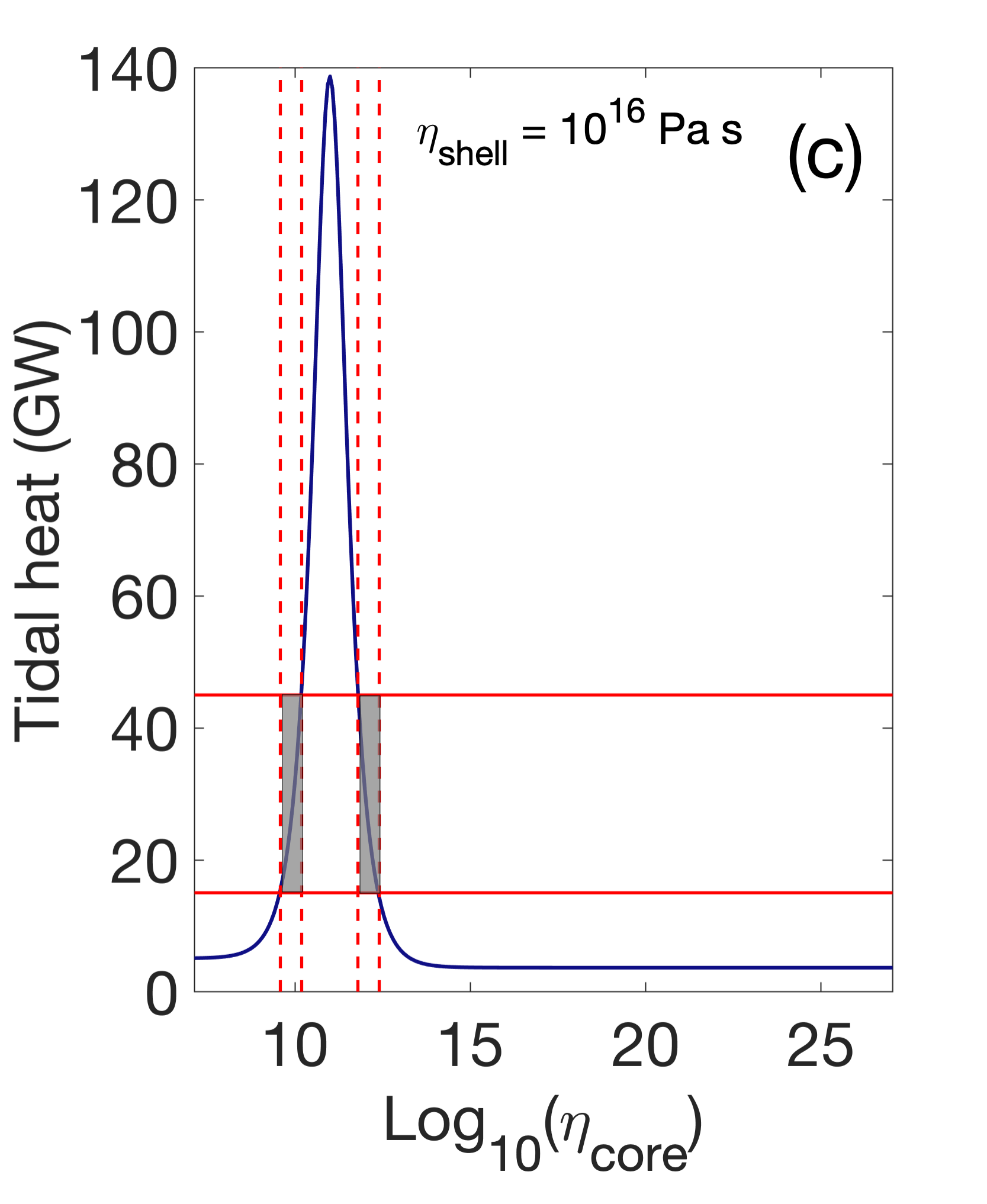}\\
\caption{(a) \textcolor{black}{Total} tidal heating (in GW) versus the effective viscosity of the ice shell and the core. The narrow regions bounded by the red lines indicate the range of heat loss that covers the estimates by \citet{howett_etal11, hemingwayMittal19, park_etal24}. (b) Tidal heating variation  with ice shell effective viscosity for a fixed $\eta_{core}=10^{20}$~Pa.s. (c) Tidal heating variation with core effective viscosity for a fixed $\eta_{shell}=10^{16}$~Pa.s. The gray areas highlight the region where tidal heating values are compatible with the estimated steady state heating. The results are presented for $D_{\rm shell}=14~$km, $D_{\rm ocean}=34$~km, $\rho_{\rm ocean}=1000~\rm kg/m^3$, $R_{\rm core}=204$~km, $\rho_{\rm shell}=930~\rm~kg/m^3$, $\mu_{\rm shell}=3$~GPa, $\kappa_{\rm shell}=10$~GPa, $\rho_{\rm core}=2175~\rm~kg/m^3$, $\kappa_{\rm core}=20$~GPa, $\mu_{\rm core}=15$~GPa. 
We assume an Andrade viscoelastic rheology.}
 \label{figure:heatviscs}
\end{figure}

\textcolor{black}{In this section, we discuss the two main categories of tidal dissipation and the ranges of the effective viscosities of the shell and core that are consistent with the inferred heating rate.}
The viscosity of consolidated rock is higher than $10^{18}$~Pa.s  \citep[e.g.,][]{hirthKohlstedt03}.
However, unconsolidated material, or even a mixture of rock and ice, may have a significantly lower effective viscosity. \textcolor{black}{ \citet{choblet_etal17} propose a range of low core viscosities that allows for substantial tidal heating in the core of Enceladus.  
Alternatively, tidal dissipation in the icy crust may explain the tidal heating the effective shell viscosity is sufficiently low, i.e., if the temperature of the ice is sufficiently high.}
The estimated tidal heating for broad ranges of shell and core viscosities is shown in Figure~\ref{figure:heatviscs}(a). \textcolor{black}{We assume that the total heat loss is  fully balanced by the  generated tidal heat computed from equation~\eqref{tidalheatingEquation}. In other words, we assume that Enceladus is currently in a thermal steady state.
We assume a tidal heating rate in the range of 15--40~GW that covers the estimates by \citet{howett_etal11, hemingwayMittal19, nimmo_etal23, park_etal24}.} 
\textcolor{black}{Each of }the four narrow regions determined by the red lines \textcolor{black}{in Figure~\ref{figure:heatviscs}(a)} indicate the ranges of the estimated tidal heating in Enceladus and the associated effective viscosities of the shell and the core. We show the variation in tidal heating with the viscosity of the shell (core) in Figure~\ref{figure:heatviscs} panels (b)~and~(c) for a fixed value of core (shell) viscosity. The dissipation peaks at certain viscosities of both the shell and core, at which the associated Maxwell relaxation time is close to the period of excitations and the tidal response transitions from anelasticity-dominated behavior to viscous-dominated behavior. Therefore, the estimated tidal dissipation rate is compatible with two regions around the peak dissipation (highlighted in gray). For the interior structure assumed here, we find that an effective core viscosity 
O(10$^{13}$)~Pa.s can satisfy the total estimated heat, consistent with \citet{choblet_etal17}.

\begin{figure}[ht]
\centering
\includegraphics[width=.3\textwidth]{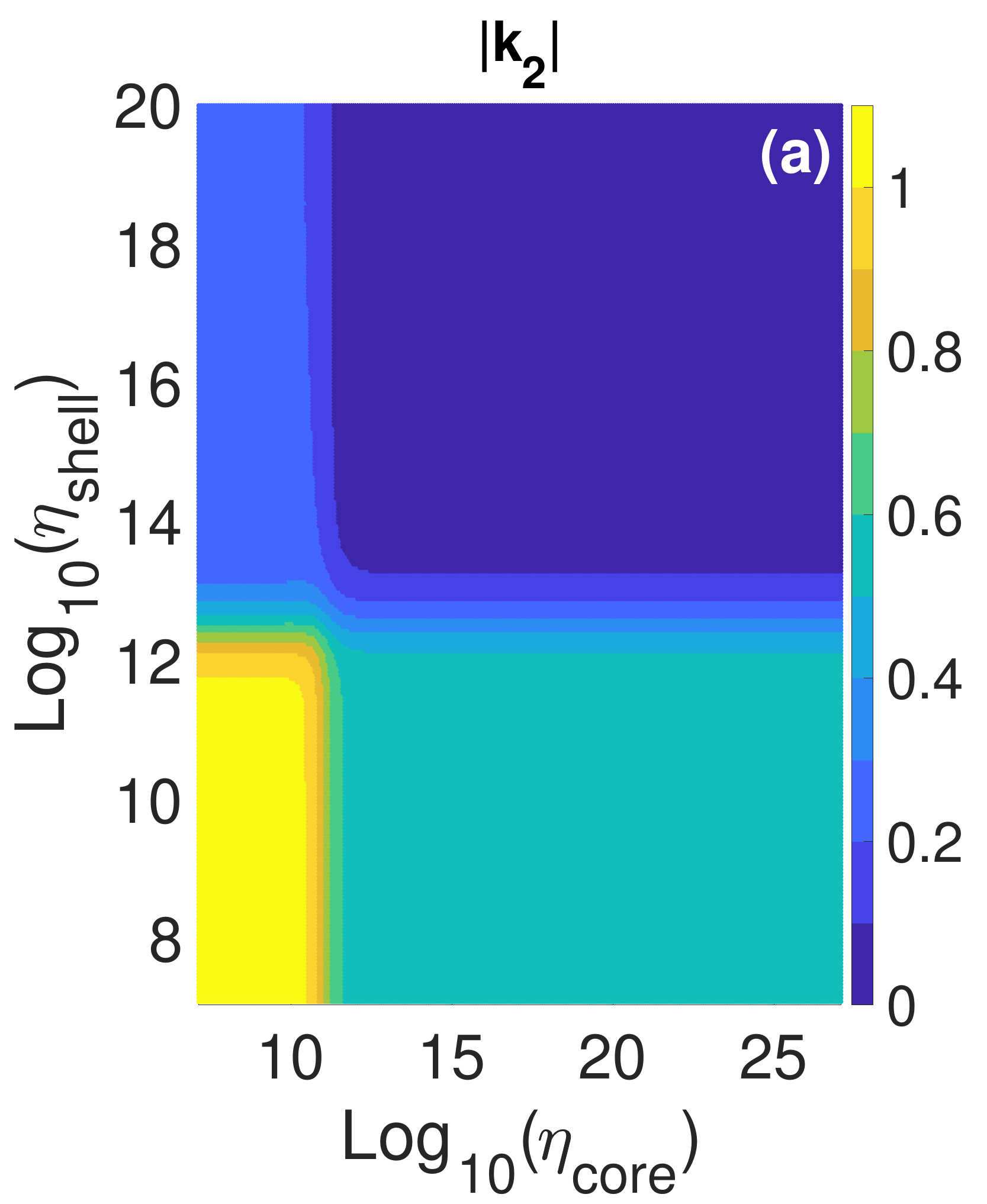}\
\includegraphics[width=.3\textwidth]{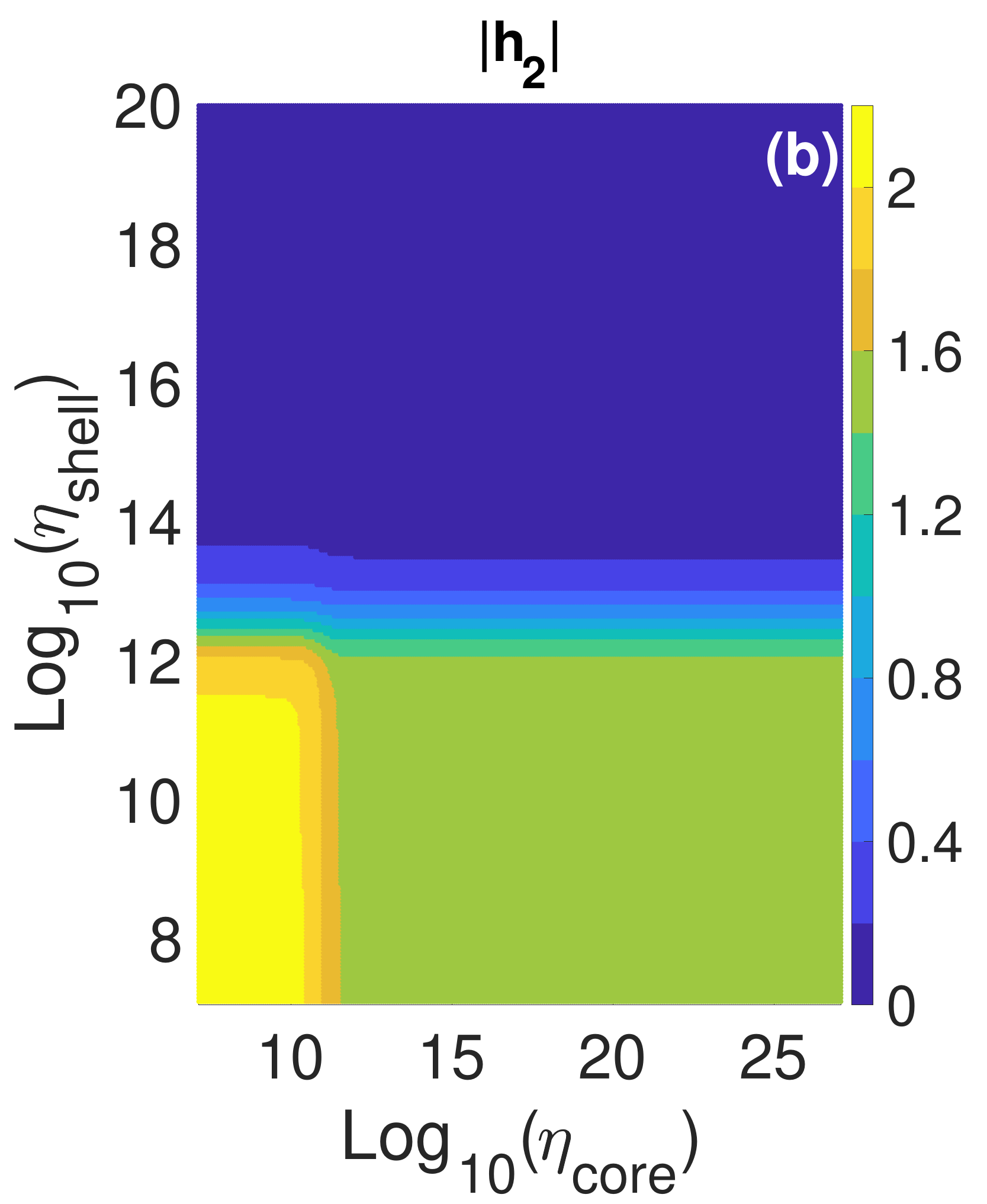}\
\includegraphics[width=.3\textwidth]{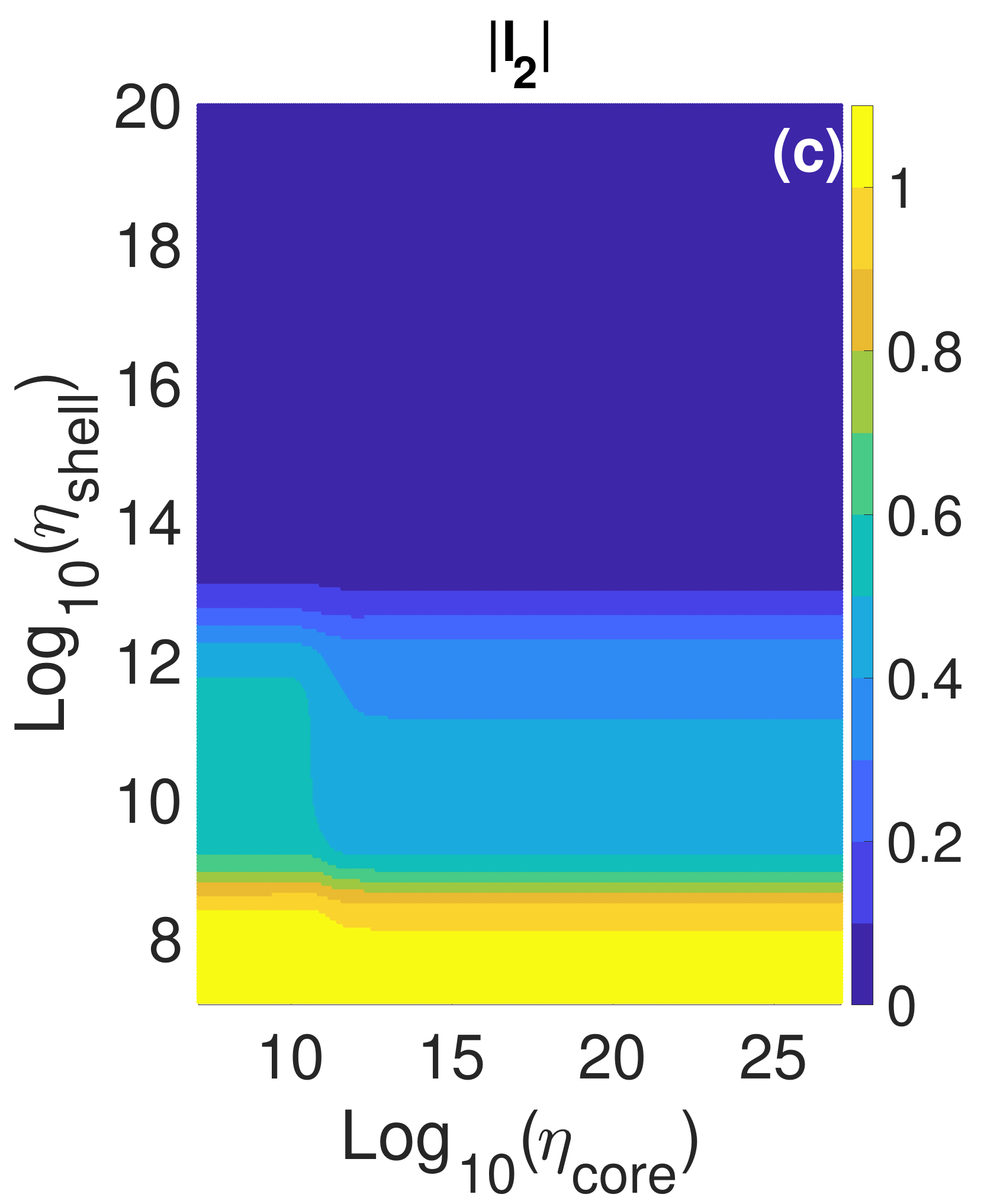}\\
\vspace{2mm}
\includegraphics[width=.3\textwidth]{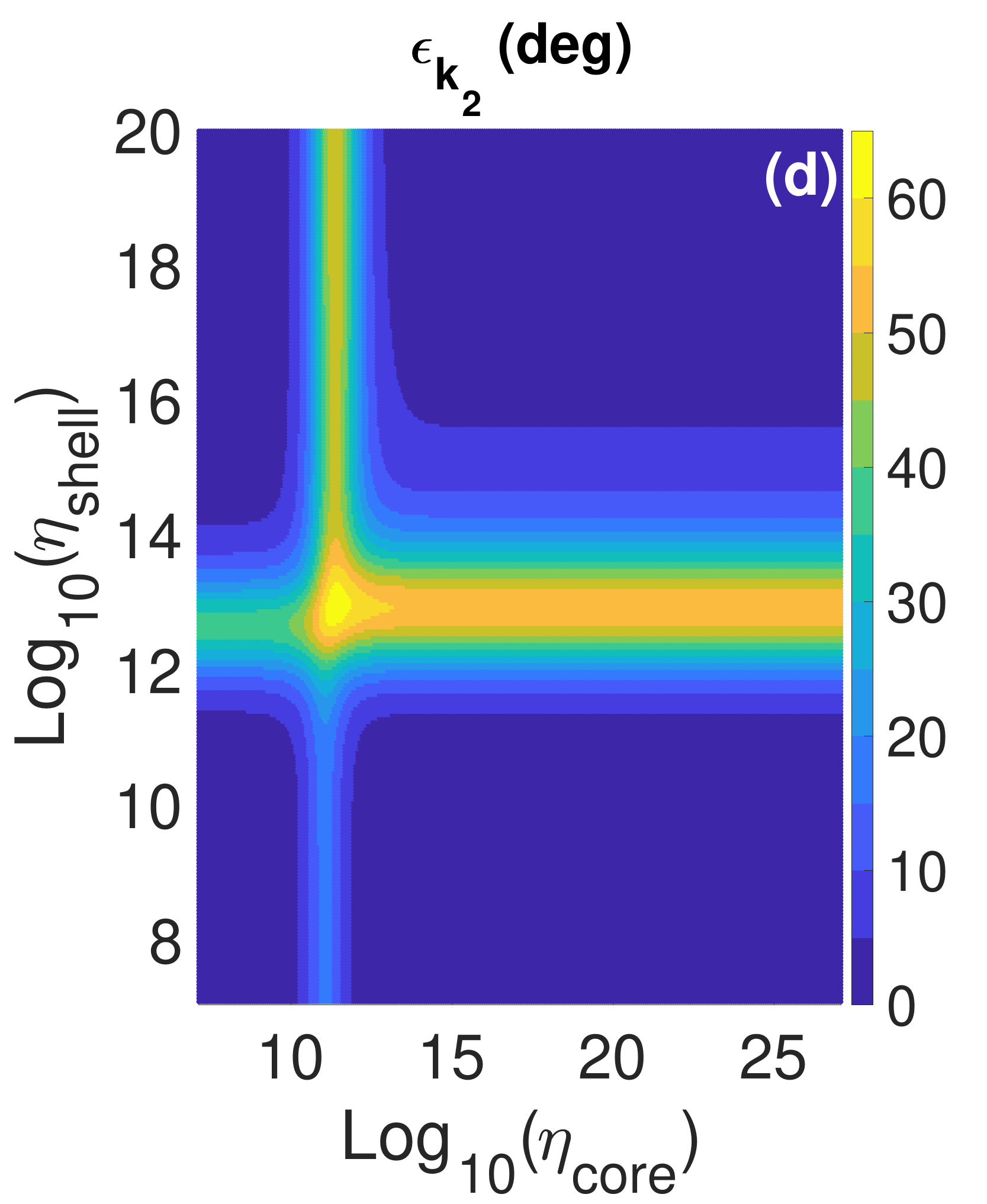}\
\includegraphics[width=.3\textwidth]{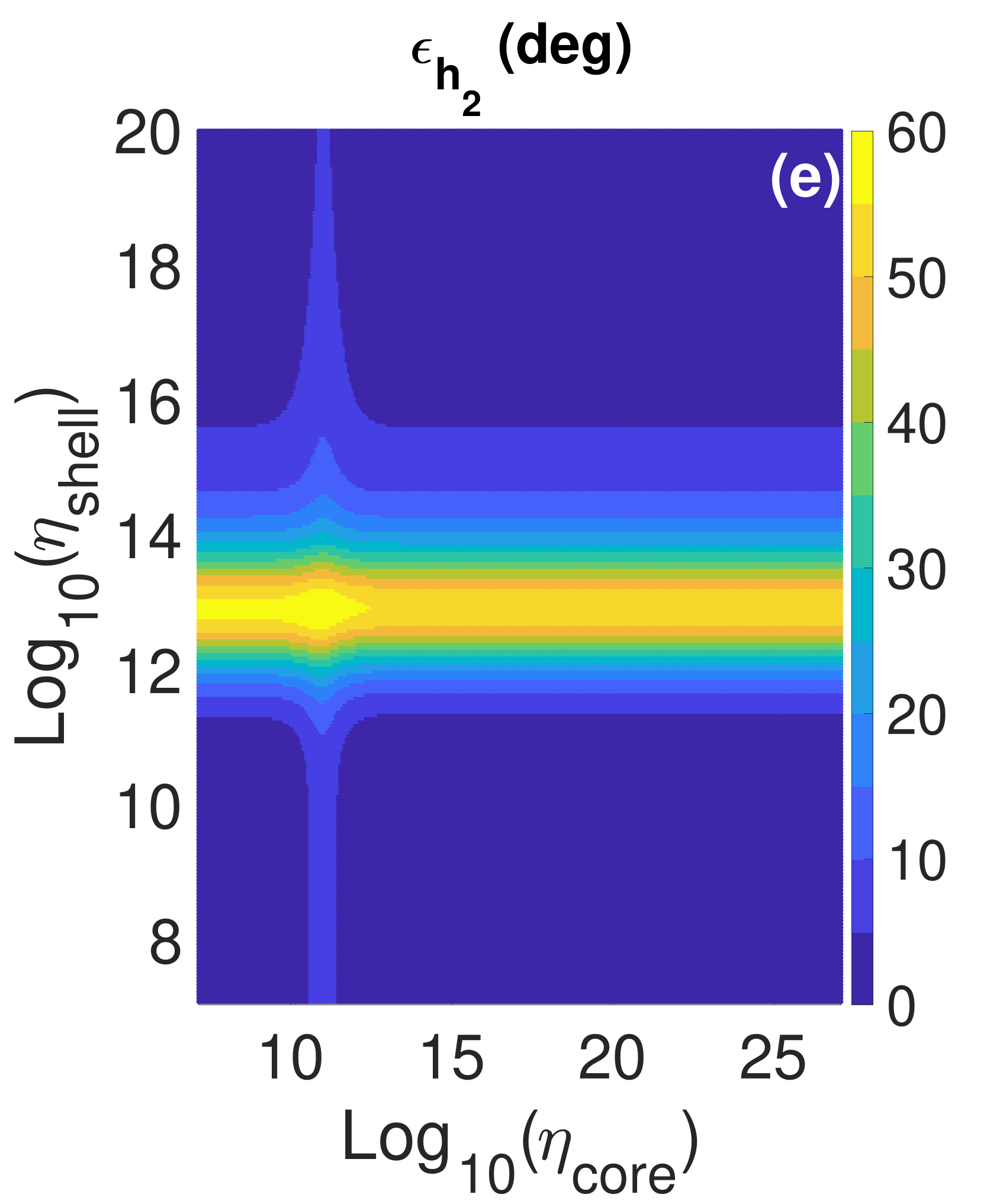}\
\includegraphics[width=.3\textwidth]{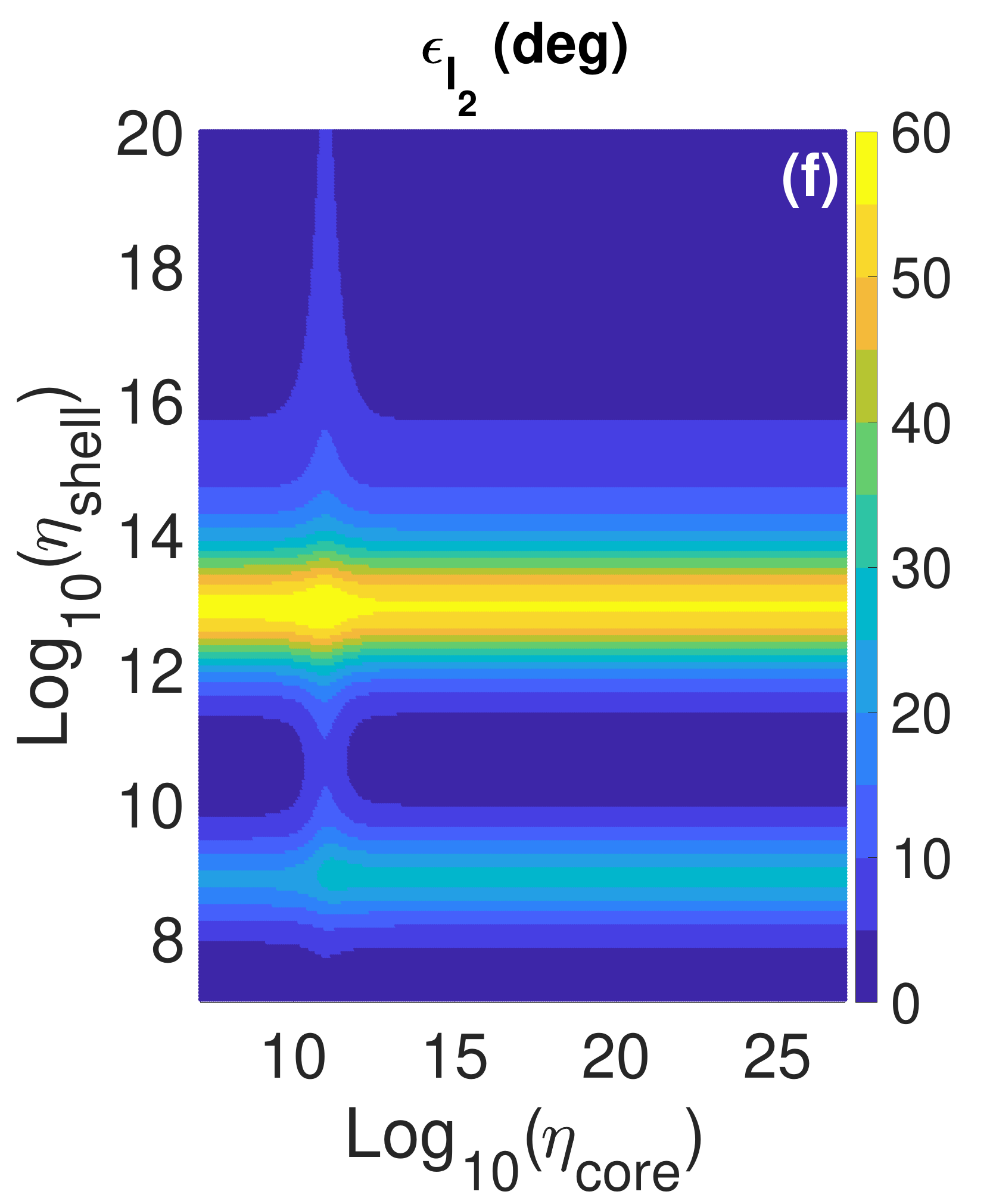}\\
\caption{\small{Tidal Love number amplitudes (a, b, c), \textcolor{black}{and} phase lags (d, e, f) for different shell and core viscosities. Same interior structure parameters as in Figure \ref{figure:heatviscs} are used.}} \label{figure:tidalresponseviscs}
\end{figure}

The two end-member scenarios manifest themselves differently in the amplitudes of $k_2$, $h_2$, and $l_2$ and their phase lags 
(Figure~\ref{figure:tidalresponseviscs}). 
When the effective viscosity of the rocky core is \textcolor{black}{$10^{10}~\rm Pa.s\lesssim\eta_{core}\lesssim10^{13}~\rm Pa.s.$,} deformation of the core will significantly affect the amplitude of ~$k_2$ and its phase lag~$\epsilon_{k_2}$ (Figure~\ref{figure:tidalresponseviscs}, panels~(a)~and~(d)). 
However, tidal deformation of the core does not substantially affect tidal displacement at the surface, resulting in only a small change in $h_2$. 
\textcolor{black}{i.e., less than 10\%. Tidal deformations in the core can have affect the surface deformations ($h_2$, $l_2$) if the viscosity of the shell is very low. This scenario is unlikely given the low temperatures \textcolor{black}{prevailing through most of the ice shell.}}

The compatibility of the \textcolor{black}{produced tidal heat with different regimes of tidal activity in the shell or in the core (Figures~\ref{figure:heatviscs} and~\ref{figure:tidalresponseviscs}) presents \textcolor{black}{complexities} in determining the distribution of tidal dissipation that requires additional constraints to resolve.}
In the case of Enceladus, and for most of the icy bodies where the ice shell is mechanically decoupled from the solid interior by an ocean, the ambiguity in the location of the tidal heating can be resolved by considering the tidal response at the surface and the associated phase lag. In the case of a mechanically coupled interior or tidally inactive interior layers, there is a strong correlation between the amplitude and phase lag of surface deformations (represented by $|h_2|$, $|l_2|$, $\epsilon_{h_2}$, $\epsilon_{l_2}$) and those of the observed gravitational signal \textcolor{black}{($k_2$ and $\epsilon_{k_2}$)}. 
A lack of such a correlation would imply that tidal activity in deeper interior layers which are decoupled from the surface---such as the core---is causing a perturbation in the body's gravitational field. 
If the ice shell is the primary locus of tidal heat generation, then it undergoes larger deformations and has a significant phase lag, a response that incorporate viscoelastic deformation, as opposed to the case where the deformations in the shell remain \textcolor{black}{purely} elastic. For a given interior structure model of Enceladus, the tidal deformations on the surface in the two cases differ by approximately a factor of two (Figure~\ref{figure:tidalresponseviscs}~panels b and c). The phase lags associated with the tidal displacements can differ by tens of degrees between the two cases. \textcolor{black}{Section~\ref{sec:roadmap} explores the details of determining the primary location of tidal dissipation.}

\textcolor{black}{
\citet{ermakov_etal21} and \citet{Genova_etal24} propose suites of geophysical measurements to study the interior of Enceladus. However, \citet{ermakov_etal21} assume $\eta_{core}>10^{20}$~Pa.s, $\mu_{core}>10$~GPa, and  \citet{Genova_etal24} assume $\eta_{core}>10^{16}$~Pa.s, $\mu_{core}>50$~GPa, the ranges of elastic and viscoelastic parameters that imply that the core is effectively negligible.} Therefore, the results of these studies represent only one of the tidal dissipation scenarios, i.e., tidal dissipation occurring in the shell without any effective contribution from the core. Here, we broaden the parameter space to allow for the possibility of significant tidal dissipation in the core.  
Note that the commonly assumed viscosity of ice, as suggested by \citet{GoldsbyKohlstedt01} ($\eta_{\rm ice} \approx 10^{14}$~Pa.s,) corresponds to ice near its melting point. Due to the strong temperature dependence of viscosity and very low temperature conditions at Enceladus that may persist deep within the shell, a uniform viscosity of $10^{14}$~Pa.s for the shell \citep{efroimsky18, ermakov_etal21, Genova_etal24} is unrealistic.

As described in Section~\ref{sec:tidesandviscoelasticity}, various viscoelastic models exist based on theoretical and laboratory results. 
Because the nominal values of the effective relaxation time of ice could be close to the orbital period (tidal forcing period) of Enceladus, we examine the choice of Maxwell and Andrade viscoelastic models by computing the tidal response components (Love numbers and heat) using both models. 
Our analysis shows that if the viscosity of the ice shell is sufficiently high due to the low temperature ($\eta_{\rm shell}\gtrsim 10^{18}$~Pa.s), its relaxation time will be significantly longer than the forcing period, implying a predominantly elastic behavior of the shell, and therefore the choice of the viscoelastic model does not significantly affect the response (See Appendix~\ref{sec:viscoelasticmodels}). 
However, if the viscosity of the ice shell is such that the relaxation time ($\tau_M~=~\eta_{\rm shell}/\mu_{\rm shell}$) is close to the forcing period, that is, $\eta_{\rm shell}~\approx~10^{13-14}$~Pa.s, Maxwell and Andrade viscoelastic models predict considerably different surface deformation ($\sim$3~m, see Appendix \ref{sec:viscoelasticmodels}).
The two viscoelastic models differ more significantly in their predicted imaginary component of the response, and thus in the generated frictional heat. The real part of the response is largely determined by the rigidity of the body. The difference in total tidal heating generated in the ice shell and the core can reach several GW.
Although we do not rule out the applicability of the Maxwell viscoelastic model \citep[see e.g.,][]{hussmann_etal16}, for further calculations in this paper, we adopt the Andrade model because of its ability to capture transitional anelastic mechanisms missing in the Maxwell model. The extent to which Enceladus's tidal response deviates from a Maxwell body to an Andrade or more complex viscoelastic body is determined by the large-scale rheological state of its interior (see, e.g., \citet{Bagheri2022TidalOverview, bierson24} for discussions on viscoelastic models). We investigate the sensitivity of Enceladus's tidal response to the choice of Andrade parameters in Appendix~\ref{sec:viscoelasticmodels}.

\clearpage

\begin{table}
\begin{centering}
\begin{tabular}{lllc}
\hline
Quantity  & Symbol & Unit & Value \\
\hline
Radius & $R_E$ & km & \hspace{3mm} 251.99$~^{(1)}$\\
Mass ($\times 10^{20}$) & $M$ & kg & $1.08101 \pm 1.08~^{(2)}$ \\
Flattening  coeff.($\times 10^{6}$) & $C_{20}$ & -- & \hspace{1mm} --5564.4 $\pm$ 37.58$~^{(1)}$\\
 &  &   & \hspace{-2mm} $-5526.1\pm35.5~^{(3)}$\\
Quadruple  coeff.($\times 10^{6}$) & $C_{22}$ & -- &\hspace{2.5mm} 1542.0 $\pm$ 14.93$~^{(1)}$\\
 &  &   & \hspace{1mm} $1575.7\pm 15.9~^{(3)}$\\
Degree-3  coeff.($\times 10^{6}$) & $J_3$ & -- & \hspace{1mm} $-180.64 \pm 33.95~^{(1)}$\\
 &   &   & \hspace{.5mm} --118.2 $\pm 23.5~^{(3)}$\\
SPT surface heat flux & $\dot E_{SPT}$ & GW & \hspace{5mm} $15.8 \pm 3.1 ~^{(4)}$\\
Estimated total surface heat flux & $\dot E_{total}$ & GW &\hspace{7mm} $18-28~^{(1)}$\\
  &  &   & \hspace{9.5mm}  $25-40~^{(5,6)}$\\
Physical libration & $\gamma$ & deg & \hspace{7.5mm}  $0.091 \pm 0.009~^{(1)}$\\
 &  &   &  \hspace{10mm}  $0.12\pm 0.014~^{(7)}$\\
 &  &   &  \hspace{8mm}   $0.155 \pm 0.014~^{(8)}$\\
\textcolor{black}{Normalized Moment of inertial} & MoI & -- & \hspace{7mm} 0.336--0.339 $^{(1)}$\\
 &  &   &  \hspace{7mm} 0.333--0.336 $^{(9)}$ \\
Orbital eccentricity & $e$ & -- & \hspace{7mm} $0.0047~^{(10)}$\\
Orbital mean motion ($\times 10^{-5}$) & $\omega$ & rad/s &  \hspace{7mm} 5.307$~^{(11)}$\\
\hline
\end{tabular}

\caption{Existing measurements for Enceladus. The subscripts indicate the references for the quoted values: $^1$\citet{park_etal24}, $^2$\citet{jacobson22}, $^3$\citet{hemingway_etal18}, $^4$\citet{howett_etal11}, $^5$\citet{nimmo_etal23}, $^6$\citet{hemingwayMittal19}, $^7$\citet{thomas_etal16}, $^8$\citet{nadezhdina_etal16}, $^9$\citet{iess_etal14}, $^{10}$\citet{nimmo_etal18}, $^{11}$\citet{robertsNimmo08}. Static gravity coefficients have been adjusted to a reference radius of 251.99~km with respect to the calculated values in \citet{park_etal24} that used a radius of 256.6~km. Heat loss rates in (1), (5), and (6), are estimated from the shell thickness.} 
\label{table:observations}
\end{centering}
\end{table}

\begin{table}
\begin{center}
\begin{tabular}{llll}
\hline
Quantity  & Symbol & Unit  & Range\\
\hline
Shell thickness &$D_{\rm shell}$ & km & $2-50$\\
Shell effective viscosity   &$\eta_{\rm shell}$  &Pa.s &  $10^{10}-10^{25}$\\ 
Shell shear modulus &$\mu_{\rm shell}$  & GPa & $2-4$ \\
Shell bulk modulus &$\kappa_{\rm shell}$ &  GPa  & 10 (fixed)\\
Ocean density  &$\rho_{\rm ocean}$  & kg/m$^3$ & $930-1400$\\
Ocean thickness  & $D_{\rm ocean}$  &km & $1-130$ \\
Ocean bulk modulus  &$\kappa_{\rm ocean}$  & GPa &  2.1\\
Core radius & $R_{\rm core}$  & km & $120-220$\\ 
Core shear modulus &$\mu_{\rm core}$ & GPa & $1-50$ \\ 
Core bulk modulus  &$\kappa_{\rm core} $ & GPa & 100 (fixed)\\ 
Core density  &$\rho_{\rm core}$  & kg/m$^3$ & $2000-2800$\\ 
Core viscosity  &$\eta_{\rm core}$ & Pa.s & $10^{9}-10^{22}$\\ 
\hline
\end{tabular}
\caption{Prior ranges for interior model parameters. We sample all the model parameters on logarithmic scale.}
\label{table:prior}
\end{center}
\end{table}

\textcolor{black}{
\section{Road-map to Constrain the interior structure and tidal heating mode}\label{sec:roadmap}}
Despite numerous studies based on \textit{Cassini} observations, several interior structure properties of Enceladus are poorly constrained. This uncertainty is due in part to the limitation of the available observations and in part to differences in the interpretation of the data for the interior structure and the underlying modeling assumptions \citep[e.g.,][]{howett_etal11, spencer_etal13, thomas_etal16, park_etal24, Schenk_McKinnon24}. Here, we explore the possible interior parameters that are consistent with the currently available observations. \textcolor{black}{Based on these constraints, we predict ranges of the currently unavailable Love numbers, and investigate the possibility of discriminating between the two main tidal heating scenarios from future measurements}. 

Currently available geophysical observations used to characterize Enceladus's interior include shape \citep[e.g.,][]{park_etal24, Schenk_McKinnon24}, gravity up to degree-3 \citep{iess_etal14,park_etal24}, physical libration amplitude \citep{thomas_etal16, nadezhdina_etal16,park_etal24}, and heat flux \citep{howett_etal11, spencer_etal13} as summarized in Table~\ref{table:observations}. We note that the rates of heat loss considered here, except by \citet{howett_etal11}, are not based on direct measurements, but are estimated from the shell thickness and assumptions on its thermal conductivity (see discussions in Section~\ref{sec:concludiscussions}). To avoid any self-consistency issues, we ensure that the obtained \textit{a posteriori} ice shell thicknesses in our MCMC simulations are within the ranges used to determine the adopted rates of heat loss. We further discuss the assumptions on the rate of heat loss in Enceladus in Section~\ref{sec:concludiscussions}.

\begin{figure}[ht]
\centering
    \rotatebox{270}{
\hspace{-2.cm}\includegraphics[width=1.5\textwidth]{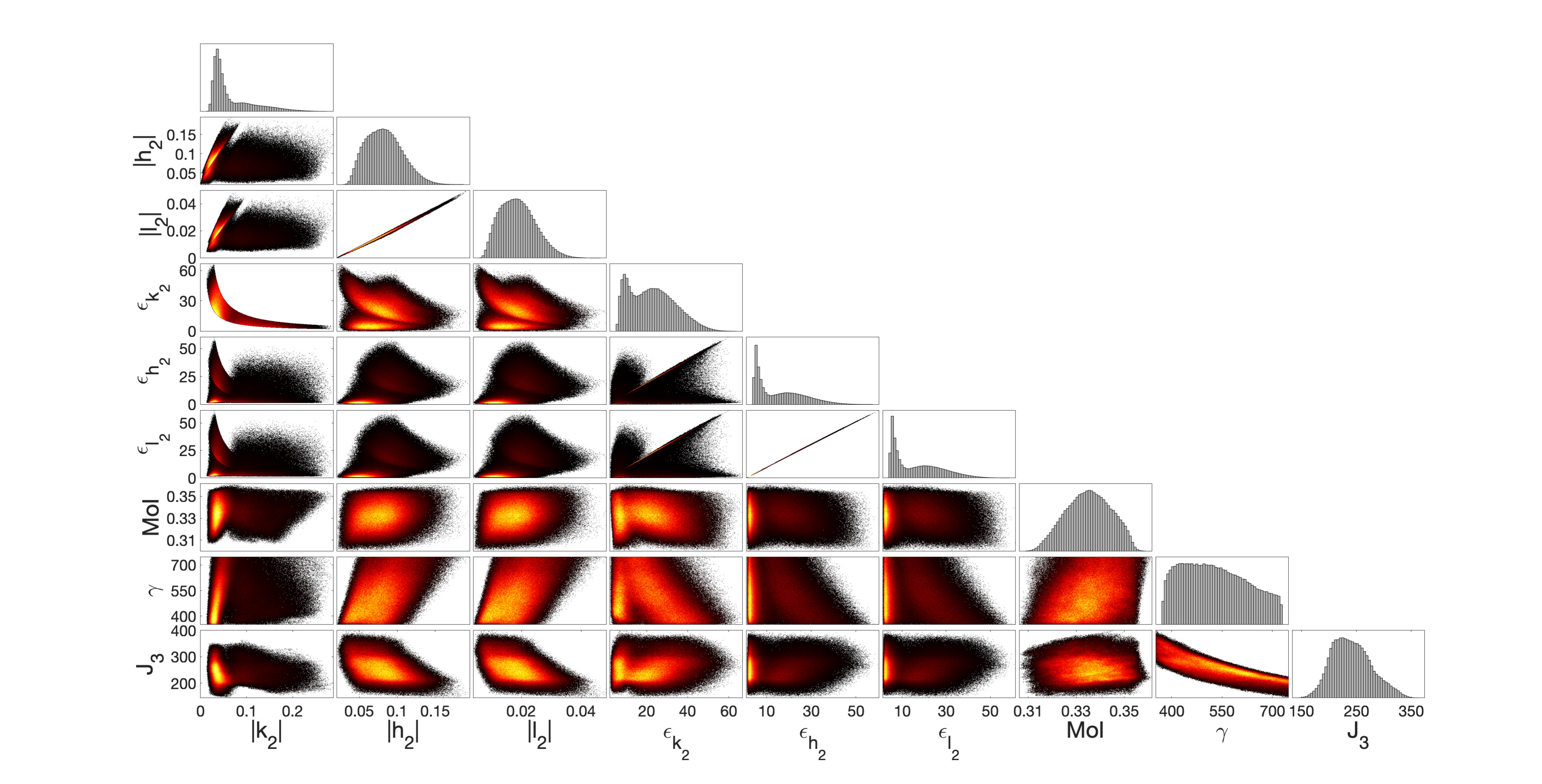}}
\vspace{-1.cm}
\caption{\small{Exploration of the potentially observable geodetic quantities that are consistent with estimates of mass, heat flux \citep{howett_etal11, hemingwayMittal19, park_etal24}, and libration amplitude spanned by estimates in \citet{thomas_etal16}, \citet{nadezhdina_etal16} and \citet{park_etal24}. Other observables in this plot are free to vary. In the 2D Probability Density Functions (PDFs), black (yellow) color indicates low (high) probability. The  observed multiple peaks in the PDFs of $|k_2|$, $|h_2|$, $|l_2|$, and phase lags reflect the different regimes of dissipation consistent with the observations (see Figure~\ref{figure:tidalresponseviscs}). Phase lags of $\epsilon_{k_2}$, $\epsilon_{h_2}$, and $\epsilon_{l_2}$ are in degrees, amplitude of the libration is in meters, and $J_3$ is divided by 10$^6$.}}
 \label{fig:obsrvblsmassheatlibgeneral}
\end{figure}

\textcolor{black}{
\subsection{Plausible ranges of observables and interior parameters}\label{sec:allplausible}}

We start with the most conservatively chosen set of available measurements to investigate all the plausible interior properties, tidal dissipation regimes, and the associated ranges of observables. We  \textcolor{black}{consider the thickness, rigidity, effective viscosity, and density of each layer to describe the interior. We assume no variations in mechanical properties in the lateral directions (we will explore the effect of shell's structural heterogeneity in Section~\ref{sec:shellheterogeneity}).} \textcolor{black}{We consider the geodetic} quantities, i.e., complex tidal Love numbers, libration, moment of inertia, and $l=3$ zonal gravity coefficient, hereafter referred to as observables.
We conduct Monte Carlo sampling based on the prior ranges of the model parameters presented in Table~\ref{table:prior} (see Appendix \ref{sec:mcmc}).
We present the correlation between the observables in Figure~\ref{fig:obsrvblsmassheatlibgeneral}. We use the total density, the physical libration amplitude in a range covering the estimates by \citet{thomas_etal16}, \citet{nadezhdina_etal16}, and \citet{park_etal24} ($360~m<\gamma<~750~m$), and the estimated heat loss rate as the only constraints. To demonstrate the widest possible parameter space, we have also considered a wide range of published estimates of the rate of heat loss that covers \citet{howett_etal11}, \citet{hemingwayMittal19}, and \citet{park_etal24} (15~GW$<\dot E<$40~GW).
The Love number amplitudes and phase lags, moment of inertia, and degree-3 static gravity coefficient vary freely as long as the model satisfies the constraints of mass, heat loss, and libration. We do not impose any constraints on the estimated moment of inertia and $J_3$ to avoid any possible inconsistency of our models with the underlying assumptions used in deriving them.


The patterns of the posterior probability density functions (PDFs) of the amplitude and phase lag of $k_2$ largely differ from the PDFs of the amplitudes and phases of $h_2$ and $l_2$ (Figure~\ref{fig:obsrvblsmassheatlibgeneral}), \textcolor{black}{consistent with Section~\ref{sec:heatgenerationclasses}}. PDFs of $|k_2|$ and $\epsilon_{k_2}$ have several peaks associated with different tidal dissipation mechanisms that can occur either in the shell or in the core. \textcolor{black}{Each of these peaks correspond to the ranges of viscosities that can explain the estimated total heat shown in Figure~\ref{figure:heatviscs}}. A different distribution pattern and peaks associated with different regimes of tidal activity are \textcolor{black}{obtained} for $h_2$ and $l_2$, and their phase lags. \textcolor{black}{This different pattern arises from the fact that $|h_2|$ and $|l_2|$ are largely insensitive} to tidal activity in the core, \textcolor{black}{and thus do not reveal any peaks associated with the dissipation in the shell}. 
\textcolor{black}{Moreover,} the posterior PDFs (Figure~\ref{fig:obsrvblsmassheatlibgeneral}) show that while the $|h_2|$ and $|l_2|$ and their phase lags  are strongly correlated with each other (dominated by the response of the shell). Their correlation with the amplitude and phase lag of $k_2$ depends on where the tidal dissipation dominantly occurs.

We also obtain the ranges of the interior model parameters and the correlation between them associated with the observables \textcolor{black}{chosen here} \textcolor{black}{(Appendix~\ref{sec:modelparametersbroadest}}).
Previously, we ruled out mean crustal thicknesses around 5--10~km  for Enceladus, as they would \textcolor{black}{imply} amplitudes of libration that are too large (see Section~\ref{sec:shellthicknesslibration}). Ice shell thicknesses~$\approx$~1--5~km would be compatible with the estimates of tidal heating and libration only if the mean effective viscosity of the shell is unrealistically low. Therefore, we conclude that the mean thickness of Enceladus's shell is larger than 10~km. 
For completeness, we include very low effective viscosities for the shell and the core, which imply deformation in the viscous-dominated regimes. As a result, the PDFs of the viscosities of the shell and core indicate multiple peaks representative of tidal dissipation dominated by both viscous deformation as well as the shear (anelastic) relaxation (See also Figure~\ref{figure:heatviscs}).
In the remaining analysis below, we consider viscosities of the ice shell and the rocky core that are close to or greater than their nominal relaxation viscosities ($\eta_{\rm shell}>10^{12}$~Pa.s and $\eta_{\rm core}>10^{11}$~Pa.s), i.e., the ice shell and the rocky core deform in the anelasticity-dominated deformation regime, as opposed to the viscous-dominated deformation regime. 

\begin{figure}[ht] 
\centering
\vspace{-2.5cm}
    \rotatebox{270}{
 \includegraphics[width=1.5\textwidth]{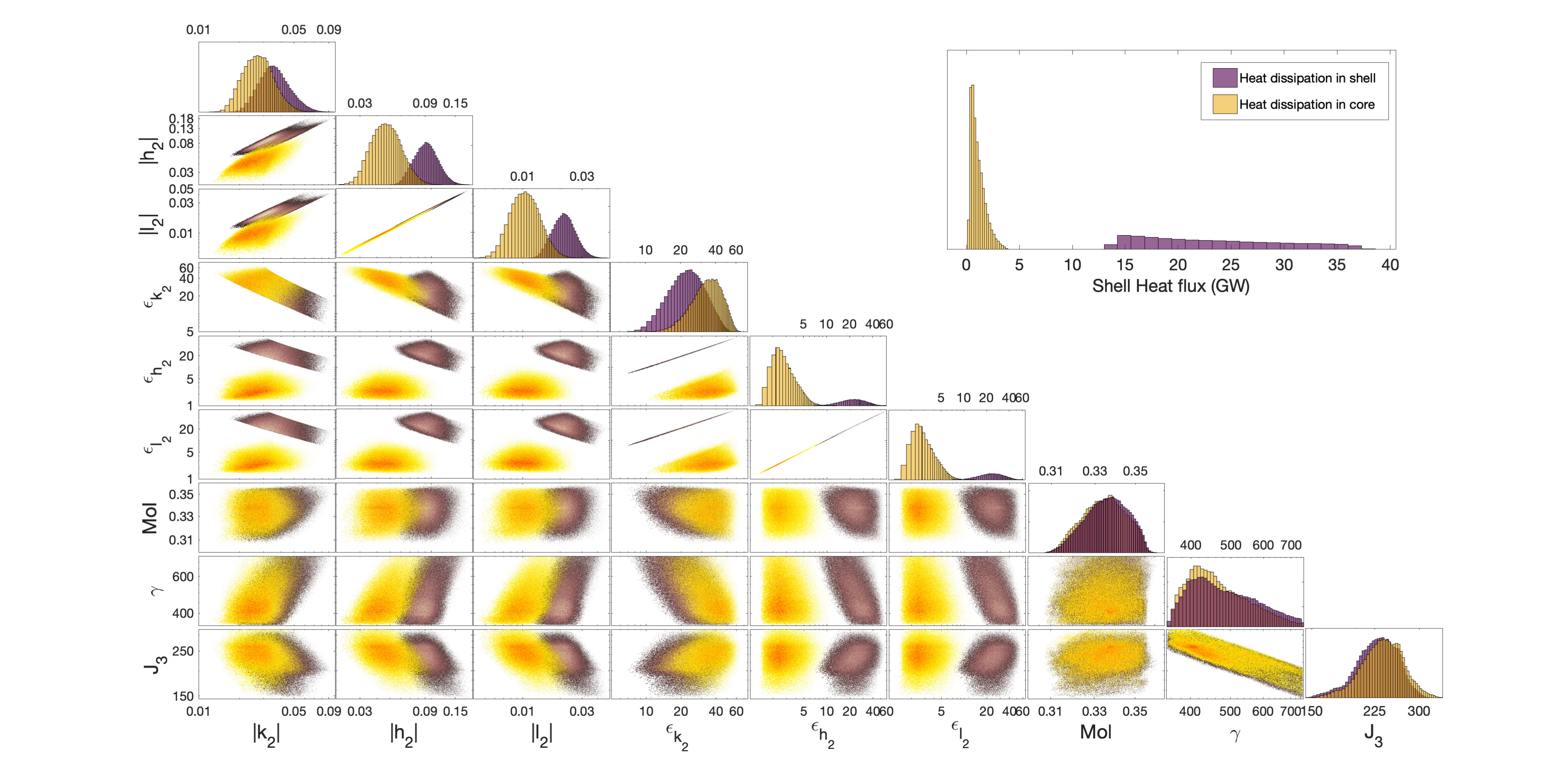}}
\vspace{-1.cm}
\caption{Exploration of the potentially observable quantities consistent with estimates of mass, heat flux and libration amplitude \textcolor{black}{for  tidal dissipation scenarios in the shell and in the core. 
The heat flux is assumed as $\rm15~GW<\dot E<40~GW$, covering the estimated ranges in \citet{hemingway_etal18,park_etal24}, and the direct measurements \citep{howett_etal11}. Libration is assumed to be between $\rm 0.082^\circ<\gamma<0.169^\circ$, covering the estimates in \citet{thomas_etal16}, \citet{nadezhdina_etal16}, and \citet{park_etal24}. Tidal heating produced in the shell in either scenario is also shown.
In the 2D PDFs, the pink-gray and \textcolor{black}{black}-yellow themes refer to dissipation in the shell and and in the core, respectively. All variables are shown in logarithmic axes.}
Unit conventions are the same as in Figure~\ref{fig:obsrvblsmassheatlibgeneral}.}
\label{fig:observblessheatlibshellcorediss}
\end{figure}

\textcolor{black}{Next,} we focus on the analysis of discriminating between two sets of geodetic observables that are associated with each of the two scenarios (Figure~\ref{fig:observblessheatlibshellcorediss}).
In this figure, and in the remainder of this paper, we distinguish between core-dominated and shell-dominated tidal dissipation scenarios by associating them with the models where at least 90\% of the total heat is produced in the shell or in the core. We compute the tidal heating in the shell using equation~\eqref{shelldiss}. 
The PDFs of $|k_2|$ and $\epsilon_{k_2}$ for the two tidal dissipation scenarios largely overlap, whereas the PDFs of $|h_2|$, $\epsilon_{h_2}$, $|l_2|$, and $\epsilon_{l_2}$ in the case of shell-dominated tidal dissipation differ significantly from those in the case of core-dominated tidal dissipation (Figure~\ref{fig:observblessheatlibshellcorediss}). 
In an inverse problem context, the degeneracy caused by the two different tidal dissipation scenarios reflected in the tidal Love numbers, as discussed earlier, leads to the non-uniqueness of the interpretation of $k_2$ and $\epsilon_{k_2}$. 

\begin{figure}[htbp]
\centering
\includegraphics[width=1\textwidth]{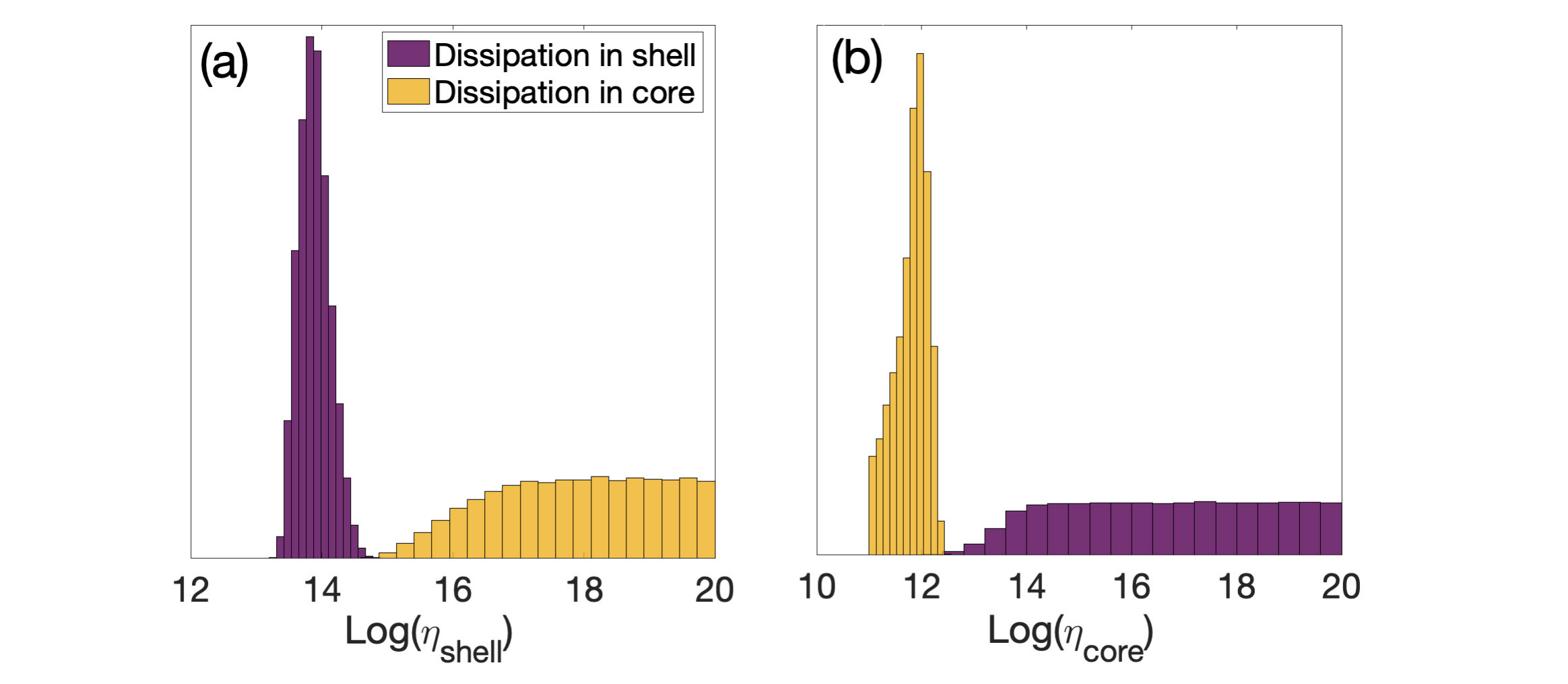}
\caption{PDFs of ice shell and core viscosities for each of the two cases of tidal dissipation occurring dominantly in the shell \textcolor{black}{(panel a)} and in the core \textcolor{black}{(panel b)}, corresponding to two sets of observables in Figure~\ref{fig:observblessheatlibshellcorediss}}
\label{fig:viscositiescorner}
\end{figure}

\textcolor{black}{
By \textcolor{black}{differentiating} between the two tidal dissipation scenarios, we obtain a tight constraint on the viscosity of the shell in the cases when the tidal dissipation is dominated in the shell. Similarly, we constrain the viscosity of the core in the case when we assume that tidal dissipation occurs predominantly in the core (Figure~\ref{fig:viscositiescorner}). Based on the set of observables used in Figures~\ref{fig:obsrvblsmassheatlibgeneral}~and~\ref{fig:observblessheatlibshellcorediss},} we find that the thickness of the ice shell, mostly constrained by the libration amplitude, is in the range of 15--40~km. 
Because we have conservatively chosen broad ranges for observables in this analysis, we do not obtained tight constraints on most of the other model parameters in this analysis. The PDFs and correlations between the interior structure parameters are presented in Appendix~\ref{sec:modelparametersbroadest}.

\vspace{2cm}

We compare the PDFs of $h_2$ and $l_2$ and their phase lags ($\epsilon_{h_2}$, $\epsilon_{l_2}$) associated with each of the two heat generation scenarios \textcolor{black}{to compute the minimum required precision of measurements to discriminate between the two dissipation scenarios (Figure~\ref{fig:observblessheatlibshellcorediss}).
We compute the minimum \textcolor{black}{required} measurement precision of $|h_2|$ and $|l_2|$ by calculating the separation between the two intervals determined by the 2-$\sigma$ confidence regions associated with the two PDFs. Therefore,} if we measure $h_2$ with a precision equal to the separation between the 2-$\sigma$ confidence intervals of the two PDFs, we obtain a confidence $\sim$95\% to identify the layer that dominates the dissipation budget. Alternatively, if we measure $h_2$ with a precision equal to the separation between the 1-$\sigma$ confidence intervals, we obtain a confidence of $\sim$68\% in identifying the layer dominating dissipation (not shown here).
\textcolor{black}{We present more detailed comparison of  $|h_2|$ and $|l_2|$ and their phase lags in Appendix~\ref{sec:modelparametersbroadest}.}

The PDFs of $|h_2|$ and $|l_2|$ associated with the two tidal dissipation scenarios using this set of observations demonstrate large overlaps (Figure~\ref{fig:observblessheatlibshellcorediss}), challenging the ability to distinguish between the dissipation in the shell or in the core.
This large overlap is \textcolor{black}{mainly} due to the large uncertainties in the observables considered in this case. This analysis implies that measurement of the maximum vertical tidal deformations with a maximum uncertainty of 62~cm can distinguish between the dominant region of tidal activity within the 1-$\sigma$ confidence level. Measurement of radial deformations with a maximum uncertainty of 56~cm can indicates the ``\textit{more likely}'' region of heat generation, but it cannot definitively discriminate between the two scenarios. 
Alternatively, measurement of the tangential displacements within an uncertainty of 28~cm is required for 1-$\sigma$ confidence.
The phase lags associated with the radial deformations ($\epsilon_{h_2}$ and $\epsilon_{l_2}$) indicate a better separation between the PDFs associated with the two regimes of tidal activity compared to the deformations ($|h_2|$, $|l_2|$). We find that measuring the phase lag angles of the tidal deformations \textcolor{black}{(either $\epsilon_{k_2}$ or $\epsilon_{l_2}$)} with an accuracy of 11$^\circ$ is required for 1-$\sigma$ confidence, and 4$^\circ$ is required for 2-$\sigma$ confidence in determining the dominant region of tidal dissipation \textcolor{black}{(see Appendix~\ref{sec:modelparametersbroadest} for more details)}. 


\subsection{Computing measurement requirements based on \citet{park_etal24}}
\label{sec:twoscenarios}
 
Next, we focus on a narrower range of observational constraints by exclusively using the analysis by \citet{park_etal24} for $\gamma$, $J_3$, moment of inertia (inferred from $C_{20}$ and $C_{22}$), total rate of heat loss (inferred from conductive shell thickness), and total density (see Table~\ref{table:observations}). \textcolor{black}{We acknowledge that estimation of the moment of inertia in \citet{park_etal24} is not completely independent of $\gamma$ and $J_3$. However, the estimated value of moment of inertia has been presented in a conservatively broad range and the effect of using it independently is negligible.} We use 3-$\sigma$ uncertainties on all of observables.
We present the correlation between \textcolor{black}{these quantities} incorporating both tidal dissipation scenarios in Figures~\ref{fig:Inversionobsrvblsboth}. 
Applying the narrower observational constraints based on \citet{park_etal24} \textcolor{black}{shows that $|k_2|$ and $\epsilon_{k_2}$ indicate a normal distribution with a \textit{single peak} that encompasses both scenarios of tidal dissipation, \textcolor{black}{thus} making them indistinguishable based only on measurement of $|k_2|$ and $\epsilon_{k_2}$.}
By contrast, $|h_2|$, $|l_2|$, $\epsilon_{h_2}$, and $\epsilon_{l_2}$ show two peaks associated with the two tidal dissipation scenarios.
By separating the two tidal dissipation regimes (see figures in Appendix~\ref{sec:modelparametersbroadest}), we also quantify the preferred values of the observables \textcolor{black}{with} the highest likelihood in the case of each of the two heat generation scenarios (Table~\ref{table:precisions}).
We note that satisfying all constraints provided in \citet{park_etal24} is possible when using 3-$\sigma$ uncertainties on the observables. Specifically, we find that the mean value of $J_3$ can be matched with the libration amplitude when considering 3-$\sigma$ uncertainty in the measured $J_3$. 

Our results can be compared to those  by \citet{Genova_etal24} and \citet{ermakov_etal21}, where only tidal heating in the shell is considered. \textcolor{black}{
Also, \citet{Genova_etal24} use only the gravity measurements without exploiting the libration measurement as a constraint, which is expected to somewhat alter the results.}
The tighter constraints obtained based on \citet{park_etal24} 
relax the precision requirements on the measurements of $h_2$ and $l_2$ required to discriminate between the two heating scenarios.
Using the observational constraints of \citet{park_etal24}, we show the predicted amplitudes and phase lags of the Love numbers in Figure~\ref{fig:k2andphasetwocases}~panels~(a)--(f). Using \textcolor{black}{these narrower ranges of} observational constraints, the PDFs of the tidal response associated with the two tidal dissipation scenarios are further separated compared to Section~\ref{sec:allplausible}, where broader ranges of values for the observables were used (Figure~\ref{fig:observblessheatlibshellcorediss}).
In this case, measuring the maximum vertical deformations with a precision of 87~cm is sufficient to differentiate between the two hypotheses of heat generation with a 1-$\sigma$ confidence. 
For a confidence level of 2-$\sigma$, measuring the maximum radial displacement of 1~cm is required. 
Alternatively, measuring the phase lag $\epsilon_{h_2}$ by 20$^{\circ}$ and 15$^{\circ}$ is sufficient to discriminate between the two tidal dissipation scenarios with 1-$\sigma$ and 2-$\sigma$ confidences, respectively. 
In addition, our analysis indicates that measurements of the maximum tangential displacements within 40~cm and 0.5~cm are required to differentiate between the two tidal dissipation scenarios providing  1-$\sigma$ and 2-$\sigma$ level confidences, respectively. The measurement precisions required for $\epsilon_{l_2}$ for the confidence levels of 1-$\sigma$ and 2-$\sigma$ are the same as those for $\epsilon_{h_2}$.


In case future measurements provide constraints on $k_2$, the measurement requirement for $h_2$ and/or $l_2$ could be relaxed. We present PDFs of displacement tidal Love numbers associated with the core- and shell-dominated tidal dissipation assuming constraints on the amplitude and phase lag of $k_2$ (Figure~\ref{fig:k2andphasetwocases} panels (g)--(j)). If the real and imaginary parts of $k_2$ are measured within 0.002, the accuracies required to measure the maximum radial deformations to obtain 1-$\sigma$ and 2-$\sigma$ confidence levels relax to 134~cm and 82~cm, respectively. In that case, the upper bounds of the uncertainty required for the horizontal deformations are 61~cm and 36~cm, respectively. In this case, the accuracies required for the phase lags of $h_2$ and $l_2$ are equal to 21$^\circ$ and 16$^\circ$ for confidence levels of 1-$\sigma$ and 2-$\sigma$, respectively.

\textcolor{black}{In} case more precise measurements of $k_2$ (both amplitude and phase) \textcolor{black}{are achieved}, i.e. within an uncertainty of 0.0002, the required measurement of radial and horizontal deformations will be further relaxed (not shown here). In that case, the requirement for the maximum radial displacement will be 150~cm and 116~cm for the 1-$\sigma$ and 2-$\sigma$ confidence levels, respectively. These values are 68~cm and 51~cm for the tangential component of the deformation, respectively. The precisions required for the phase lags of $h_2$ and $l_2$ are equal to 24$^\circ$ and 22$^\circ$ to obtain confidence levels of 1-$\sigma$ and 2-$\sigma$, respectively. \textcolor{black}{Due to the large overlap between the PDFs of $k_2$ between the two scenarios, we cannot differentiate between them even with high accuracy measurement of $k_2$ alone.} We summarize the precision required for measuring the radial and horizontal displacements based on different scenarios discussed here in Table~\ref{table:uncertaintiesneeded}. Note that the measurements of the $h_2$ and $l_2$ as well as the $\epsilon_{h_2}$ and $\epsilon_{l_2}$ are correlated with each other and, therefore, measuring \textcolor{black}{either of these quantities} is sufficient for discriminating between tidal dissipation scenarios. We perform our analysis based on both the derivation in this paper (equation~\ref{finalheatinshellwithlove}) and that in \citet{beuthe19}. We find that the difference in the results is less than 5\%. This close approximation indicates the applicability of equation~\ref{finalheatinshellwithlove} in capturing the tidal dissipation in the shell without requiring \textit{a priori} constraint on the shell's rheology. 


\begin{table}
\begin{centering}
\begin{tabular}{ll lll}
\hline
Observable & Unit & Dissipative core &   Dissipative shell & Expected Uncertainty\\
\hline
$|k_2|{^*}$ & -- & \centering 0.0240 $\pm$ 0.0102  & 0.0317 $\pm$ 0.0130 & 0.002 (0.0002) \\
$|h_2|$  & -- &  \centering 0.0459 $\pm$ 0.0294  & 0.0848 $\pm$ 0.0359 &  0.0172 \\ 
 $|l_2|$  & -- & \centering 0.0104 $\pm$ 0.0072  & 0.0200 $\pm$ 0.0090 & --\\
$ \epsilon_{k_2}{^*}$ & deg & \centering  39.35 $\pm$ 19.72 & 28.91 $\pm$ 17.25 &  4 (0.4)  \\
$\epsilon_{h_2}$ & deg & \centering 2.95 $\pm$ 2.69 & 28.73 $\pm$ 17.18 &  -- \\
   $\epsilon_{l_2}$ & deg &\centering 3.10 $\pm$ 2.85  & 29.71 $\pm$ 17.89 & -- \\
 $J_3(\times10^{-6})$ & -- & \centering  260.4 $\pm$ 36.9 & 259.3 $\pm$ 38.06 & 0.4 (0.04) \\ 
 $\gamma$& m & \centering 413.4$ \pm$ 65.9 & 408.1 $\pm$ 68.6 & 4.0 (1.3)  \\ 
  Total heat & GW & \centering  18--28 & 18--28 & -- \\ 
 Density &  kg/m$^3$ &  \centering 1611 $\pm$ 1.6 & 1611 $\pm$ 1.6  & --  \\ 
\hline
\end{tabular}
\caption{Favored values and uncertainties (3-$\sigma$) of observables calculated using MCMC sampling based on the \textit{available inferred observables} given by \citet{park_etal24}. Also, expected uncertainties from \textit{future explorations} are provided (see Section~\ref{sec:Feasibility}) which are used in the inversion in Section~\ref{sec:syninversion}. The values in the parentheses in the expected measurements corresponds to a scenario with higher accuracy, as discussed in the text. Expected uncertainties are associated with 2-$\sigma$ error. $^*$Note that if we do not separate the two dissipation regimes, the $|k_2|$ and $|\epsilon_{k_2}|$ have single peaks at 0.027 and 34$^\circ$, respectively.}
\label{table:precisions}
\end{centering}
\end{table}
\begin{table}
\begin{center}
\begin{tabular}{lccccl}
\hline
Current Knowledge & $|k_2|$ & $\max \Delta R  $ (cm)  &   $ \max \Delta E$ (cm)  &  $\epsilon_{h_2}$ and $\epsilon_{l_2}$ (deg)\\
\hline
All studies & -- & N/A(62) & N/A(28) & 4 (11)    \\
\citet{park_etal24} & -- & 1 (87) & 0.5 (40)  & 15 (20)   \\
\citet{park_etal24} & 0.002 &  82 (134) &  36 (61) & 16 (21) \\
\citet{park_etal24} & 0.0002 &  116 (150) & 51 (68)  & 22 (24)  \\
\hline
\end{tabular}
\caption{Range of the required precision for measuring $h_2$, $l_2$, $\epsilon_{h_2}$, $\epsilon_{l_2}$ for different scenarios of available and future measurements. 
``Current knowledge" implies the use of $\gamma$, $J_3$, and heat loss from the referenced studies. 
``All studies" refers to the use of estimates covering \citet{howett_etal11}, \citet{thomas_etal16}, \citet{nadezhdina_etal16} , \citet{hemingwayMittal19}, and \citet{park_etal24}. In the two bottom cases, measurement uncertainties on $k_2$ are assumed and other requirements are computed correspondingly. The requirements are given for obtaining 2-$\sigma$ confidence on the dominating region of the heat dissipation (with those associated with the 1-$\sigma$ confidence provided in parenthesis).
N/A indicates that given the corresponding uncertainties, obtaining the 2-$\sigma$ confidence is not possible.}
\label{table:uncertaintiesneeded}
\end{center}
\end{table}

\renewcommand{\thefigure}{7}

\begin{figure}[ht]
\centering
\vspace{-2.5cm}
    \rotatebox{270}{
\includegraphics[width=1.5\textwidth]{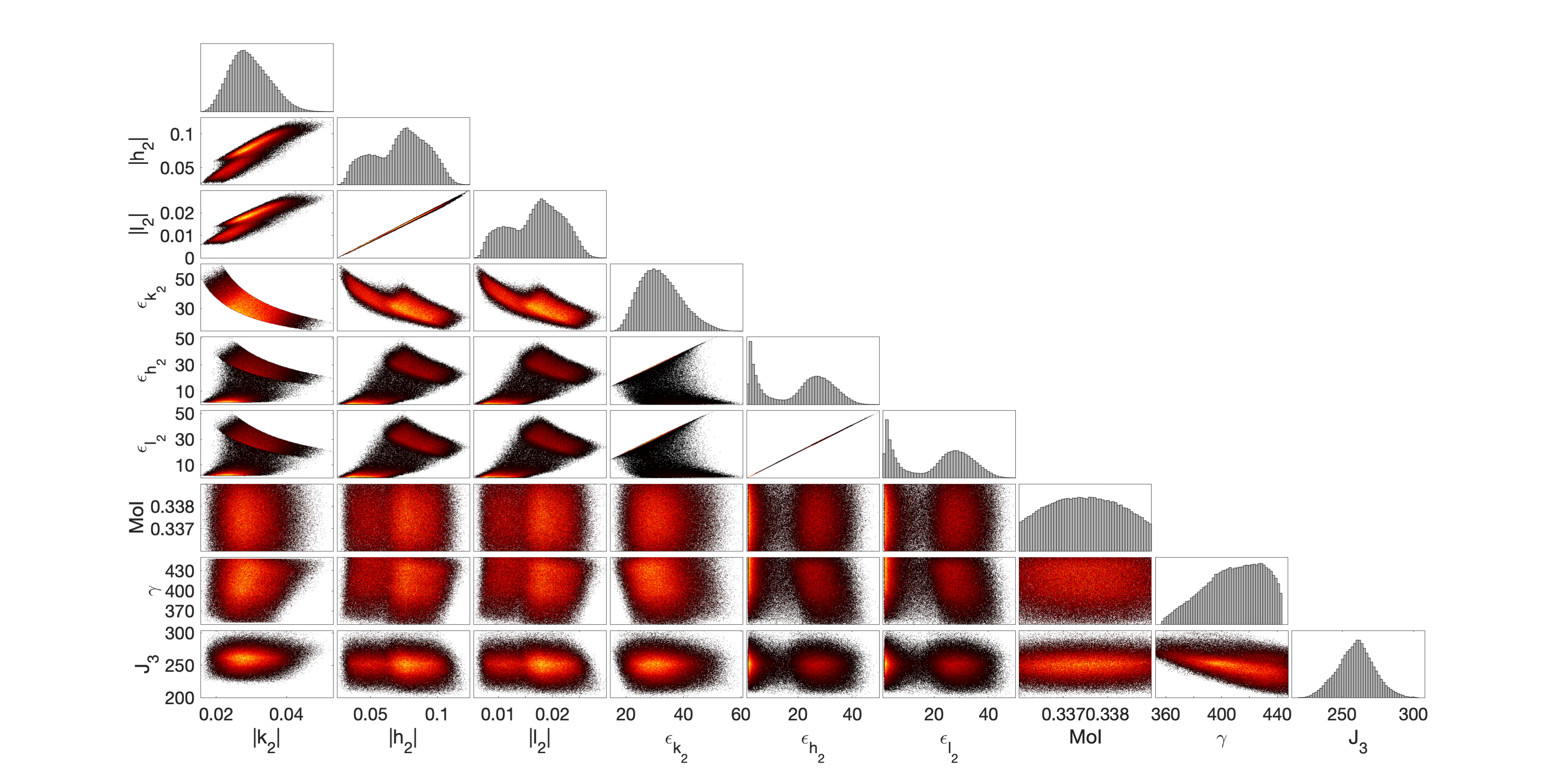}}\\
\vspace{-1.5cm}
\caption{\small{Exploration of observables constraints from $J_3$, moment of inertia, libration amplitude, and rate of heat loss based on \citet{park_etal24}. We present PDFs of predictions for $|k_2|$, $|h_2|$, $|l_2|$, $\epsilon_{k_2}$, $\epsilon_{h_2}$, $\epsilon_{l_2}$. Both scenarios of tidal dissipation are considered here. The plots of the PDFs for the $k_2$ amplitude and phase lag are notably different than those of the $|h_2|$ and $|l_2|$ amplitudes and phases. 
Plotting and unit conventions are the same as in Figure~\ref{fig:obsrvblsmassheatlibgeneral}.}}
\label{fig:Inversionobsrvblsboth}
\end{figure}

\clearpage


\renewcommand{\thefigure}{\arabic{figure}}
\setcounter{figure}{7}

\begin{figure}[ht]
 \includegraphics[width=1.\textwidth]{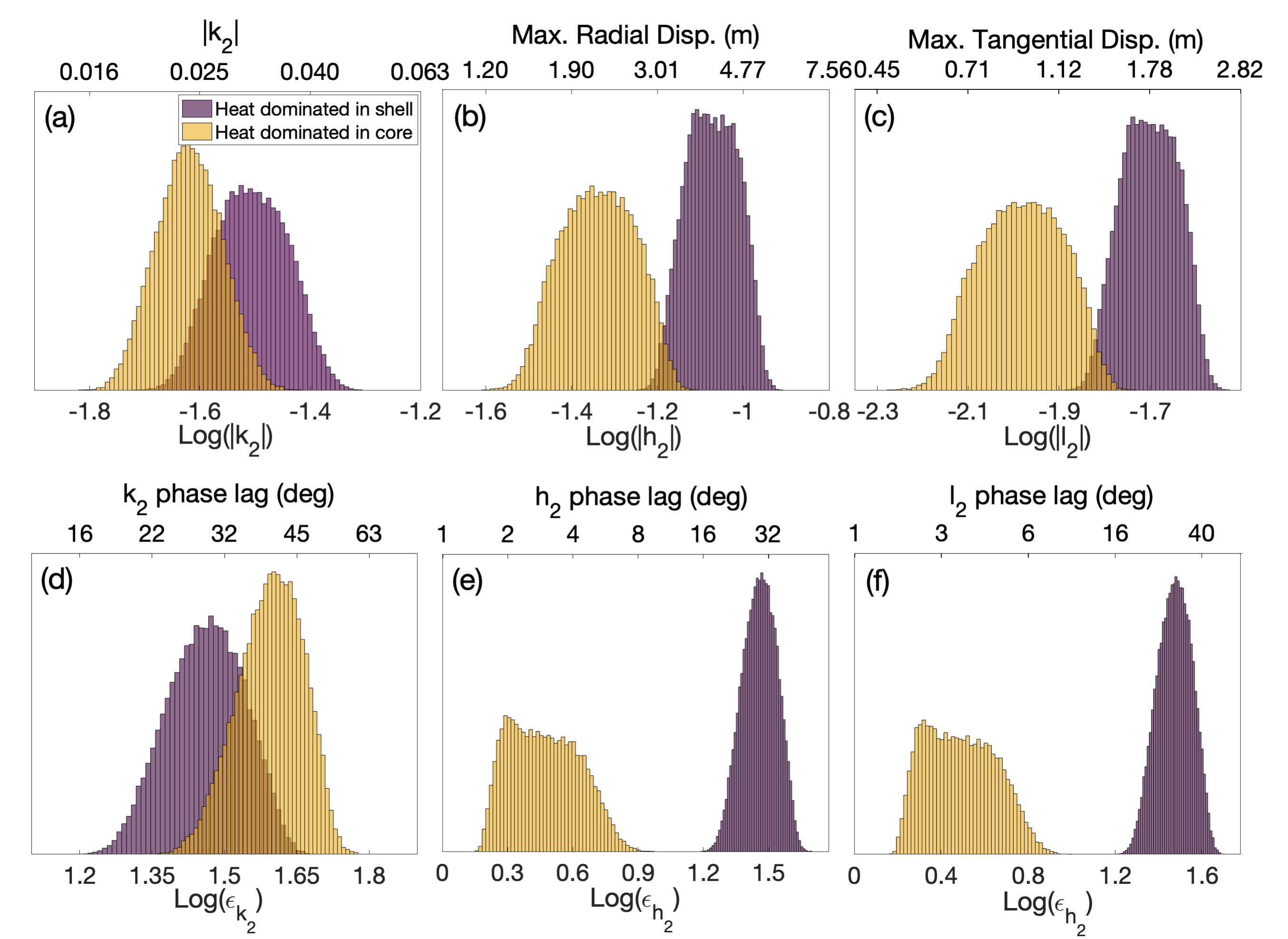}\\
\includegraphics[width=1.\textwidth]{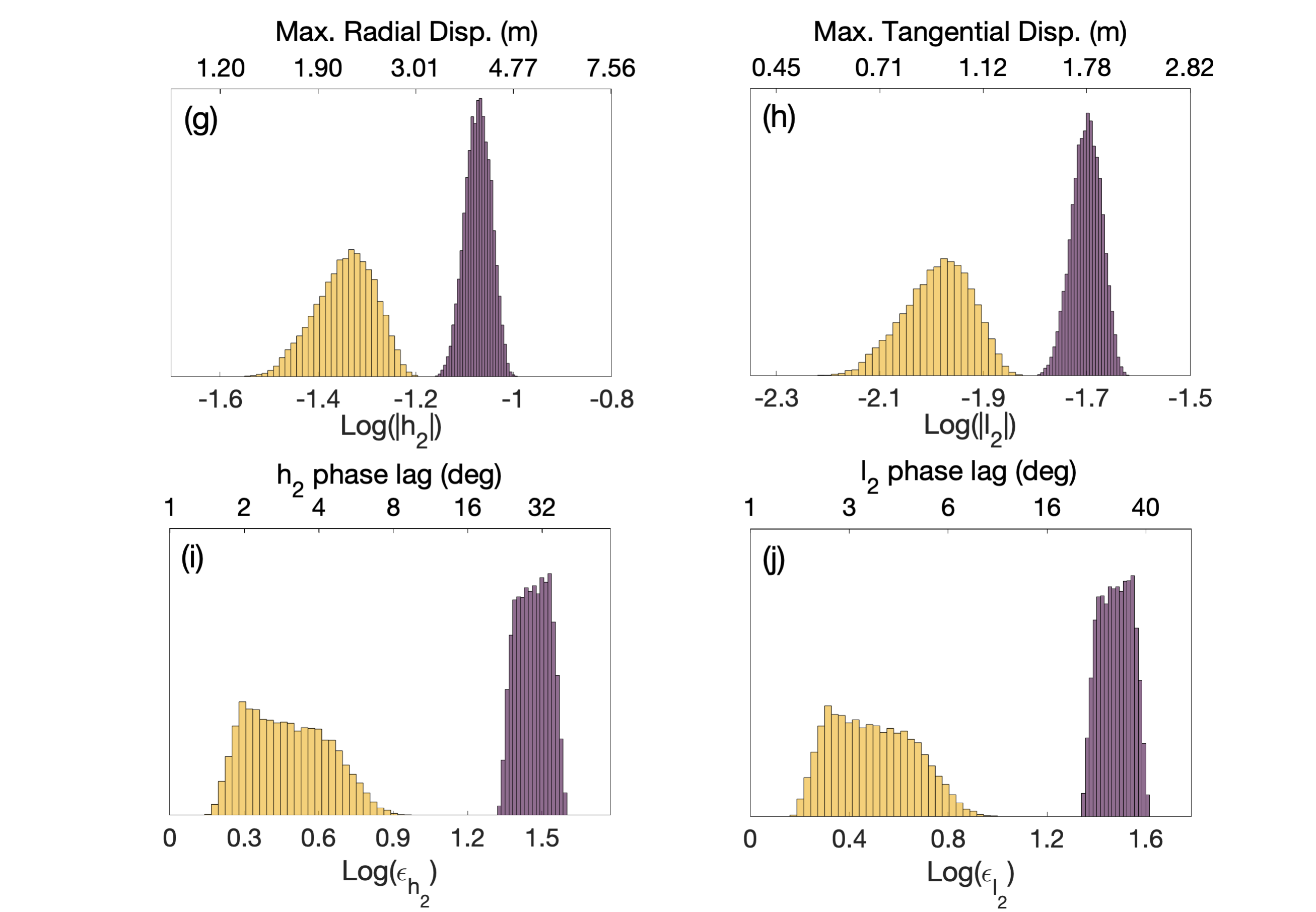}
\caption{\textcolor{black}{Panels (a)--(f): Tidal Love numbers and phase lags for heat dissipation in the shell and in the core based on the constraints by \citet{park_etal24}. 
Panels (g)--(j): Love numbers and phase lags for two dissipation regimes assuming that the real and imaginary parts of $k_2$ are measured with an uncertainty of 0.002. 
}}
 \label{fig:k2andphasetwocases}
\end{figure}

\renewcommand{\thefigure}{9a}

\vspace{-1cm}
\begin{figure}[ht]
\hspace{-2cm}\includegraphics[width=1.3\textwidth]{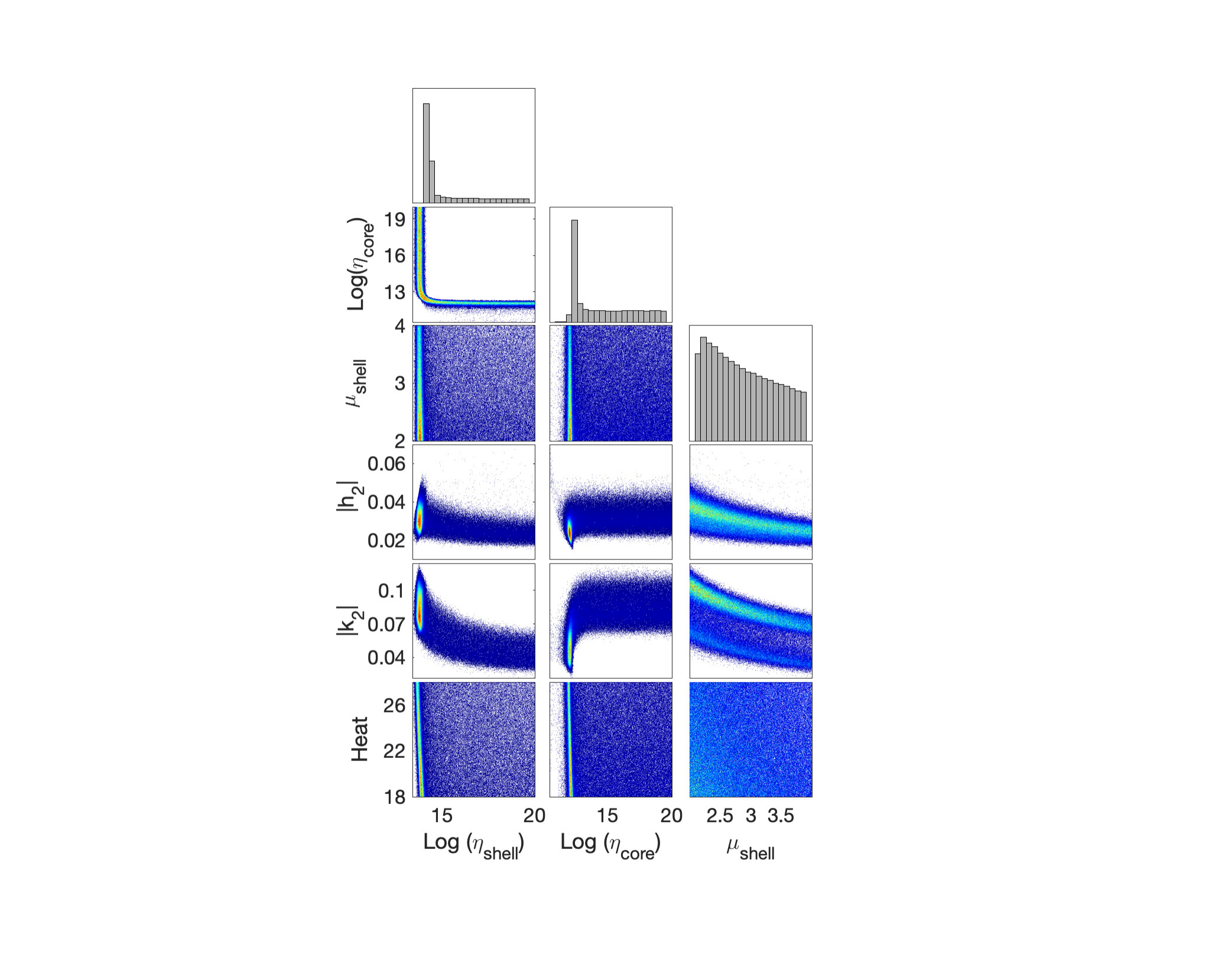}
\vspace{-1cm}
\caption{\textcolor{black}{Sensitivity of  key parameters in determining the heat dissipation consistent with the observables in Figure~\ref{fig:Inversionobsrvblsboth} based on the analysis of the observables by \citet{park_etal24}. Rate of heat loss and $\mu_{shell}$ are shown in in GW and GPa, respectively. Blue and red colors demonstrate lowest and highest likelihoods, respectively.}}
\label{fig:Heatsensitivity}
\end{figure}

\renewcommand{\thefigure}{9b}

\vspace{-3cm}

\begin{figure}[ht]
\centering
\hspace{-1cm} \includegraphics[width=1.1\textwidth]{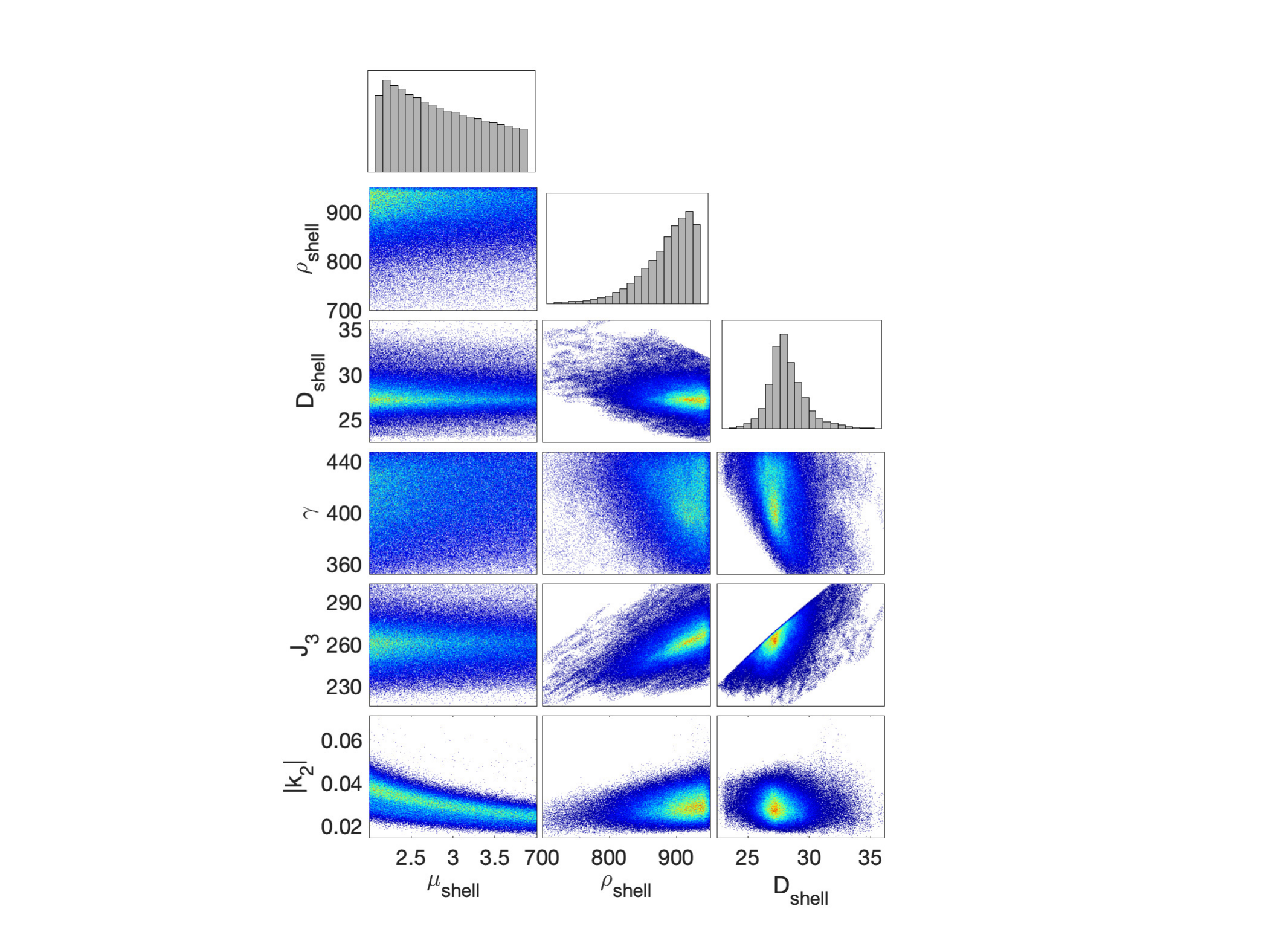}
\caption{\textcolor{black}{Sensitivity of key parameters in determining the ice shell properties associated with observables in Figure~\ref{fig:Inversionobsrvblsboth} based on the analysis of the observables by \citet{park_etal24}. $J_3$ is shown divided by $10^{6}$, $\gamma$, $D_{shell}$, and $\rho_{shell}$ are shown in meters, km, and kg/$\rm m^3$, respectively. Blue and red colors demonstrate lowest and highest likelihoods, respectively.}}
\label{fig:librationsensitivity}
\end{figure}

\renewcommand{\thefigure}{\arabic{figure}}
\setcounter{figure}{9}

\clearpage

\begin{table}
\begin{centering}

\begin{tabular}{llll}

Quantity & Unit &  tidally active core & tidally active shell \\
\hline 
$D_{\rm shell}$ & km& \centering $22-34$ & $22-34$\\ 
$\eta_{\rm shell}$ & Pa.s & $6.3\times 10^{14}-10^{20}$ & $2.5 \times 10^{13}-1.3\times 10^{14}$\\ 
$\mu_{\rm shell}$ & GPa & $2-4$ &  $2-4$ \\
$\rho_{\rm shell}$ & kg/m$^3$ &    $700-950$ &  $700-950$  \\
$\rho_{\rm ocean}$  & kg/m$^3$ &  $920-1250$ & $920-1250$ \\
$R_{\rm core}$ & km &   $183-213$ & $183-213$ \\ 
$\mu_{\rm core}$ & GPa &  $4-70$ &   $5-80$\\ 
$\rho_{\rm core}$ & kg/m$^3$ &    $2160-2400$ & $2160-2400$\\ 
$\eta_{\rm core}$ & Pa.s &   $4.4\times 10^{11}$--$2.3\times 10^{12}$ & $3.16\times 10^{13}-10^{22}$ \\ 
\hline
\end{tabular}
\caption{Ranges of the model parameters from two MCMC analysis satisfying the \textit{currently available measurements} based on the analysis of observables in \citet{park_etal24}.} 
\label{table:mdlprmtrscurrentdata}
\end{centering}
\end{table}

Using the ranges of $\gamma$, $J_3$, MoI, and heat inferred by \citet{park_etal24},
we also obtain considerably improved constraints on the interior parameters compared to the case where we use broader uncertainties on the observations (See Figures \ref{fig:mdlprmtesmassheatlibgeneral}~and~\ref{fig:Inversionmdlprmtrsboth} in Appendix~\ref{sec:modelparametersbroadest}). However, several model parameters remain weakly constrained by the currently available observations. The ranges of the obtained plausible interior parameters are listed in Table~\ref{table:mdlprmtrscurrentdata}. We present PDFs of all the model parameters associated with each of the cases in Appendix~\ref{sec:modelparametersbroadest} (Figures~\ref{fig:mdlprmtrsinversionboth90shell} and \ref{fig:mdlprmtrsinversionboth90core}).

\textcolor{black}{To obtain further insights into the correlation between the observables and model parameters,} we investigate the correlation between total heat and the observables and the interior parameters to which they are most sensitive (Figures~\ref{fig:Heatsensitivity} and \ref{fig:librationsensitivity}). We show the ranges of the effective viscosities of the shell and the core, \textcolor{black}{as well as} their correlation to the shell rigidity and $h_2$ and $k_2$ in Figure~\ref{fig:Heatsensitivity}. The total tidal heating is largely correlated to the viscosity of the region where the tidal dissipation is dominated.  The rigidity of the shell influences the amplitude of both $k_2$ and $h_2$.
We also explore the correlation between the physical properties of the shell (thickness, density, and rigidity) and the observables that are most sensitive to them, i.e., $\gamma$, $J_3$, and $k_2$ (Figure~\ref{fig:librationsensitivity}). Shell thickness and density influence both $\gamma$ and $J_3$, although the effect of the density on $J_3$ is stronger than on $\gamma$. $D_{shell}$ also strongly correlates with $k_2$. 
The effect of the shell rigidity on the libration amplitude is small compared to the thickness and density which determine the shell's moment of inertia. This sensitivity suggests that the shell's response to the libration forces is mainly determined by  inertial forces, consistent with the discussions in Section~\ref{sec:shellthicknesslibration} and Appendix~\ref{sec:librationrigidity} and the results obtained by \citet{hemingwayNimmo_24}. The analysis shows that although libration is currently an effective constraint on the shell thickness, for obtaining very high accuracy on the shell and ocean thickness, precise measurements of the gravity field alongside the shape are more impactful. \textcolor{black}{The} lower sensitivity of libration is largely due to its dependence on the moment of inertia and flexibility of the shell which include effects from thickness, density, and rigidity of the shell.

\subsection{Constraining the interior of Enceladus with future observations}\label{sec:syninversion}
We use the preferred values of the observables obtained in Section~\ref{sec:twoscenarios} along with \textcolor{black}{the uncertainty on the measurements that could be available from future \textcolor{black}{observations} (see Section~\ref{sec:Feasibility})} to constrain the main interior structure parameters. 
\textcolor{black}{In order to break the degeneracy in identifying the location of the tidal dissipation regimes, it is necessary to adopt the measurements of both $k_2$ and $h_2$ and perform two separate inversions associated with either of the dissipation \textcolor{black}{scenarios}. We adopt the minimum required precision for measuring $h_2$ \textcolor{black}{according to the analysis} in Section~\ref{sec:twoscenarios} to resolve the degeneracy in determining the location of the tidal heating.}
\textcolor{black}{We present the uncertainties on the observables adopted in the inversion alongside the central values obtained in Section~\ref{sec:twoscenarios} in Table~\ref{table:precisions}}. We use the favored values given in this table for $J_3$, contrary to the mean value of 181.066$\times$10$^{-6}$ (adjusted for a reference radius of 251.99~km used here as compared to \citet{park_etal24}). Our preferred value $J_3$ is within the 2-$\sigma$ uncertainty range of that suggested by \cite{park_etal24}. We perform inversion using MCMC sampling and synthetic observations for two classes of measurement uncertainty given in Table~\ref{table:precisions} and discussed in Section~\ref{sec:Feasibility}. Each of our MCMC simulations includes approximately 600,000 accepted models which ensures \textcolor{black}{sufficient exploration of the parameter space}. The posterior distributions of the model parameters are demonstrated in Figures~\ref{fig:synthinversionshelldiss} and~\ref{fig:synthinversioncorediss} for the two sets of inversions (only shown for the higher uncertainty case). Results based on both cases of higher and lower uncertainties are summarized in Table~\ref{table:posteriorsprmtrs}.


For the case of lower measurement precisions in Table~\ref{table:precisions}, the results show that the suite of observations can constrain the thickness of the shell and the radius of the rocky core to within \textcolor{black}{$\pm$2.0~km, and $\pm$5~km, respectively (2-$\sigma$ uncertainty)}. When tighter constraints on the observables are adopted, the mean thickness of the shell can be constrained within approximately 1.2~km.
The ocean density is constrained within $\pm$40~kg/m$^3$ and $\pm$30~kg/m$^3$ in the cases where the tidal dissipation is dominated in the core and the shell, respectively. This low precision is roughly equivalent to a knowledge of seawater salinity to within 30~psu \citep[Earth's ocean salinity is 35~psu or parts per thousand,][]{Millero_2008}. An improved resolution of approximately $\pm$15~kg/m$^3$ can be obtained if the higher precision scenario mentioned in Table~\ref{table:precisions} is used. The corresponding 15~psu salinity precision would allow for distinguishing the nominal 10~psu salinity of \citet{postberg_etal09} from an \textcolor{black}{Earth-like} or greater salinity.

The viscosity and rigidity of the core are not well constrained if tidal dissipation is dominated in the shell because the range of $\eta_{\rm core}$ precludes a notable effect on the global tidal response of the core. This finding is consistent with the results of \citet{ermakov_etal21} and \citet{Genova_etal24}. 
The density of the core is constrained within 65~kg/m$^3$ and 50~kg/m$^3$ in case tidal dissipation occurs in the core and in the shell, respectively. If higher-precision measurements are acquired, the density of the core can be constrained within approximately 35~kg/m$^3$. 

In the case of a \textcolor{black}{tidally active} shell, the viscosity of the ice shell is constrained in the range of 4.4$\times$10$^{13}$--9.4$\times$10$^{13}$~Pa.s 
to satisfy the tidal heating constraint. 
In this tidal dissipation scenario, the shear modulus of the shell is constrained by the tidal Love numbers to be within 2.3--3.4~GPa. 
When tidal dissipation is dominantly in the core, the shell \textcolor{black}{viscosity} remains largely unconstrained. We constrain the effective viscosity of the core in this scenario to be 9.1$\times$10$^{11}$--1.4$\times$10$^{12}$~Pa.s. \textcolor{black}{We obtain somewhat} tighter constraints on the core viscosity $\sim$(9.5$\times$10$^{11}$--1.2$\times$10$^{12}$~Pa.s), if we have higher precisions on the measurements.
 In contrast to the viscosity of the core, its rigidity remains largely unconstrained.







\renewcommand{\thefigure}{\arabic{figure}}
\setcounter{figure}{9}

\renewcommand{\thefigure}{10a}

\begin{figure}[ht]
\hspace{-.5cm}\includegraphics[width=1.\textwidth]{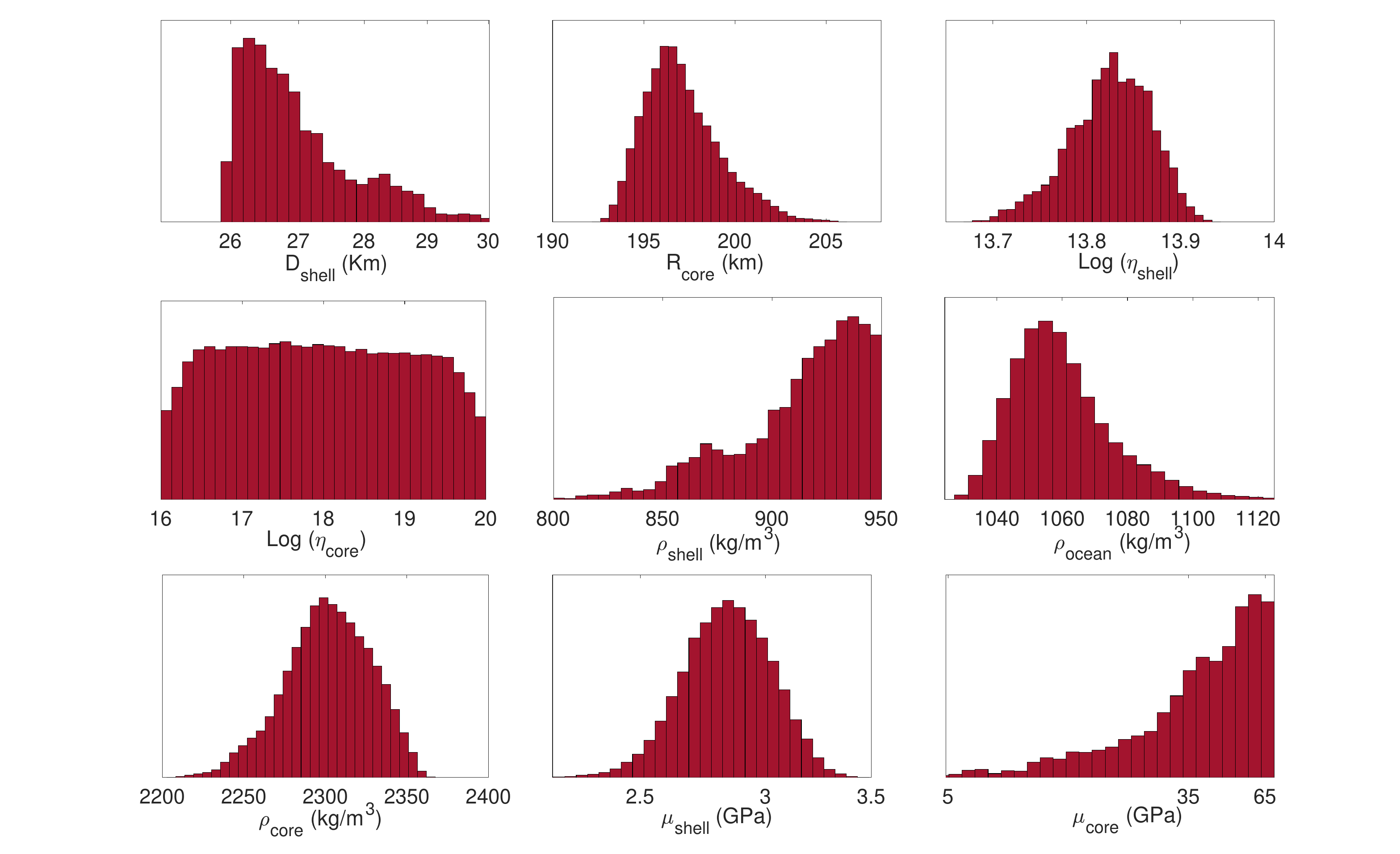}\\
\caption{\textcolor{black}{Results of the MCMC inversion for the case where the tidal dissipation is dominantly in the \textit{ice shell} and adopting the measurement uncertainty given in Table~\ref{table:precisions}. The shear moduli are plotted in log axes. Results are presented assuming the larger set of uncertainties shown in Table~\ref{table:precisions}.}}
 \label{fig:synthinversionshelldiss}
 \end{figure}

\renewcommand{\thefigure}{10b}

\begin{figure}[ht] 
\hspace{-.5cm}\includegraphics[width=1.\textwidth]{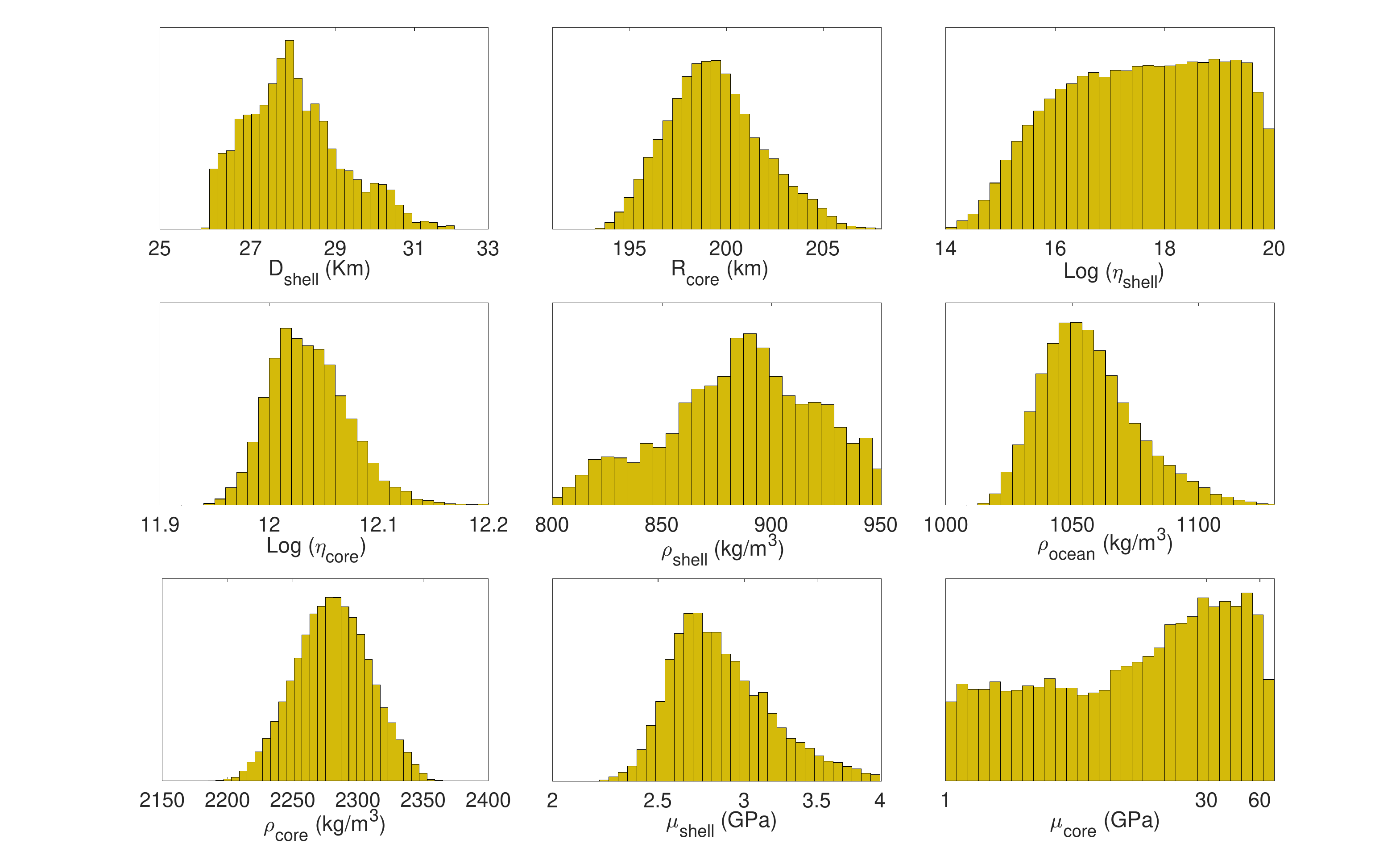}\\
\caption{\small{Results of the MCMC inversion for the case where the tidal dissipation is dominantly in \textit{the core}, provided the uncertainty on the measurements as shown in Table~\ref{table:precisions}. The shear moduli are plotted in log axes. Results are presented assuming the larger set of uncertainties shown in Table~\ref{table:precisions}.}}
\label{fig:synthinversioncorediss}
\end{figure}

\renewcommand{\thefigure}{\arabic{figure}}
\setcounter{figure}{10}

\begin{table}
\begin{centering}
\begin{tabular}{l l c c}
Quantity & Unit &  Tidal activity in the core &  Tidal activity in the shell \\
\hline 
$D_{\rm shell}$  & km & \centering 26.1--31.5 (26.1--28) &   26--30 (26--28.4)\\
$\eta_{\rm shell}$  & Pa.s & \centering 1.1$\times$10$^{14}$--10$^{20}$ (10$^{15}$--10$^{20}$) &  [4.4--9.4]$\times$10$^{13}$ ([5--8.3]$\times$10$^{13}$)\\ 
$\mu_{\rm shell}$  & GPa & \centering 2.25--4~(2.65--4)  &   2.3--3.4 (2.65--3.1) \\
$\rho_{\rm shell}$ & kg/m$^3$ &   \centering 800--950~(900--950) &  810--950 (860--950)  \\
$\rho_{\rm ocean}$ & kg/m$^3$ &  \centering 1025--1145~(1020--1060) & 1030--1120 (1040--1080) \\
$R_{\rm core}$ & km &  \centering 192--207~(193--200) & 193--205 (193--202) \\ 
$\mu_{\rm core}$ & GPa &  \centering 1-65~(1--65) & 2--70 (5--70)\\ 
$\rho_{\rm core}$ & kg/m$^3$ & \centering 2190--2360~(2260--2355) & 2210--2360 (2240--2355) \\ 
$\eta_{\rm core}$ & Pa.s &  \centering 9.1$\times$ 10$^{11}$--1.4$\times$10$^{12}$ (9.5$\times$10$^{11}$--1.2$\times$10$^{12}$) & 10$^{16}$--10$^{20}$ (10$^{16}$--10$^{20}$) \\ 
\hline
\end{tabular}
\caption{Posterior ranges of the model parameters  from two exemplars of MCMC inversions using synthetic \textcolor{black}{future measurements} with expected accuracies presented in Table~\ref{table:precisions}. Ranges are presented for two scenarios of measurement uncertainties.}
\label{table:posteriorsprmtrs}
\end{centering}
\end{table}

\subsection{Structural heterogeneity in the ice shell}\label{sec:shellheterogeneity}

\begin{figure}[htbp]  
    \centering
    \includegraphics[width=\linewidth]{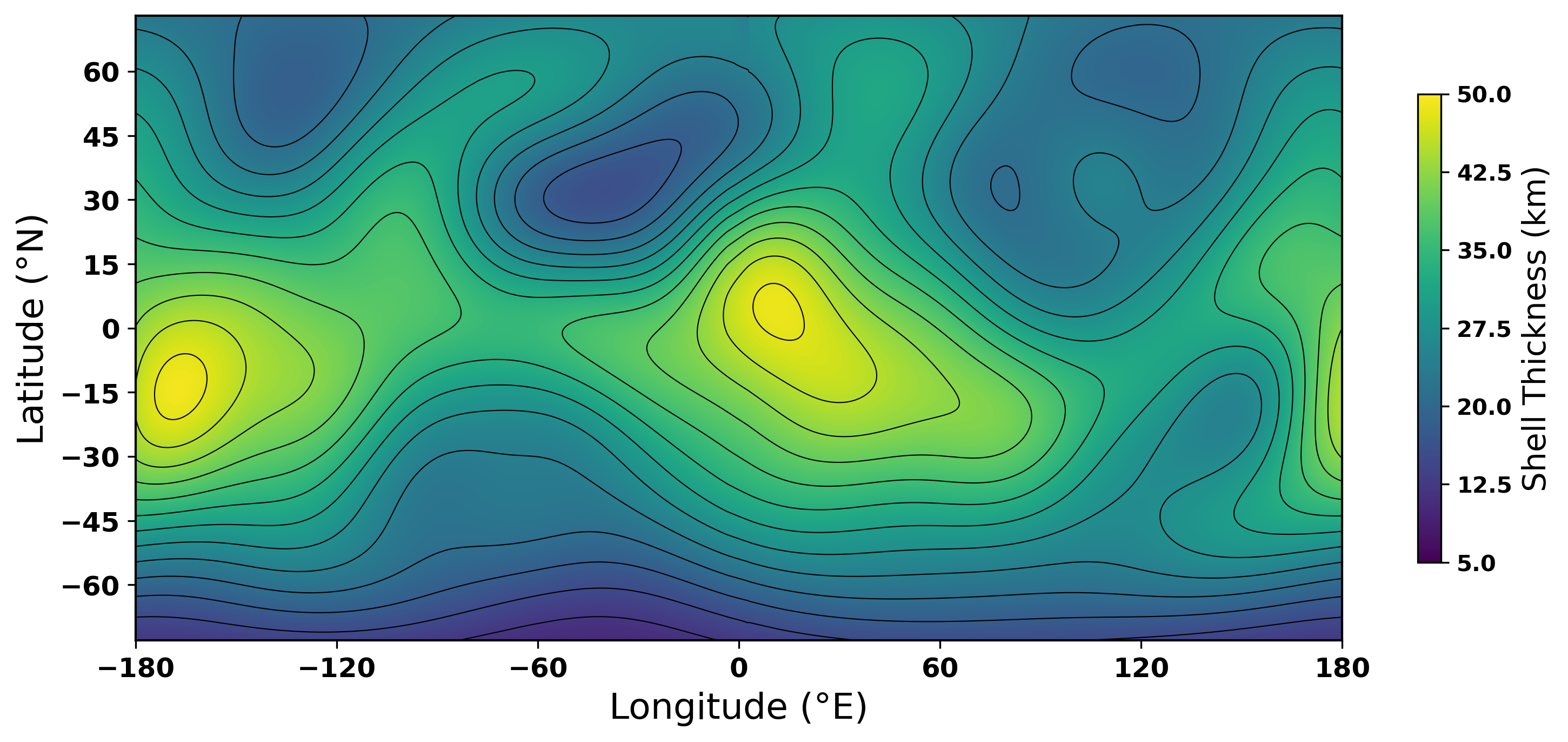}  
    \caption{\textcolor{black}{Variation of Enceladus's ice shell thickness based on a modified  shape model from \citet{park_etal24} assuming isostasy. Contours denote intervals of 2~km.}}
    \label{fig:3dshellmap}  
\end{figure}

\textcolor{black}{The required measurement precisions computed in Section~\ref{sec:twoscenarios} are based on a body \textcolor{black}{without any lateral variation of properties} (equations~\ref{eq:deltaD}). However, shape and gravity observations suggest that Enceladus's ice shell deviates from spherical symmetry, which in turn can change its tidal response \citep[]{bvehounkova_etal17, berne_etal23, berne_etal23b}.}
Here, we compute the tidal displacements of a heterogeneous ice shell and compare it to the response of a shell with a constant thickness. \textcolor{black}{The two cases have the same mean} thicknesses. We compute the crustal thickness variations by \textcolor{black}{applying a modified topography model from \citet{park_etal24}} evaluated up to $l = 8$ to the outer surface of geometries and altering the shape of the ice-ocean boundary to mirror the non-hydrostatic components of topography at the outer surface. \textcolor{black}{The topography model presented in \citet{park_etal24} contained artifacts including a depression and a mound, as a result of ignoring to incorporate the available limb data. The present analysis is based on this updated shape model. An erratum by \citet{park_etal24} is in preparation to publish the revised shape model.}
The relationship between structure at the ice-ocean boundary and surface topography follows \textcolor{black}{an Airy-type isostasy} \citep{hemingway_Matsuyama17} (see e.g. equation 19 in \cite{Berne2024}). \textcolor{black}{ For this shape model}, the thickness of the shell \textcolor{black}{ranges} between 5 and 50~km  (Figure~\ref{fig:3dshellmap}). 


The interaction between tidal forcing and thinning in the south polar region drives a quadrupole displacement field that increases $k_{22}$ and $h_{22}$ compared to $k_{20}$ and $h_{20}$ \citep{berne_etal23}. This displacement can increase $k_{22}$ and $h_{22}$ by up to 20$\%$, if Enceladus's mean ice shell thickness is thinner than 25~km, i.e. the size of structural heterogeneity is comparable to the \textcolor{black}{mean thickness of the ice shell. 
Consistent with  Figure~\ref{figure:tidalresponseviscs} and the discussion in Section~\ref{sec:heatgenerationclasses}, a dissipative core has a generally small effect ($<$10$\%$) on $h_{2m}$, but it can substantially alter $k_{2m}$ ($<\sim $300$\%$) compared to a non-dissipative core (see Appendix \ref{sec:Heterogeneityapp}). For thin shells, variations in the shell thickness and the tidal activity of the core act together and influence the overall $k_{2m}$ in a superimposed manner. 
\textcolor{black}{For a thin shell,} if the core is dissipative, its effect on $h_2$ is comparable to the effect of the shell thickness heterogeneity (details in Appendix~\ref{sec:Heterogeneityapp}).}

\begin{figure}[htbp]  
    \centering
    \includegraphics[width=1.\linewidth]{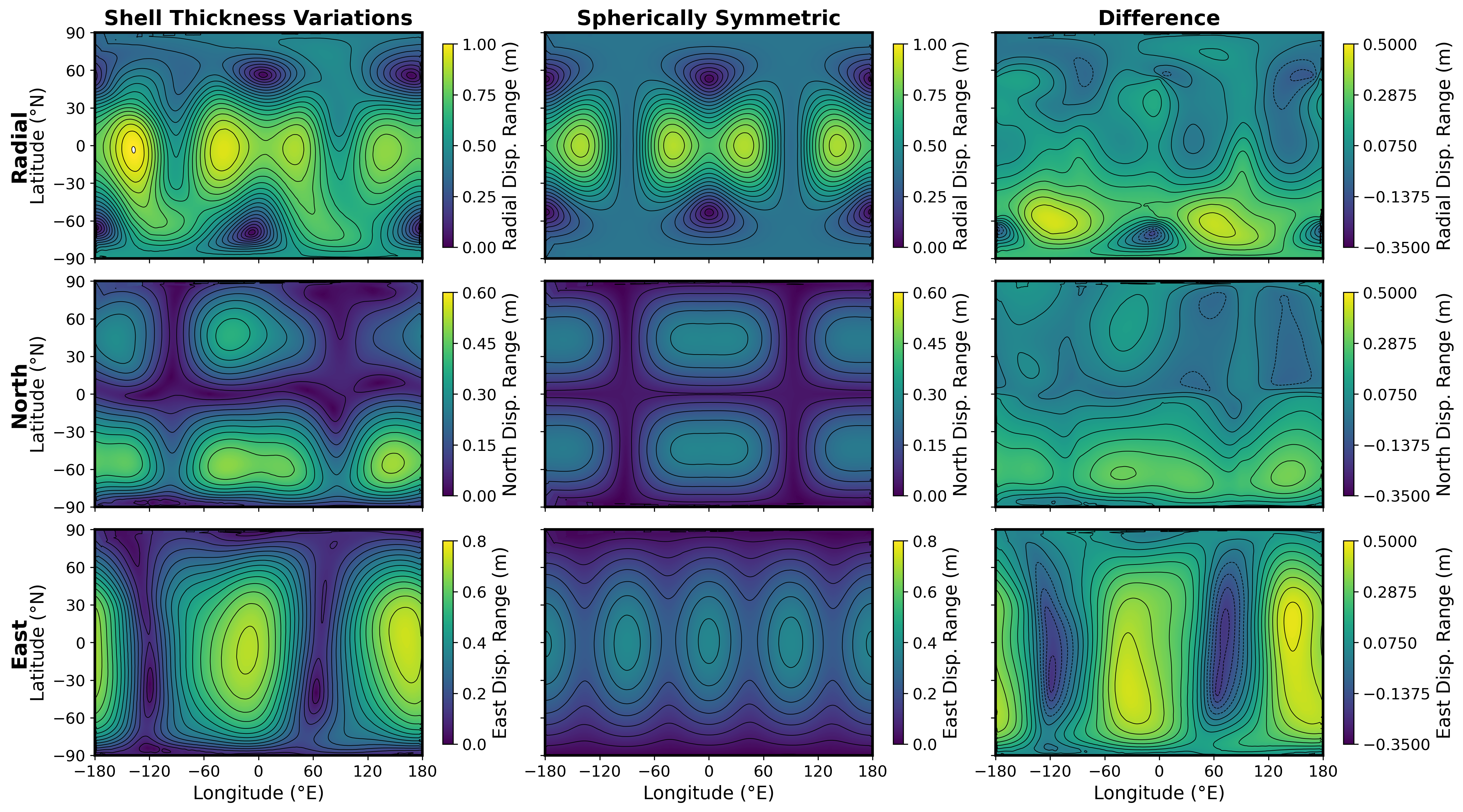}  
    \caption{Amplitude of tidal deformation in radial, northerly, and easterly directions for two cases of an ice shell with variable thickness as shown in Figure~\ref{fig:3dshellmap} and a \textcolor{black}{constant shell thickness} with the same average thickness equal to 30~km. Here, it is assumed that the tidal dissipation is happening in the rocky core and the viscosity of ice shell is 10$^{20}$~Pa.s, implying elastic deformations in the shell. Contours denote intervals of 0.05~m}  
    \label{fig:surfacedisp3D}  
\end{figure}

We assess the effect of the ice shell's structural heterogeneity on the maximum radial and horizontal tidal deformations on the surface compared to those derived in Section~\ref{sec:twoscenarios}. 
The overall maximum surface deformation can differ in comparison to the \textcolor{black}{case of constant thickness shell} used in previous studies. 
We compute the maximum displacements in the radial, northerly, and easterly directions for an ice shell with variable thickness and an ice shell of constant thickness with the same average thicknesses equal to 30~km (Figure~\ref{fig:surfacedisp3D}). In this figure, we assume that tidal dissipation occurs in the core and the viscosity of the ice shell is high enough to assume \textcolor{black}{a} purely elastic behavior. 
The ice shell with a heterogeneous thickness demonstrates variations in the radial and horizontal directions reaching 50~cm compared to a shell of constant thickness (Figure~\ref{fig:surfacedisp3D}). 
Larger deviations are observed for all components around the South pole, as expected, due to the significant thickness variations from the mean thickness \citep{berne_etal23b}. The global maximum radial displacement for the case of \textcolor{black}{a shell with variable thickness} is approximately 20~cm higher than the \textcolor{black}{constant thickness} case. The significant deviation of surface deformation demonstrated in these plots implies that for Enceladus, $h_2$ and $l_2$ can provide rough estimates of the actual surface displacements as commonly used for spherically symmetric bodies, and for a precise calculation of the displacement at any point on the surface, the effect of structural heterogeneity \textcolor{black}{has to be accounted for}. %
Since the globally maximum radial and horizontal displacements for a body of \textcolor{black}{constant shell thickness} are smaller than those for a shell with thickness variations, the required accuracies presented in Section~\ref{sec:twoscenarios} derived by assuming \textcolor{black}{variation only in radial direction} are conservative and thus remain valid.

\subsection{Feasibility of required measurements}\label{sec:Feasibility}

The static and time-varying gravity field of Enceladus can be recovered by tracking the motion of a spacecraft \citep{park_etal16, park_etal20, park_etal24}. The shape and topography of Enceladus can be derived using spacecraft measurements such as imaging \citep{park_etal19, park_etal20} or altimetry \citep{smith_etal01}. The accuracy of recovered parameters typically depends on the measurement sensitivity, orbiting geometry, and spacecraft operating conditions. With reasonable assumptions, we can compute the expected precision of recovered Enceladus gravity and tidal parameters calculated in Section~\ref{sec:twoscenarios} presented in Table~\ref{table:precisions}  \citep{smith_etal01, park_etal16, park_etal19, park_etal20, park_etal24}.

Previously, \citet{park_etal20} simulated a scenario in which a spacecraft is in a stable orbit around Enceladus with approximately 60$^{\circ}$ inclination and periapsis and apoapsis altitudes of 150 km and 200 km, respectively. Considering a typical Deep Space Network (DSN) X-band tracking capability, $k_{2}$ and $J_{3}$ could be easily recovered with accuracies of $<0.002$ and $0.4\times10^{-6}$ \citep{park_etal20}, respectively. Considering a similar orbit geometry, but with a reasonable mission duration and a more capable Ka-up/Ka-down tracking system, $k_{2}$ could be recovered with an accuracy of $<0.0002$ and $0.04\times10^{-6}$ \citep{park_etal20}, respectively.

If an imaging system with high resolution and accurate pointing capabilities is used, temporal changes in surface features can be measured with an accuracy better than approximately half a pixel using a stereo reconstruction technique \citep{park_etal20}. Assuming an imaging system with 1-meter resolution and $<1$-meter pointing accuracy, the libration amplitude can be measured with $<1$-meter accuracy. \textcolor{black}{This is partly possible because the reconstructed orbit accuracy using DSN data will be at the 1-m level.} Furthermore, tidal deformation can be measured with an accuracy $<50$ cm (1-$\sigma$). This scenario assumes that images of the same surface feature are collected under varying observation geometries and at different phases of Enceladus relative to Saturn.

An onboard altimeter would be ideal for directly measuring radial \textcolor{black}{deformation \citep[e.g.,][]{steinbrugge_etal15, dirkx_etal19, thor_etal20}}. By assuming a similar orbiting geometry and sampling strategy with an altimetry precision of $<$5~m, the radial deformation could be recovered with an accuracy $<25$~cm (1-$\sigma$) using surface tie points (i.e. repeat observation of the same surface point at different orbital phases). 

\section{Conclusions and discussion}\label{sec:concludiscussions}

Determining the interior properties of Enceladus remains challenging given the currently available observations.
We evaluated the extent to which geodetic measurements can constrain Enceladus’s interior structure and the mode of its tidal heating. We presented constraints on the interior structure of Enceladus by revisiting the latest analysis of geodetic data \textcolor{black}{acquired} by \textit{Cassini}. Based on these analyses, we provide estimates of the plausible ranges of the currently unavailable tidal response, i.e., $k_2$, $h_2$, and $l_2$, and their phase lags. 
The tidal Love numbers of Enceladus compatible with the currently available observations span a wide range depending on the interior properties.
We showed that a potential measurement of the $k_2$ and phase lag can be explained by tidal heat generated \textcolor{black}{either} in the shell or in the core, representing a case of degeneracy that requires \textcolor{black}{additional} constraints to be resolved. 
Our analysis shows that measuring \textcolor{black}{either the radial or tangential tidal displacements can discriminate between core-dominated and shell-dominated tidal dissipation cases. Measuring the associated phase lags can also differentiate between the two scenarios.}
If both real and imaginary parts of $k_2$ is measured within 0.002, measuring the maximum vertical and horizontal displacements within 82~cm and 40~cm are sufficient to obtain a 2-$\sigma$ confidence for discriminating between the two heat generation scenarios. The phase lags of $h_2$ \textcolor{black}{and $l_2$} differ significantly between the two cases of tidal dissipation. In the case of a core-dominated tidal dissipation and an elastic ice shell, \textcolor{black}{both} $\epsilon_{h_2}$ \textcolor{black}{and $\epsilon_{l_2}$} are very small, while for a shell-dominated tidal dissipation, \textcolor{black}{both} $\epsilon_{h_2}$ \textcolor{black}{and $\epsilon_{l_2}$} can reach tens of degrees (10$^{\circ}$--50$^{\circ}$).
The requirements for measuring $h_2$ and $l_2$ are \textcolor{black}{further} relaxed \textcolor{black}{in case} refined constraints on $k_2$ are obtained. If real and imaginary parts of $k_2$ is constrained within 0.0002, measuring the vertical and/or tangential tidal displacements within 116~cm and 51~cm, respectively, is sufficient to discriminate between the two classes of tidal dissipation. \textcolor{black}{We note that while in the case of most of the moons such as Europa and Ganymede measuring the tidal phase lags is probably necessary to constrain the heat generation mechanisms \citep{hussmann_etal16}}, here we demonstrate that the anelastic deformations \textcolor{black}{at Enceladus} can be large enough to differentiate between the main heat generation scenarios. \textcolor{black}{Yet, obtaining tighter constraints on the rheology of the interior layers demands measuring the phase lags, in addition to the amplitudes of the Love numbers.}

We also investigated the effect of the structural heterogeneities of the ice shell on the tidal response. Our results show that in the case of a largely elastic shell the surface displacements can significantly deviate from those of a shell \textcolor{black}{of constant thickness}. These deviations can reach more than $\sim$50~cm for the thinnest parts of a heterogeneous shell compared to a shell \textcolor{black}{of constant thickness} with the same global mean thickness.
This observation is important for mission concept design purposes and for determining the locations at which the displacements are measurable.
We quantified the effect of a tidally active core on the tidal Love numbers of Enceladus with a heterogeneous ice shell. We found that if the core is tidally active to the extent needed to explain the steady-state heat loss rate, its effect on the observed $k_2$ in the presence of a heterogeneous shell can be comparable or larger than the effect of the ice shell heterogeneities.

We conducted inversions using synthetic observables and demonstrated their efficacy for probing the main interior structure parameters, using a set of feasible measurements and uncertainties. As one of the important factors in determining habitability,  we demonstrated that high- and low-density ocean scenarios can be differentiated using the geodetic measurements.
Such different interpretations originate mainly from the underlying assumptions on the provenance of the plume material, the potential change in their composition during their transport from the ocean, and the undetermined extent of overturning circulation and possible ice-ocean interactions \citep{Zhu_2017, soderlund19, kangFlierl2020,Lobo_2021,Kang_2022b, meyer_etal25, Ames_etal25, bouffard_etal25}. These interpretations differ by more than 40~psu \citep[see][]{{postberg_etal09, glein2018geochemistry, gleinWaite20, Fifer_2022, postberg_etal23}}.
The constraint obtained on the ocean density from our geophysical measurements is not strong, but it can differentiate between a low-density nearly pure ocean and a high-salinity high-density ocean. This constraint can be strengthened when precise measurements, mainly from the static and time variable gravity field, are obtained.
\textcolor{black}{We also constrain the core density and rheology which can help distinguish between a low-density porous core and a higher-density rocky core and help us understand its composition, formation, and thermo-chemical evolution.}

\textcolor{black}{The analyses presented here imply underlying assumptions as elaborated here.
We base the estimated tidal dissipation used in our analysis on the heat loss rates from previous studies (i.e., \citet{hemingwayMittal19, howett_etal11, nimmo_etal18, howett_etal25}). Assuming that the estimated heat loss is equal to the internal heat production by tides implies that Enceladus is currently in thermal steady state. However, the thermal state and the orbital state of Enceladus, which tightly affect one another, are not well understood. If Enceladus is currently experiencing an episodic or periodic cycle of melting and freezing of its hydrosphere, as suggested by \citet{ONeillnimmo_2010, nimmo_etal18, goldreich_etal25}, the rates of heat production and the heat loss can be different. Although, these scenarios are hindered by requiring that we happen to be observing Enceladus in a special time of its overall evolution. Alternatively, the resonance locking hypothesis implies a nearly steady state and that the high tidal dissipation at Enceladus is sustained over long timescales. This scenario can best explain the observed Saturnian system as discussed in detail in \citet{nimmo_etal18, nimmo_etal23}}.
Through obtaining independent observations of total heat loss and tidally produced heat and comparing them, we can understand whether Enceladus is in a thermal and orbital steady state.

We also note that he measurement requirements presented in this paper are computed under the assumption that the total \textcolor{black}{heat produced} is partitioned by at least a ratio of 90\% and 10\% between the core and the shell. These cases are the most likely scenarios, considering the patterns observed in Figure~\ref{figure:heatviscs} where very limited values of ice and rock viscosities represent comparable dissipation in the shell and in the core. However, in a general case where partitioning of heat between the shell and the core are comparable, 
quantifying the tidal dissipation rate in the two parts of the body is possible by measuring the complex $k_2$ and $h_2$, and the density structure of the body (equation~\ref{finalheatinshellwithlove} derived here). In this case, the uncertainty of quantifying the heat partitioning depends on the uncertainties on the parameters and observables involved in~equation~\eqref{finalheatinshellwithlove}.

We also point out that the computed values of shell and core viscosity should be considered as the ``effective" viscosity \textcolor{black}{of the layer}. The viscosity profiles can depend on the variation of parameters such as porosity, temperature, and the water saturation percentile with depth. However, even if other mechanisms than viscoelastic dissipation are producing the inferred tidal heating, their effect on the tidal response will remain the same, \textcolor{black}{and thus our analysis remains valid}. 

The methodology presented here does not consider the possibility of tidal dissipation in the ocean, as a result of resonantly enhanced tidal flow as discussed in \citet{hayMatsuyama17, matsuyama_etal18, tyler20}. A small $h_2$ amplitude and phase lag can rule out the occurrence of tidal dissipation in the shell, but may be associated with tidal dissipation occurring either in the ocean or in the core. However, significant ocean tidal heating at Enceladus is very unlikely because it would require a very thin ocean (less than a few kilometers thick) to cause the resonance \citep{hayMatsuyama17, matsuyama_etal18}. Such a thin ocean is incompatible with the constraints we obtain on the interior structure from the available estimates of the libration, gravity field, and total mass of Enceladus.

The necessary measurements suggested here can be achieved by a future mission that provides additional constraints on the static gravity field, topography, libration, tidal Love numbers, and rate of surface heat loss.
The orbital state can be \textcolor{black}{studied} by the ground-based or spacecraft astrometry.
\textcolor{black}{The methodology presented here is general and can be used to investigate the modes of tidal heating in other icy satellites such as Europa, Ganymede, and Titan, through geophysical data which will become available by the upcoming Europa Clipper, Juice, and Dragonfly missions.}

\section{Acknowledgments}
A. Bagheri acknowledges the support from the Swiss National Science Foundation by grant number P500PT\_214435. A. Berne is supported by the Future Investigators in NASA Earth and Space Science and Technology (FINESST) Program (80NSSC22K1318). We acknowledges insightful discussions with Francis Nimmo, Michael Efroimsky, Mikael Beuthe and Sander Goossens. 
We also acknowledge the two anonymous reviewers for their constructive suggestions which helped improvement of the paper.

\clearpage

\begin{appendix}

\section{Viscoelastic models}\label{sec:viscoelasticmodels}

Different models are proposed to capture the viscoelastic behavior of materials such as Maxwell, Andrade, Extended Burgers, and Sundberg-Cooper. The models differ in their representation of the transient regime between largely elastic to viscous dominated dissipation. If the frequency of excitations is close to the nominal Maxwell relaxation time of the body ($\tau_M = \eta/\mu$) or some layers of the body (i.e. $\tau_{M}(core) = \eta_{core}/\mu_{core} $ or $\tau_{M}(shell) = \eta_{shell}/\mu_{shell}$), the effect on the predicted response from different viscoelastic models can differ. However, if the excitation frequency (tidal period) is not close to the Maxwell relaxation time (e.g., in the case of Titan with a tidal period of $\sim$16 days), the choice of viscoelastic models does not change the result. For Enceladus, the orbital frequency is relatively close to the Maxwell relaxation time of the ice with its mechanical properties observed under Earth conditions. Therefore, Maxwell's model may not be appropriate for predicting Enceladus's tidal response.

\begin{figure}[ht]
\centering
\includegraphics[width=.48\textwidth]{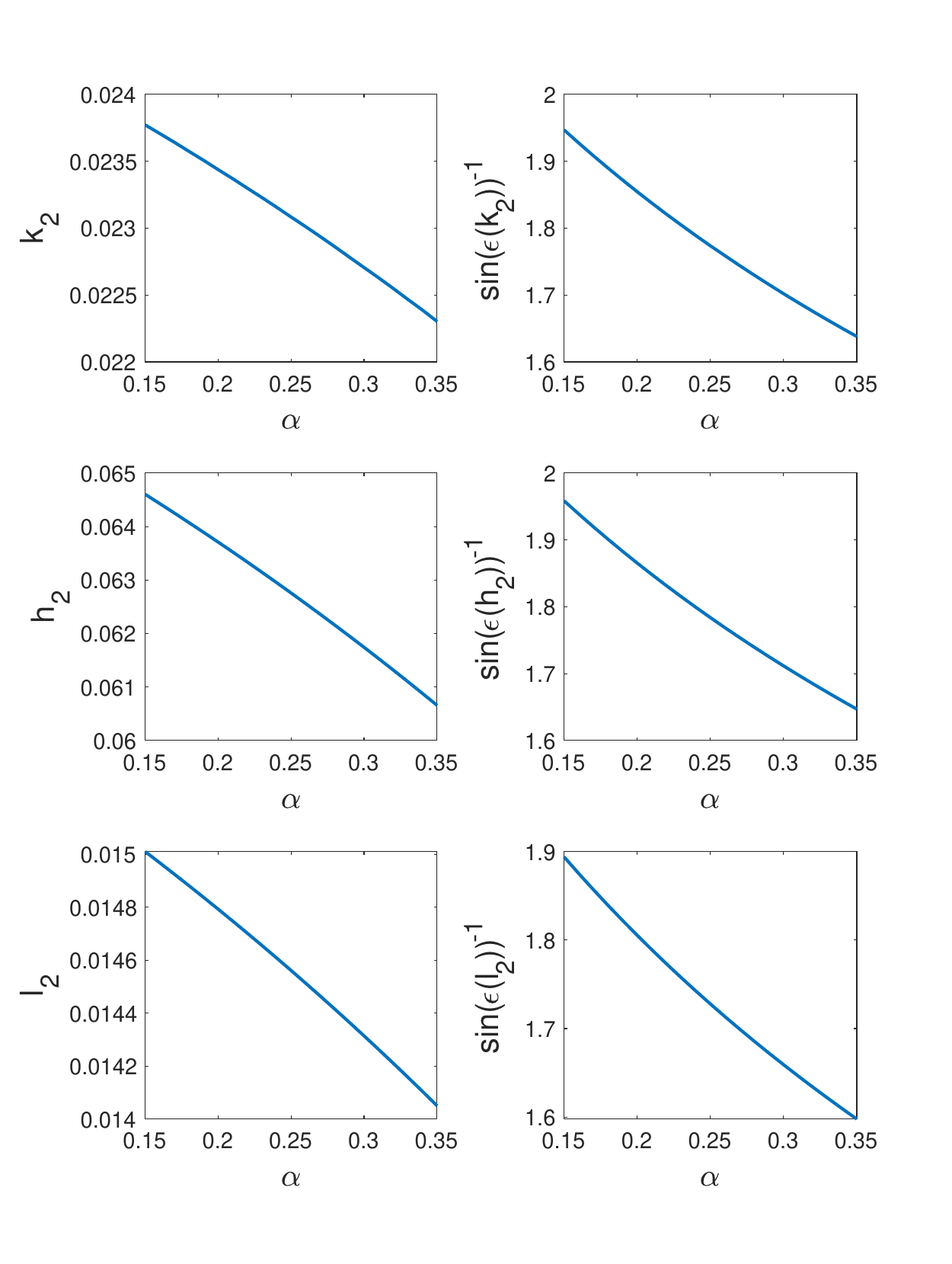}
\includegraphics[width=.48\textwidth]{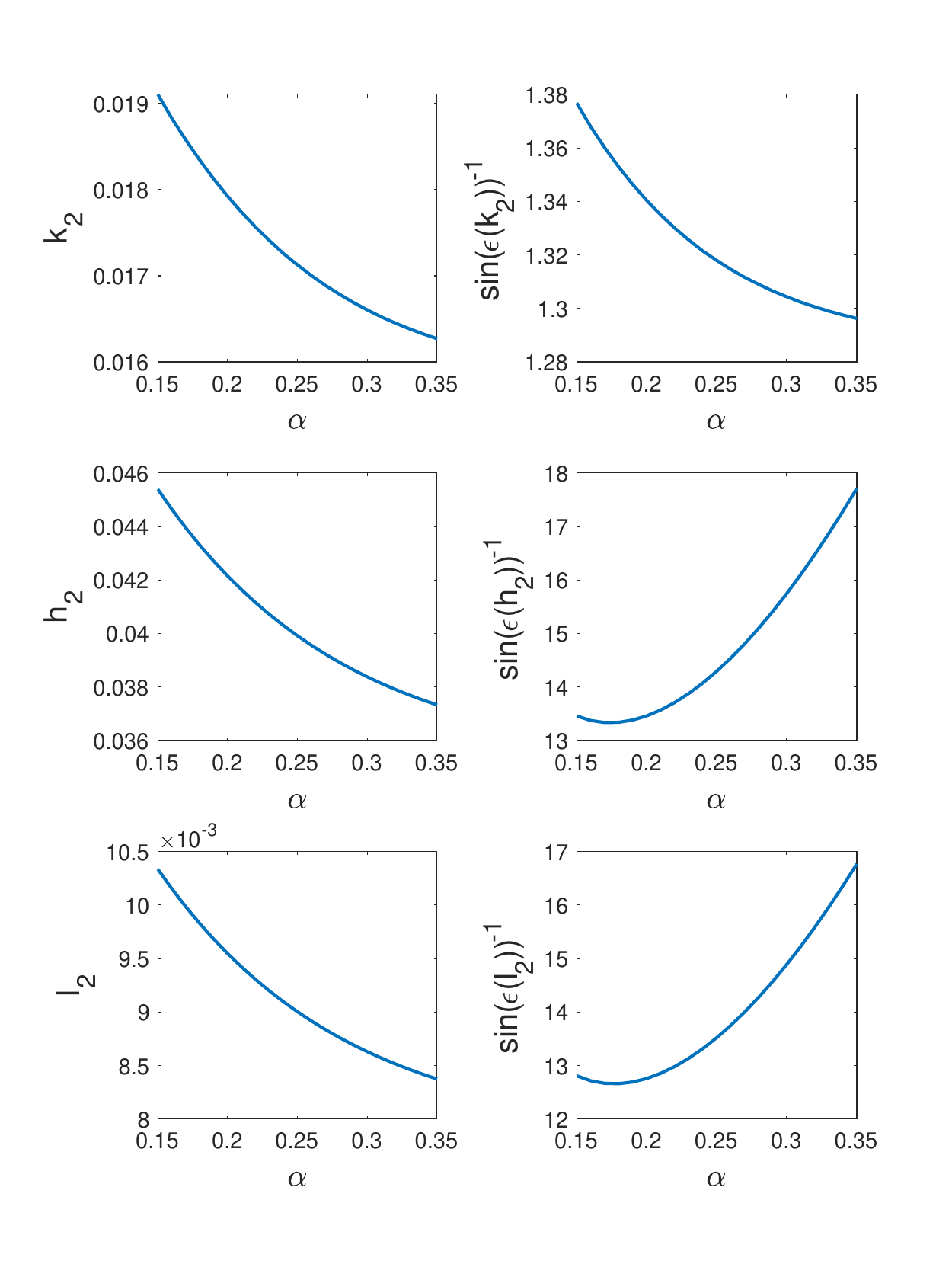}
\caption{Effect of changing the Andrade parameter $\alpha$ on the tidal response of Enceladus. \textcolor{black}{\textcolor{black}{Left and Right} panels show the tidal response parameters for models of tidal dissipation dominantly in the shell  and in the core, respectively.}}\label{fig:andradeparameter}
\end{figure}

Because periodic tidal deformations are small, we assume that the stress‐strain relationship is linear, such that the response of the material to shear forcing is described in the time domain by the creep function $J(t)$. The creep function for Maxwell's model is written as:
\begin{equation}
J(t) =  J_U +  \frac{t}{\eta},
\end{equation}
whereas for the Andrade viscoelastic model, the creep function is given by \citep{Andrade62}:
   \begin{equation}
J(t) =  J_U + \beta t^\alpha+ \frac{t}{\eta},
\end{equation}
where $\alpha$ and $\beta $ describe a time-scale dependence of the anelastic relaxation strength. The associated dynamic compliance, which
represents the response to a sinusoidally time-varying
stress, has real and imaginary parts given by the
following functions \citep{findley_Etal13}:
\begin{align}
&    \Re{(\hat J(\omega))} = J_U + \beta \Gamma(1+n)\omega^{-n}\cos(n \pi/2), \\
&  \Im{(\hat J(\omega))} = -\beta \Gamma(1+n)\omega^{-n} \sin(n \pi/2) - 1/\eta \omega,
\end{align}
where $\Gamma(1+n)$ is the Gamma function and $\omega$ is the angular frequency. Here, $\alpha$ (frequency dependence) and $\beta$ are empirical parameters that are not well constrained. $\alpha$ is usually assumed to be in the range 0.1--0.4 based on laboratory experiments \citep[e.g.,][]{JacksonFaul10} and $\beta$ is suggested to be approximately $\mu^{\alpha-1}/\eta^{\alpha}$ \citep{castillorogez_etal11}. Currently, these parameters are not well constrained for most planetary bodies \citep[see e.g.][]{nimmo_faul13, bagheri_etal19, bierson24, petricca_etal25}. 
Given the current lack of such constraints for Enceladus, we follow \citet{Andrade62} and \citet{FaulJAckson15}, and use $\alpha=0.2$ and $\beta=1$. We investigate the effect of variation in the Andrade parameters in Figure~\ref{fig:andradeparameter}.  The approximation for $\beta \approx \mu^{\alpha-1}/\eta^\alpha$ \citep{castillorogez_etal11} is employed. We find that the change in the real parts of the Love number as a function of the changes in the Andrade parameters within the range of conventionally assumed end-member values remains less than 10\% if the dissipation is dominated in the shell and less than 20\% if the dissipation is dominated in the core.

The expressions for Maxwell's model can be obtained by setting $\beta =0$ in the above equations. The relaxed (effective) shear modulus is thus calculable as \citep{findley_Etal13}:
\begin{equation}
        \mu = [J_1^2(\omega) + J_2^2(\omega)]^{-1/2},
\end{equation}
and shear dissipation is
\begin{equation}
    Q^{-1}_B = J_2(\omega)/J_1(\omega),
\end{equation}
\textcolor{black}{where $J_1$ and $J_2$ are the real and negative imaginary parts of the dynamic compliance $\hat J(\omega)$.}
\begin{figure}[ht]
\begin{center}
\includegraphics[width=.32\textwidth]{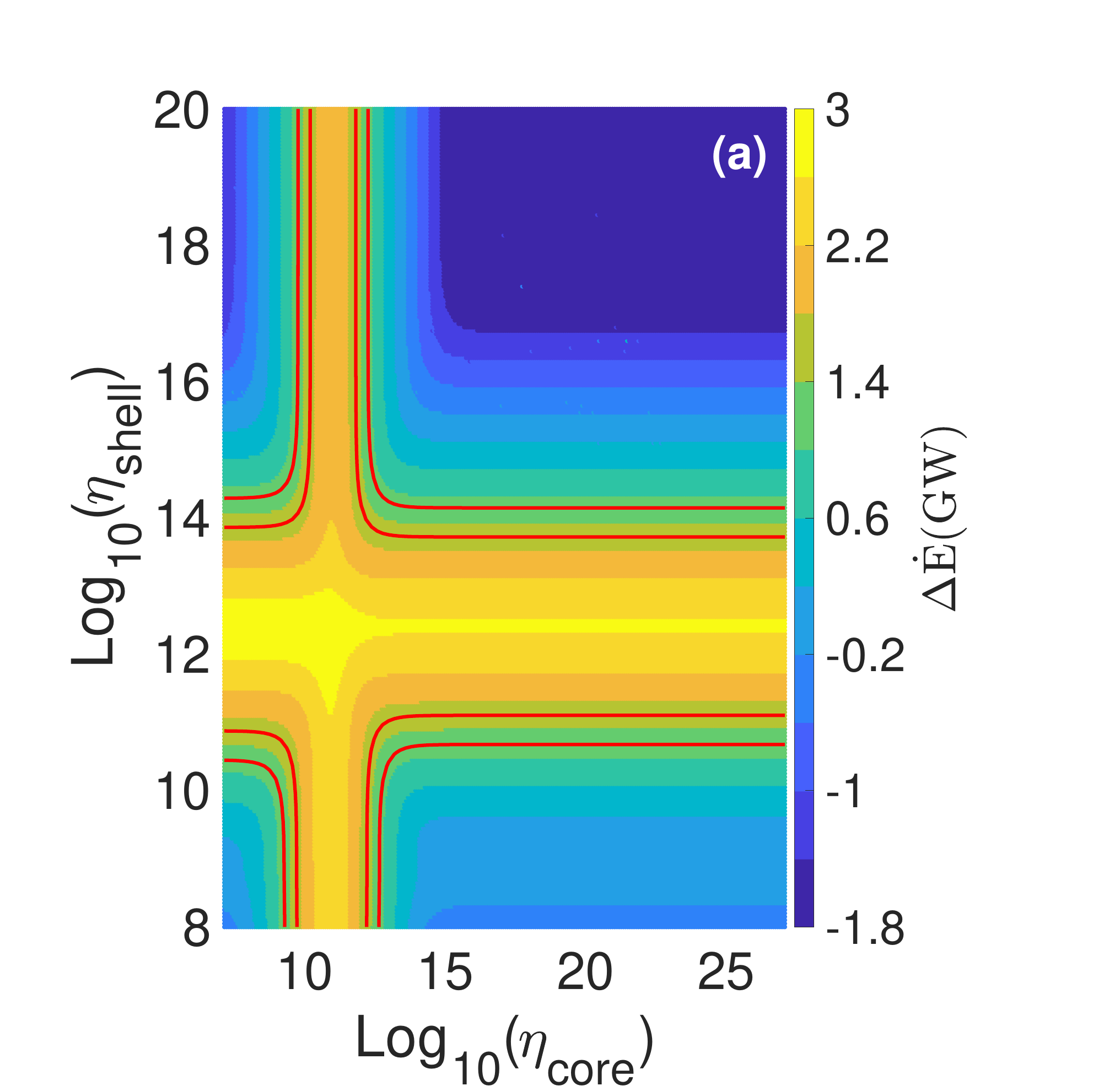}
\includegraphics[width=.32\textwidth]{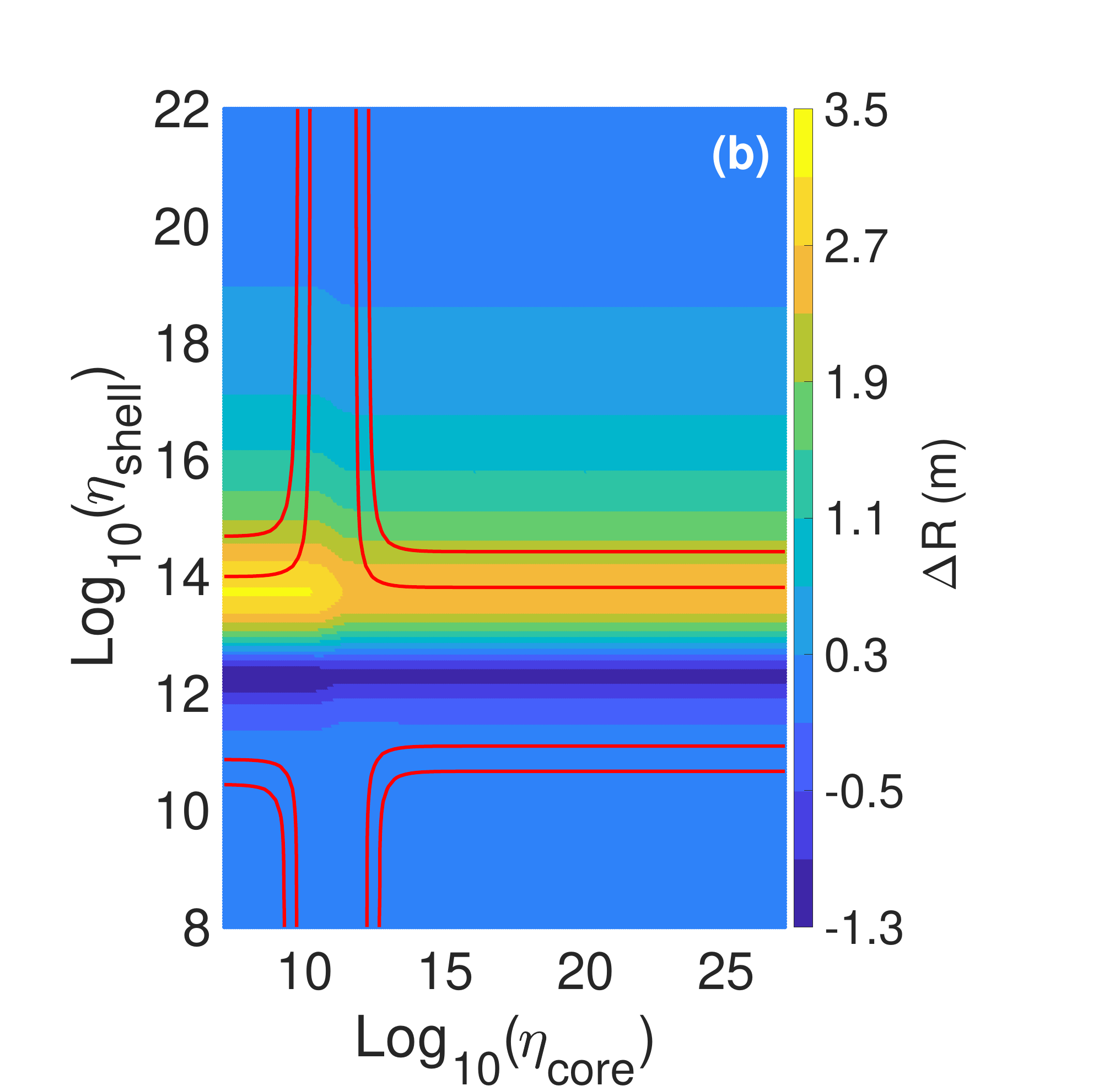}
\includegraphics[width=.32\textwidth]{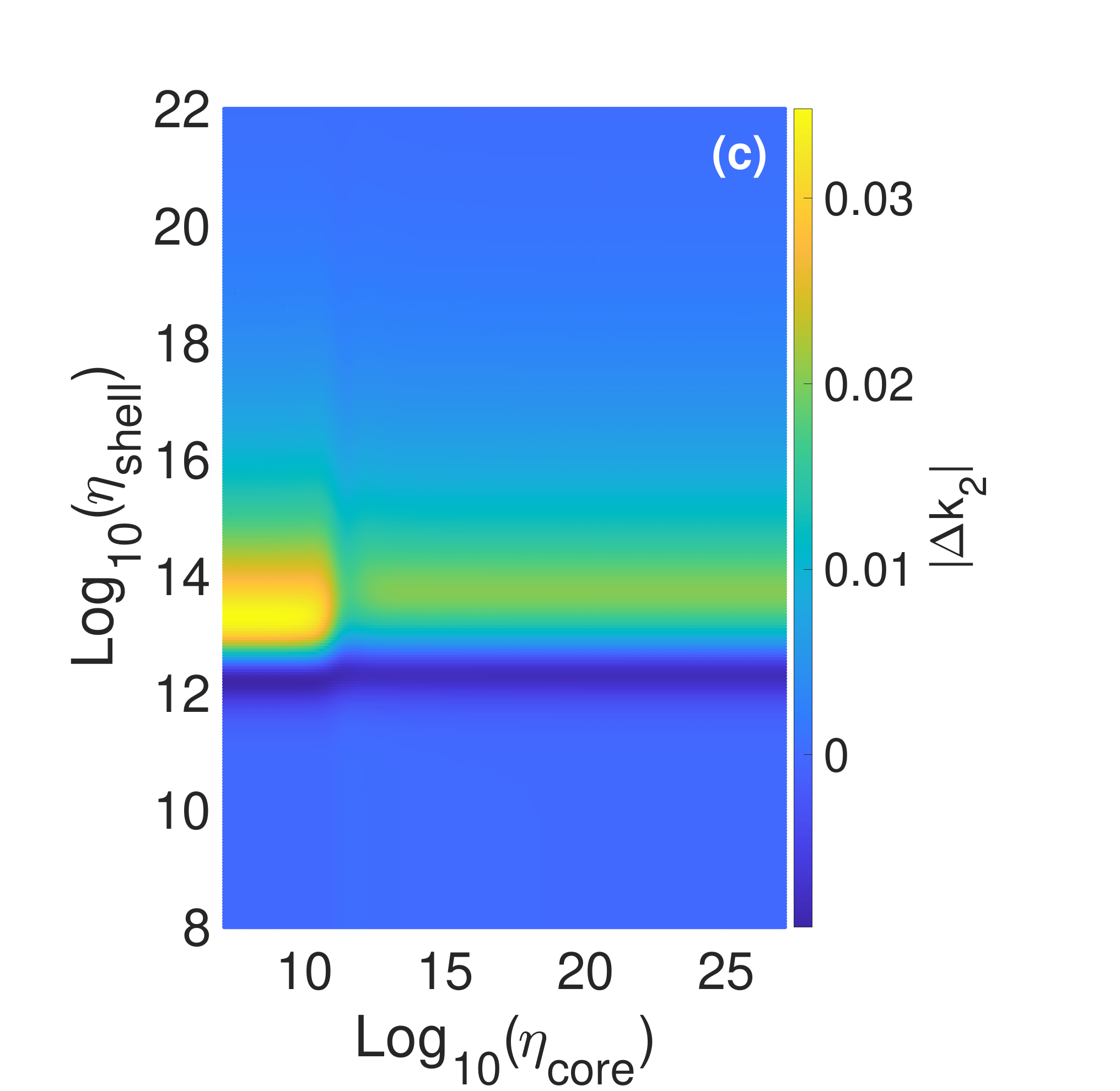}\\
\includegraphics[width=.32\textwidth]{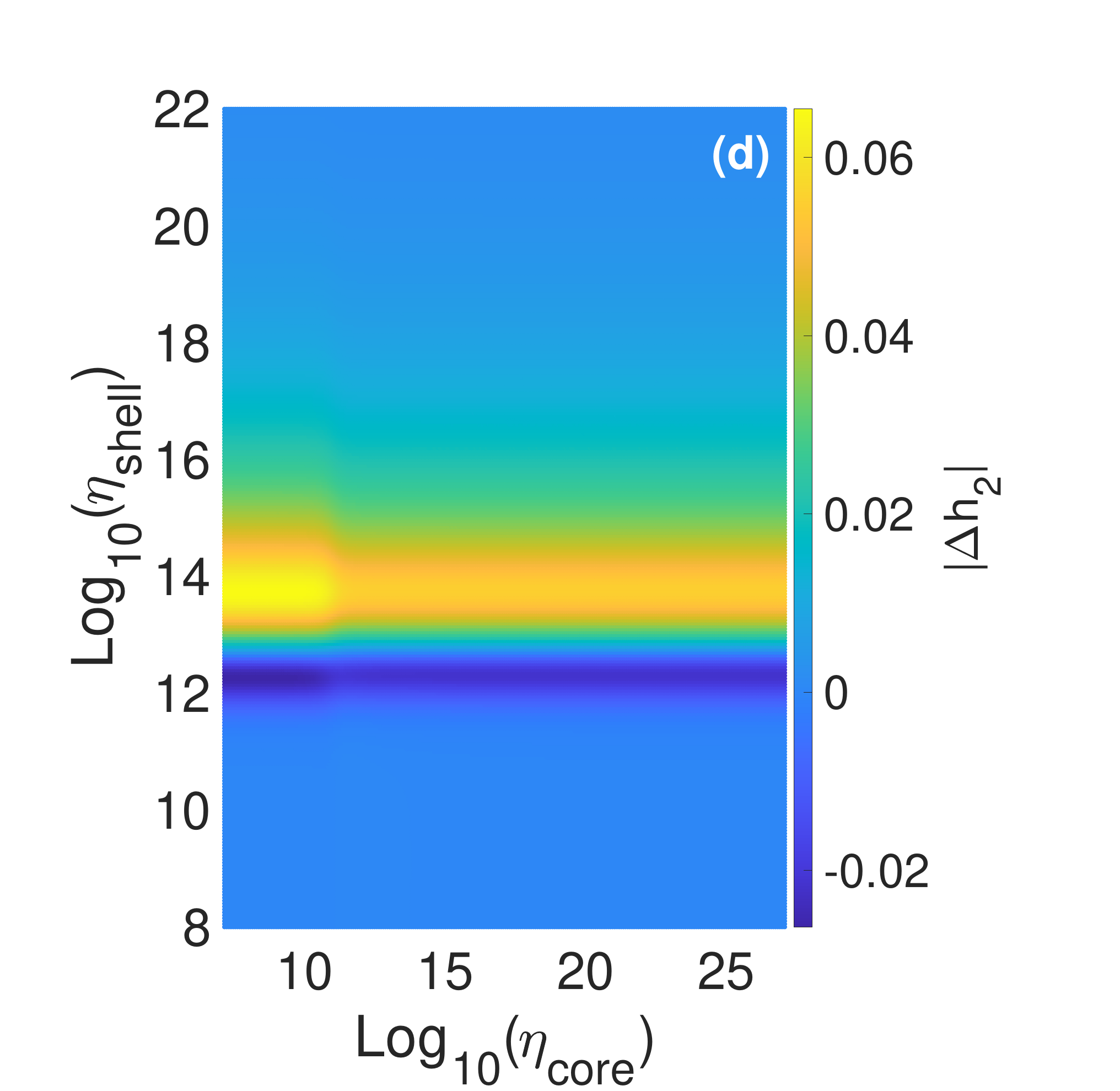}
\includegraphics[width=.32\textwidth]{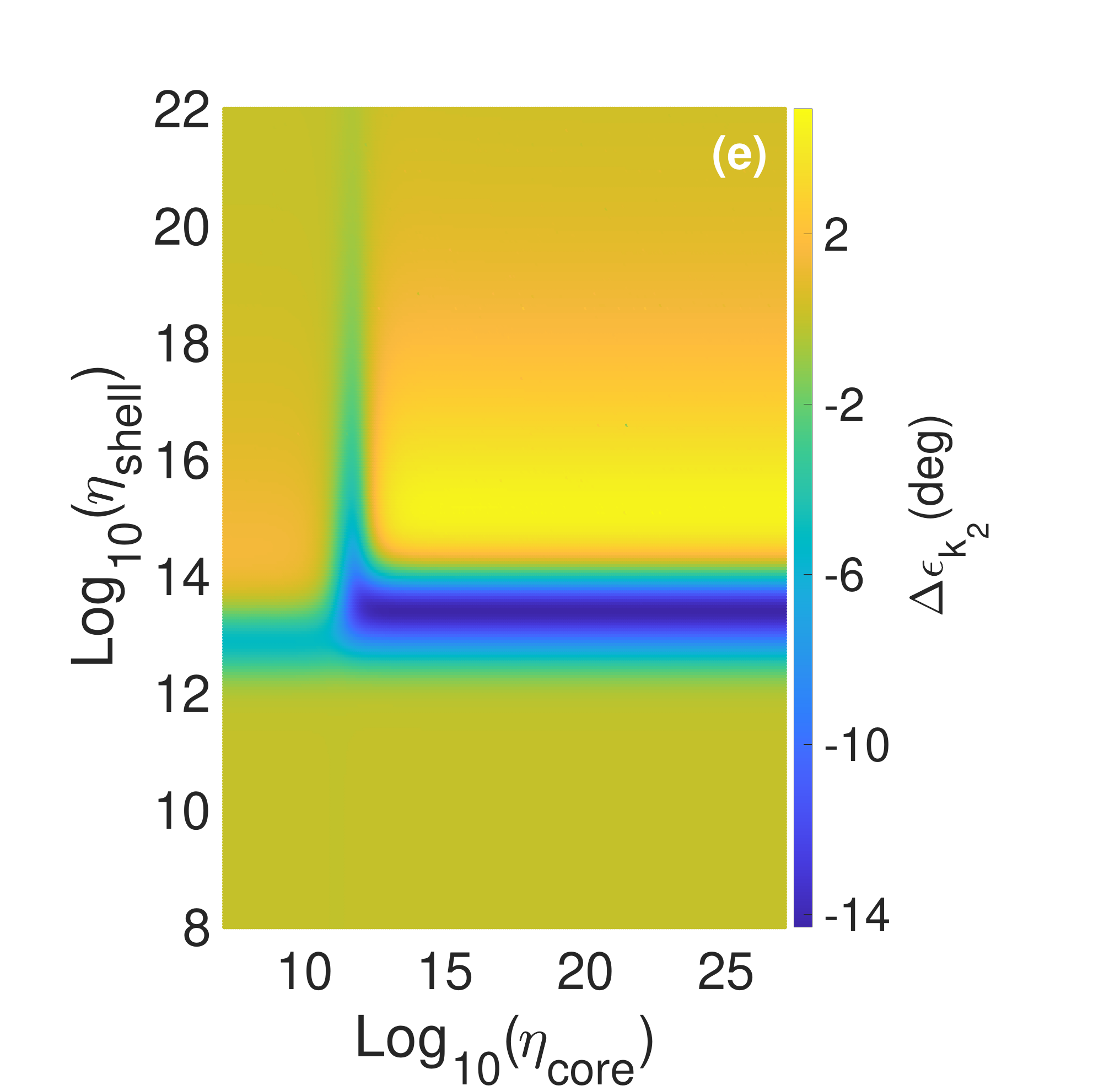}
\includegraphics[width=.32\textwidth]{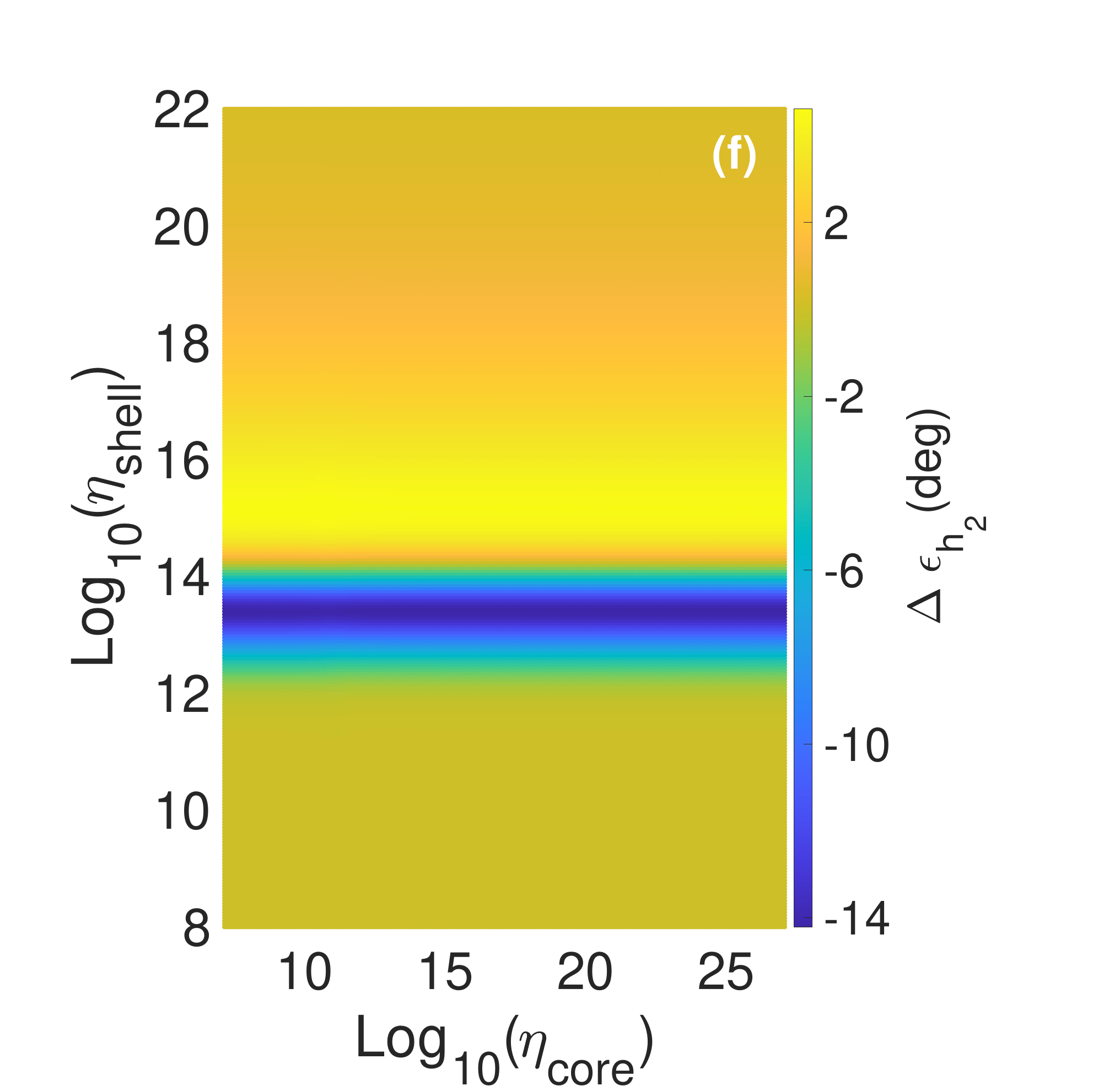}
\caption{\small{Difference between different tidal observables as due to the use of the Maxwell and Andrade rheology: (a) tidal heating ($\Delta \dot E$), (b) maximum radial displacements ($\Delta R$), $k_2$ (c) and $h_2$ (d) Love numbers ($\Delta k_2$ and $\Delta h_2$) and associated phase lags ($\Delta \epsilon_{k_2}$ and $\Delta \epsilon_{h_2}$) (e, f). The four narrow regions constrained by the red lines in panels (a) and (b) indicate the estimated tidal heat (see Figure~\ref{figure:heatviscs}).}}
\label{fig:maxdifferencetidal}
\end{center} 
\end{figure}

We show the differences between the tidal response parameters calculated based on Maxwell and Andrade viscoelastic models (for both the ice shell and the core), including the amplitudes of $k_2$ and $h_2$, their phase lags, and the resulting tidal heating (Figure~\ref{fig:maxdifferencetidal}). $h_2$ is relatively insensitive to the choice of the viscoelastic model for the rocky core due to the isolating effect of the ocean.
In contrast, $k_2$ is sensitive to the choice of the viscoelastic model both for the ice shell and, to some extent, for the rocky core. The phase lag of $k_2$ can differ by as much as $\sim$14$^{\circ}$ between the Maxwell model and the Andrade model.

\clearpage

\clearpage
\section{Derivation of partitioning of tidal heating  between the shell and the core}\label{sec:heatpartioningderivation}

We can compute the tidal heating in the specific region of the satellite, such as the core or the shell, by integrating the radial shear dissipation in that part of the body as a result of tidal perturbations. 
Note that equation \eqref{tidalheatingEquation} cannot be used for the core or the surrounding layers simply by replacing $k_2/Q$ or $\Im(k_2)$ of the body by the tidal Love numbers of the core or the layers computed independently of the rest of the body. Here, we use equations~(33) and (37) in \citet{tobie_etal05} to derive the contribution of the core to the global tidal heating:
\begin{equation}\label{radialheatapp}
    H_{tidal}(r) = -\frac{21}{10} \frac{ \omega^5 e^2 R_E^4 }{r^2}H_{\mu} \Im{(\mu(r))},
\end{equation}
where $H_{tidal}$ is the radial distribution of the dissipation rate per unit volume, $r$ is the radius from the center, $H_{\mu}$ is the sensitivity kernel for shear dissipation given in equation~(33) in \citet{tobie_etal05}, and $\Im{(\mu(r))}$ is the imaginary part of the complex shear modulus which is assumed here to be constant throughout each layer. By integrating this expression from the center of the satellite to the core-ocean boundary, we obtain:

\begin{equation}\label{coredissapp}
    \dot E_{core} = -\frac{42 \pi}{5}  \omega^5 e^2 R_E^4 \int_{0}^{R_{core}} H_{\mu} \Im{(\mu(r))dr}.
\end{equation}
The integral terms can be written in terms of the imaginary part of the \textit{piecewise} tidal Love number of the core following equation 35 in \citet{tobie_etal05}:

\begin{equation}\label{imtidalapp}
     \int_{0}^{R_{core}} H_{\mu} \Im{(\mu(r))dr} = \frac{5R_{core}}{4\pi G}\Im{(\hat k_2^{core}}).
\end{equation}
Here, $\hat k_2^{core}$ is the piecewise gravitational tidal Love number associated with the core. For a homogeneous core, $k_2^{core}$ can be computed from the deformation at the top of the core \citep[e.g.,]{vanhoolst_etal13, Shao_Nimmo22}:

\begin{equation}\label{coreloveapp}
  \Im{(\hat k_2^{core}}) =  \frac{4\pi G\rho_{core}R_{core}^4}{5R_E^3} \Im{(\Delta R_{core})^4}
\end{equation}
The tidal heating in the shell is computed by subtracting the contribution of the core from the total heat:
\begin{equation}\label{shelldissapp}
    \dot E_{shell} = \dot E - \dot E_{core}.
\end{equation}
Tidal heating in the shell can be computed separately with a similar approach:

\begin{align}\label{shelldissipationonlyapp}
&    \dot E_{shell} = -\frac{42 \pi}{5}  \omega^5 e^2 R_E^4 \int_{R_{BO}}^{R_{E}} H_{\mu} \Im{(\mu(r))dr} \notag  \\
&  \hspace{.92cm}   = -\frac{42 \pi}{5}  \omega^5 e^2 R_E^4 \bigg( \int_{0}^{R_{E}} H_{\mu} \Im{(\mu(r))dr} - \int_{0}^{R_{BO}} H_{\mu} \Im{(\mu(r))dr} \bigg) \notag  \\
&  \hspace{.93cm}   -\frac{21}{2G}\omega^5e^2R_E^4  \bigg( R_E  \Im{(k_2)} - R_{core} \Im{(\hat{k_2}^{core})} \bigg).
\end{align}
Here, $R_E$ and $R_{OB}$ refer to the radius at Enceladus's surface at the bottom of the ocean. equations~\eqref{coredissapp},~\eqref{shelldissapp}, and~\eqref{shelldissipationonlyapp} provide a method to calculate the partitioning of global tidal heating for a given interior structure model. These equations require \textit{a priori} information, assumptions, or independent constraints on the rheological and structural properties of the core to compute the shear dissipation in the core. 
\noindent

This method can be used in a forward model \textcolor{black}{by assumptions on the core properties} or in the context of an inverse problem based on a sampling approach. However, we consider requirements on the geophysical measurement that can not directly depend on the interior model parameters. Thus, we extend our analysis to derive the tidal heating on the basis of potentially observable quantities. Quantifying the partitioning of the heat in the icy body is equal to calculating the contribution of the core and the shell to the imaginary part of the gravitational Love number $\Im{(k_2)}$, as indicated from equation~\eqref{tidalheatingEquation}. \cite{beuthe19} uses membrane mechanics to obtain the contribution of each of the interior layers to the imaginary part of the tidal Love number:

\begin{equation}
    \Im{(k_2)} = \left| \frac{k_2+1}{k_2^o + 1} \right|^2\Im{(k_2^o)} - \xi_2 |h_2|^2 \Im{(\Lambda_T)},
\end{equation}
where $    \xi_2 = 3\rho_o/5\rho_b$, $\rho_o$ and $\rho_b$ are the densities of the ocean and the whole body, $k_2$ and $h_2$ are the tidal Love numbers of the whole body, and $k_2^0$ is the fluid-crust gravitational Love number (assuming the crust is fluid), and $\Lambda_T$ is the shell spring constant introduced by \citet{beuthe19}.
The contribution of the shell and the core to the total tidal heating can then be written as:

\begin{align}\label{heatingwithbeuteapp}
 &   \dot E_{\text{core}} = - \frac{5}{2} \frac{ \omega R}{G} \left|\frac{k_2 + 1}{k_2^0 +1} \right|^2 \Im(k_2^0) \langle | U_2^{(T)}  |^2 \rangle, \\
 & \dot E_{\text{shell}} = \frac{5}{2} \frac{\omega R}{G} \xi_2 \,|h_2|^2  \Im(\Lambda_T) \,  \langle | U_2^{(T)} |^2 \rangle,
\end{align}
where, 
\begin{equation}
    \langle |U_2^{(T)}|^2 \rangle = (\frac{21}{5})(\omega R)^4 e^2.
\end{equation}
The fluid crust gravitational Love number can be computed from equation D.2 in Appendix D and Table 9 in \citet{beuthe19}. Once the tidal heating rate in the shell is computed, the heat produced in the core can be computed by subtracting the heating in the shell from the total heat. 
Following \citet{beuthe15}, the term $\Lambda_T$ in equation~\eqref{heatingwithbeuteapp} is the complex spring constant of the shell and is computed as:
\begin{equation}
    \Lambda_T = \Lambda_M + \Lambda_\rho + \Lambda_\chi + \Lambda_\omega.
\end{equation}
The term $\Lambda_M$ refers to the contributions by the membrane spring constant that has the dominant effect, and $\Lambda_\rho$, $\Lambda_\chi$, and $\Lambda_\omega$ are corrections due to the density contrast between the shell and the ocean, compressibility correction, and dynamical correction, respectively. Here, we consider the effect of the two leading terms, $\Lambda_M$ and $\Lambda_\rho$, the latter two terms having a negligible contribution \citep{beuthe15}. The contributions of the membrane spring estimate and density contrast are \citep{beuthe15}:
\begin{align}
&    \Lambda = f_\mu \hat{\mu} \epsilon, \\
&    \Lambda_\rho = f_\rho \frac{\delta \rho}{\rho} \epsilon.
 \end{align}
Here, $\delta \rho$ is crust-ocean density contrast, $\epsilon = d/R$ and $f_\mu$ and $f_\rho$ are:
\begin{align}
 &   f_\mu = \frac{8 (1+{\nu})}{(5+{\nu})}, \\
 &   f_\rho = \frac{1+5{\nu}}{5+{\nu}},
\end{align}
and, 
\begin{equation}
\hat{\mu} = \frac{ \mu}{\rho g R},
\end{equation}
where $\mu$ is the shear modulus, $\rho$ is the density, and $\nu$ is the Poisson's ratio of the shell. $\nu$ can be written in terms of the shear and bulk moduli:
\begin{equation}
    \nu = \frac{3K-2\mu}{6K+2\mu}.
\end{equation}
This formulation provides a means for obtaining the tidal heating in the shell by measuring the tidal Love number $h_2$ (real and imaginary parts) and independent constraint on the imaginary part of the complex shear modulus of the shell, $\Lambda_T$, as shown in equation~\eqref{heatingwithbeuteapp}. 
Here, we extend the formulation to develop expressions for the partitioning of heat between the shell and the core, which do not require \textit{a priori} knowledge of the shell rheology. 
equation 125 in \citet{beuthe15} provides an expression between the displacement and potential Love numbers, the rigidity of the shell, and the density structure of the body:

\begin{equation}
    k_2 + 1 = (1+\Lambda_T)h_2 - 5\frac{\delta \rho}{\rho}\epsilon.
\end{equation}
The rigidity constant of the shell can be written in terms of the other parameters as:
\begin{equation}\label{rigiditywithlovenumbersapp}
\Im{(\Lambda_T)} = \Im \bigg( \frac{1}{h_2} {(k_2+1 + 5 \frac{\delta \rho}{\rho} \epsilon)} \bigg),
\end{equation}
Combining equations~\eqref{heatingwithbeuteapp} and \eqref{rigiditywithlovenumbersapp}, we can rewrite the expression of tidal heating in the shell as:

\begin{equation}\label{finalheatinshellwithloveapp}
 \dot E_{\text{shell}} = \frac{105}{10} \frac{(\omega R)^5}{G} \xi_2 \,|h_2|^2  \bigg[ \Im \bigg( {\frac{1}{h_2} \big( k_2+1 + 5 \frac{\delta \rho}{\rho} \epsilon \big)} \bigg)\bigg] \,  e^2.
\end{equation}
This formulation provides the tidal heating in the shell by measurements of the real and imaginary parts of the tidal Love numbers ($k_2$, $h_2$, $\epsilon_{k_2}$, $\epsilon_{h_2}$), and information on the density structure. 
The densities of the layers of the body can be constrained using the static gravity field and libration, and are easier to constrain compared to the rheology of the layers as required in equation~\eqref{heatingwithbeuteapp}.
Tidal heating in the core can be computed by subtracting the tidal dissipation in the shell from the total heat. This derivation is general and can be used for all other icy satellites for which the necessary geodetic observations are or will become available.

\section{Sensitivity of libration to interior properties}\label{sec:librationrigidity}

\begin{figure}[ht] 
\begin{center}
\vspace{5mm}
\includegraphics[width=.65\textwidth]{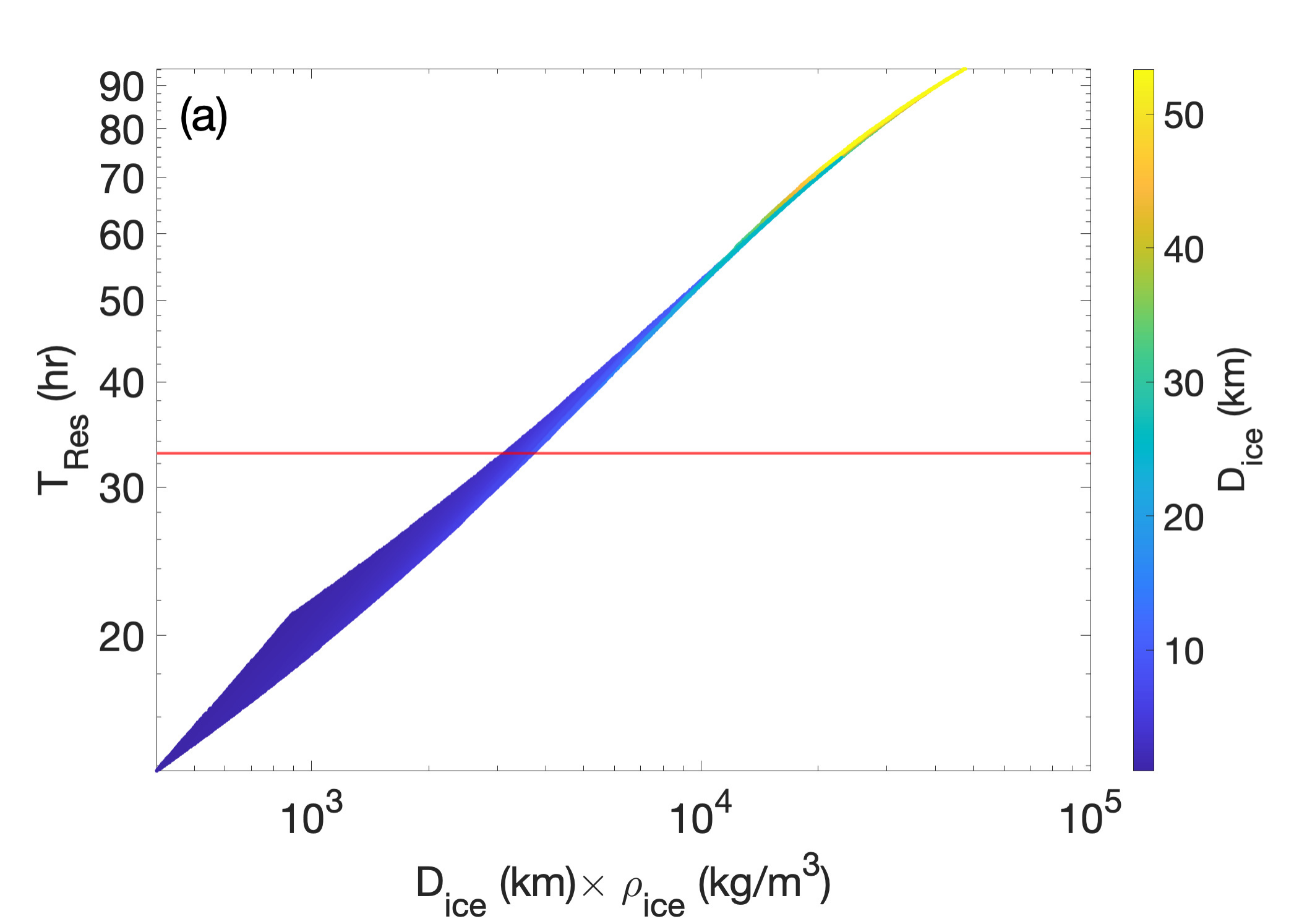}
\includegraphics[width=.65\textwidth]{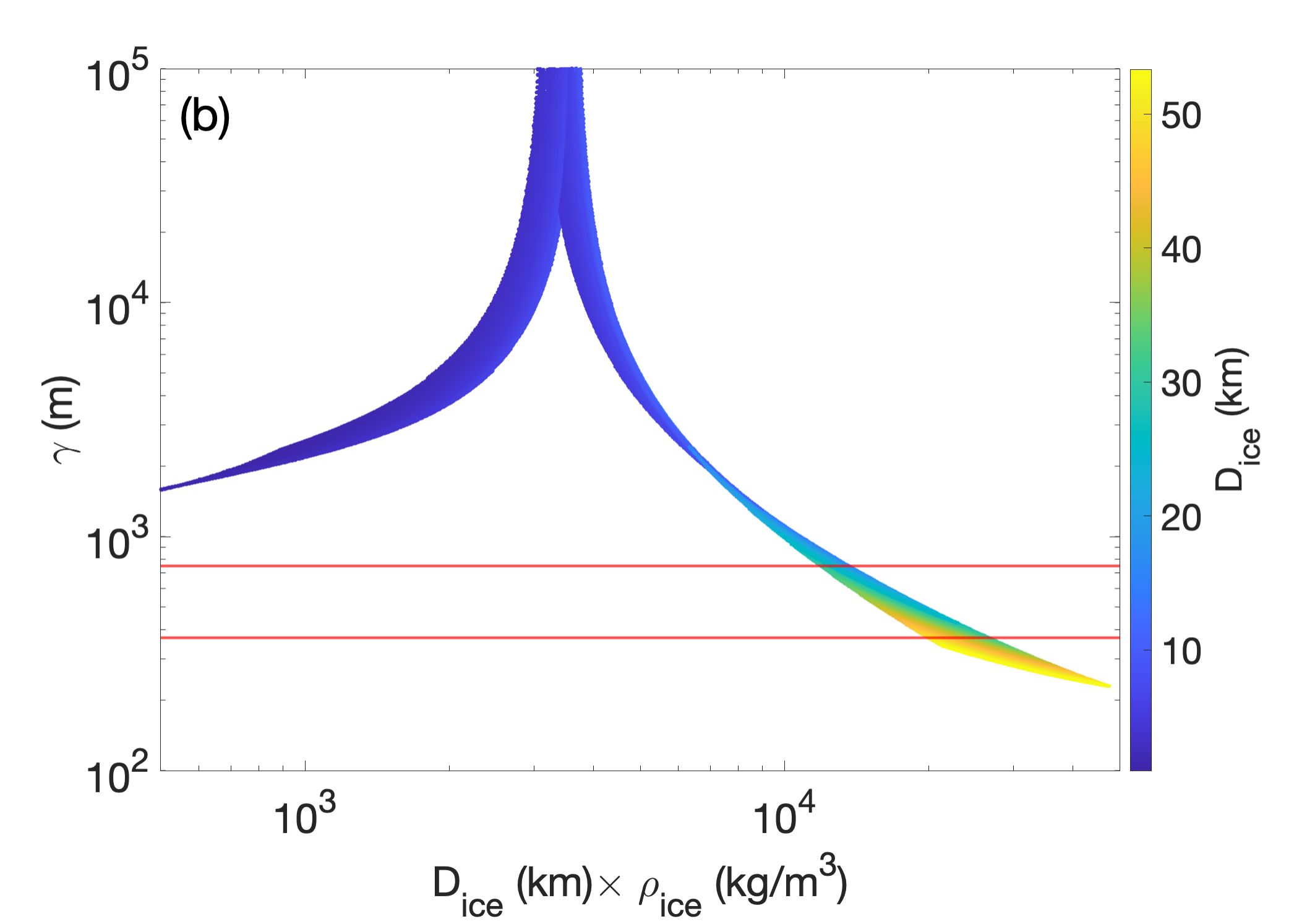}

\caption{(a) Natural libration period and (b) the amplitude of forced physical libration of Enceladus as functions of the product of shell thickness and density as a representative of its shell MoI. (c,d) Same quantities as a function of the ratio of shell thickness and shell density. The plots are produced for fine grids of shell thickness and densities in ranges [1-60] km and [400-900] kg/m$^3$. Colorbar indicates the associated shell thickness.}
\label{fig:librationfreeperiodproductratio}
\end{center}
\end{figure}

As highlighted in Section~\ref{sec:shellthicknesslibration}, if the ice shell is sufficiently thick as inferred for Enceladus, the libration amplitude is mostly determined by the moment of inertia and therefore, the thickness and density of the shell.
We further explore the variations of the shell libration amplitude and the period of free libration as functions of the product of the shell density and shell thickness as a representative of its moment of inertia (Figure~\ref{fig:librationfreeperiodproductratio}). As shown in the figure, the libration amplitude has dramatically large values for a certain value of  $\rho_{shell}\times D_{shell}$ where the natural frequency of the shell equals the orbital period of Enceladus. 

\begin{figure}[ht] 
\begin{center}
\vspace{5mm}
\includegraphics[width=.65\textwidth]{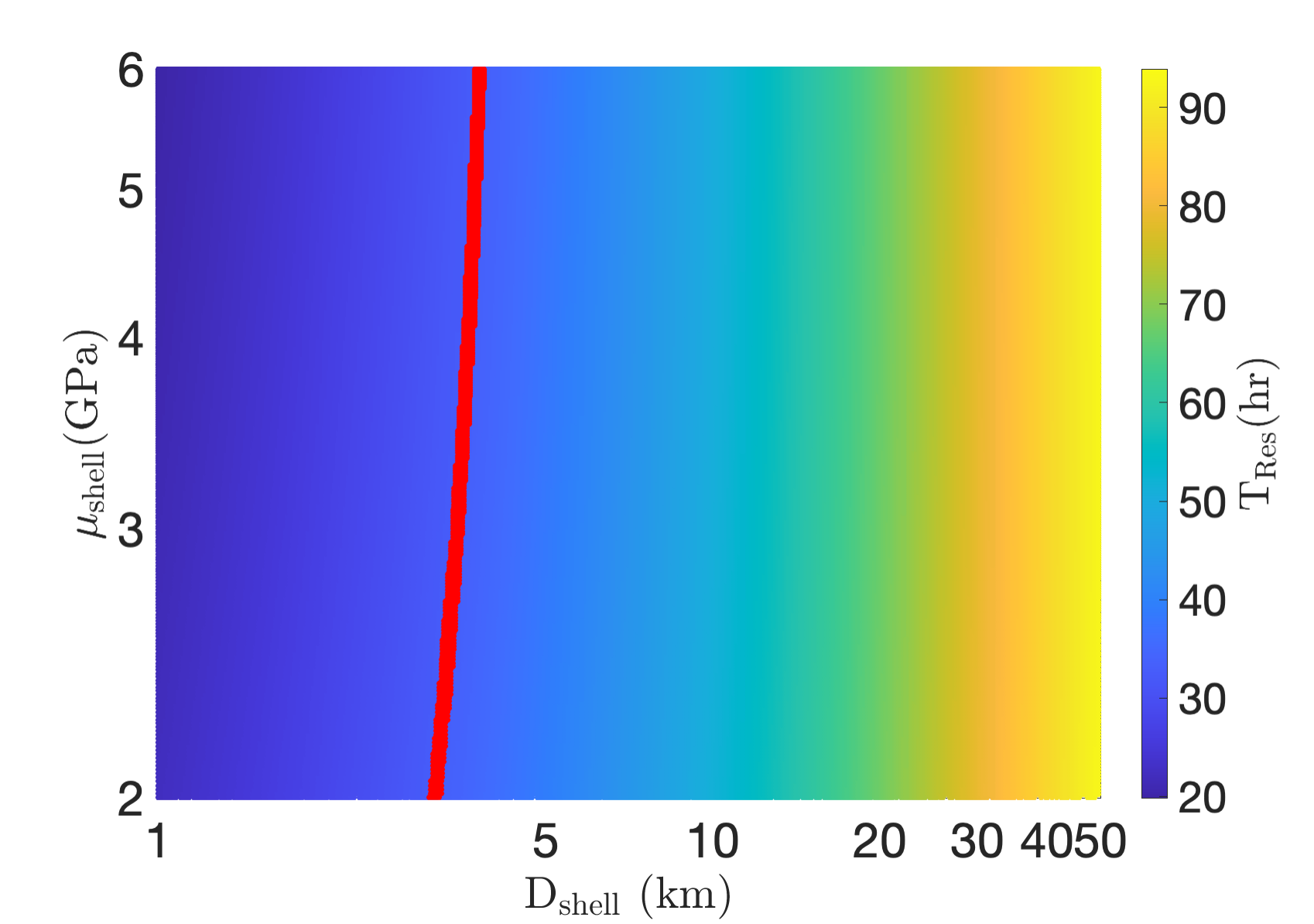}
\includegraphics[width=.65\textwidth]{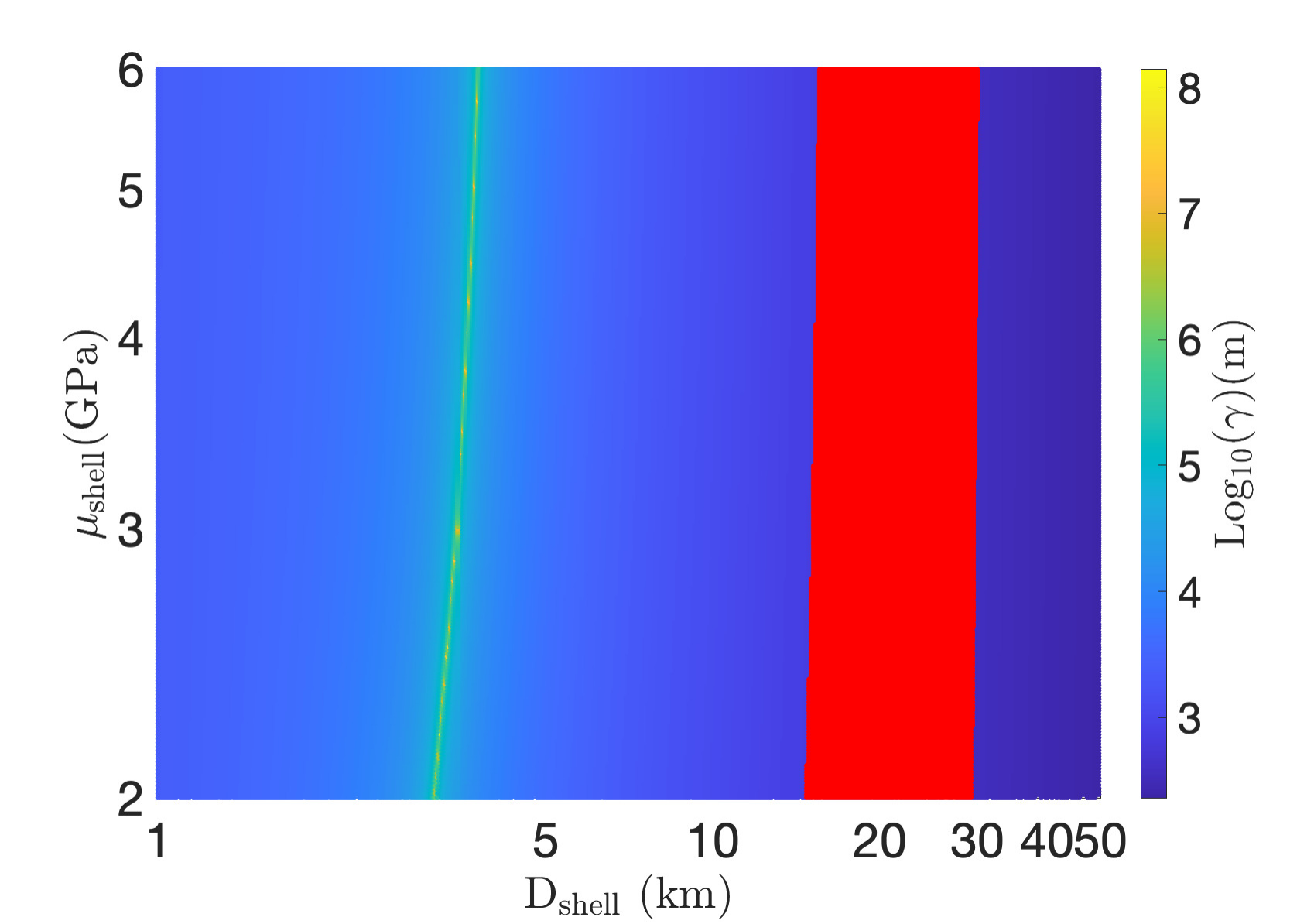}
\caption{Natural libration period (hr) and the amplitude of forced physical libration of Enceladus as functions of the shell thickness and rigidity. }
\label{fig:mu_D_libration}
\end{center}
\end{figure}

Elastic forces would be important only if the shell is very thin or its rigidity is very low. The effects of these parameters are also reflected in the natural (resonant) libration of the shell. We compare the effects of the shell rigidity against the shell thickness on the natural free libration period and the libration amplitude (Figure~\ref{fig:mu_D_libration}). As expected, libration is significantly more sensitive to the shell thickness compared to its rigidity. This different sensitivity is because for the shells that are not very thin compared to the radius of the body (such as those of Europa and Titan), the librational response is dominated by the inertial forces and the shell is effectively rigid. Note that as shown in equation~\eqref{eq:naturalomega}, two free libration frequencies exist for a body. Here we only show the natural frequency which is closer to the orbital period of Enceladus with the second free libration frequency is significantly longer for any reasonable shell properties and thus not relevant ($>100~\rm hr$). Orbital period of Enceladus and the range of the inferred libration amplitudes has been demonstrated in the plots.

Because the rigidity of the shell does not play an important role in the librational response of the body, the viscosity of the shell has also a very small effect as well. Only in the case of a very low-viscosity shell with very low effective rigidity that can flexible respond to the tidal forces, the effect of the shell's viscoelastic response becomes important (Figure~\ref{fig:librationsensitivity}, panel a). As expected, the viscosity of the shell does not affect the libration even if highly tidally deformable.

\begin{figure}[ht] 
\begin{center}
\vspace{5mm}
\includegraphics[width=.65\textwidth]{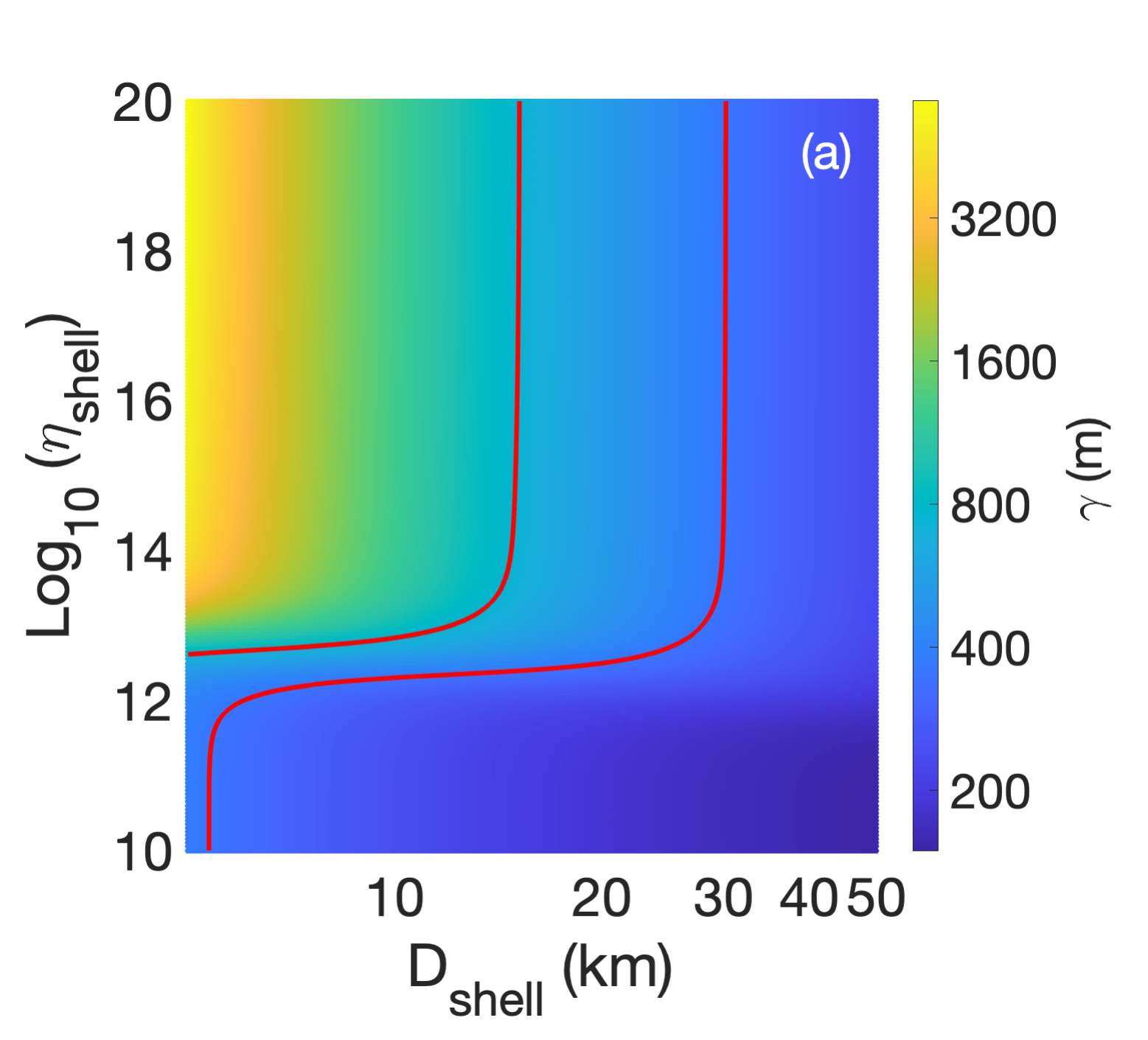}
\includegraphics[width=.65\textwidth]{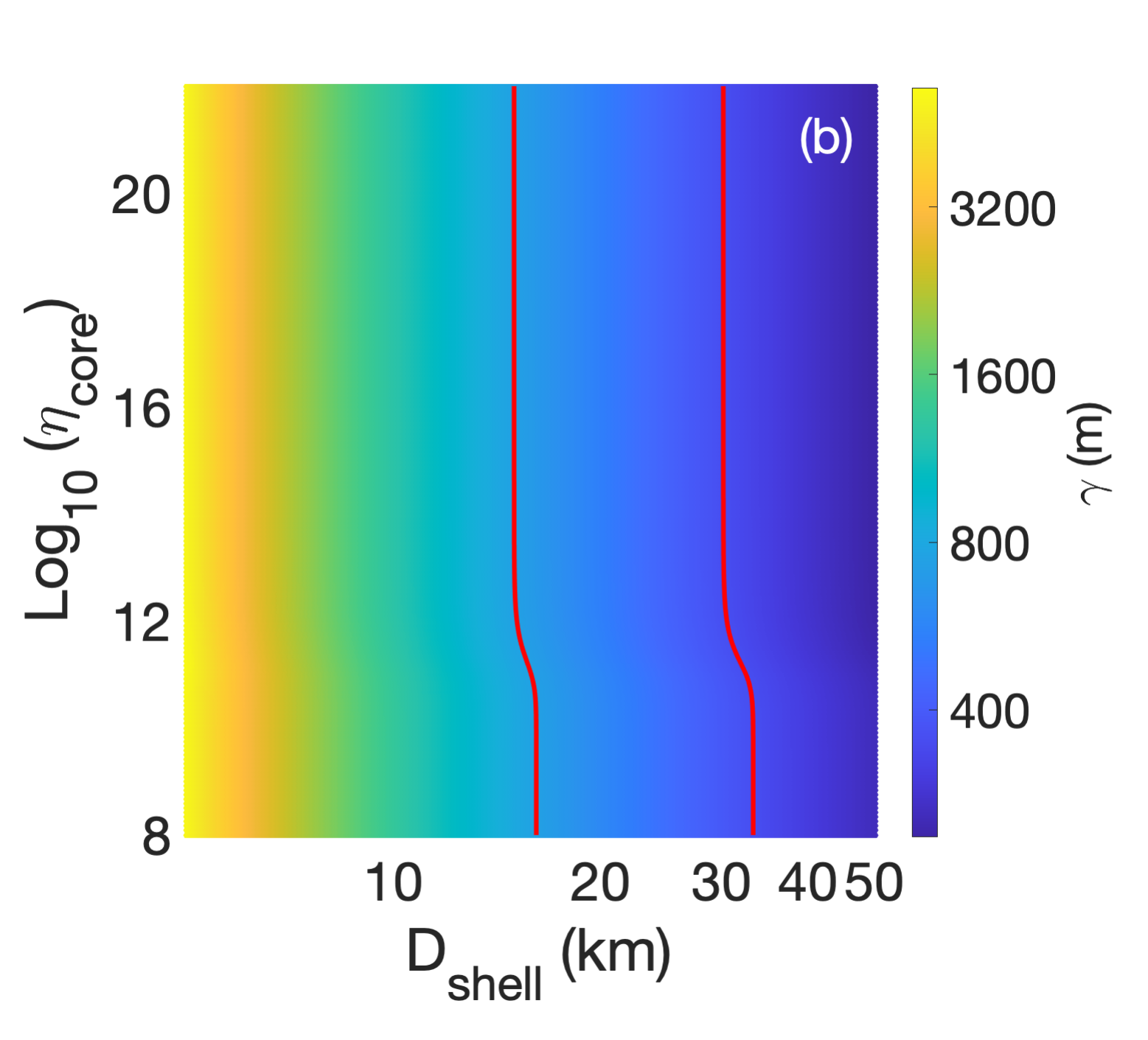}
\caption{Variation of the amplitude of the libration as a function of ice shell thickness and the effective viscosities of shell (a) and the core (b). In panel (a), the core viscosity is assumed equal to $\rm10^{20}~Pa.s$. The region between the red lines shows the libration amplitude spanning the same range in Figure~\ref{figure:librationnatural}. Other interior structure parameters are the same as in Figure~\ref{figure:tidalresponseviscs}. 
The shell thickness and amplitude of libration are plotted in logarithmic scales.}
\label{figure:librationviscosity}
\end{center}
\end{figure}

\clearpage

\section{Markov Chain Monte Carlo Internal Structure Inversion}\label{sec:mcmc}
We adopt the Metropolis algorithm to sample the posterior distribution in the model space \citep{mosegaard_Tarantola95}. This algorithm, which samples the model space in a random manner, ensures that models that fit the data better and are simultaneously consistent with prior information are sampled more frequently. The Metropolis algorithm samples the model space with a sampling density proportional to the posterior probability density and, therefore, ensures that low-probability areas are sampled less excessively. Assuming that data uncertainty is Gaussian distributed and that observational uncertainties and calculation errors among the data sets considered are independent, the likelihood function can be written as

\begin{equation}
    \mathcal{L}(\textbf{m}) \propto \Pi \exp \bigg(-\frac{|d_{obs}^i-d_{cand}^i (\textbf{m})|^2}{2\sigma_i^2}\bigg)
\end{equation}
where $i$ runs over the observables, $\textbf{m}$ is the vector composed of the model parameters, and $d_{obs}$ and $d_{cand}$ are observed and computed candidate data in the algorithm, respectively, and $\sigma_i$ is data uncertainty.

\section{Probability Density Function (PDF) of interior model parameters and correlation}\label{sec:modelparametersbroadest}

\begin{figure}[ht] 
\centering
    \rotatebox{270}{
\hspace{-2.cm} \includegraphics[width=1.5\textwidth]{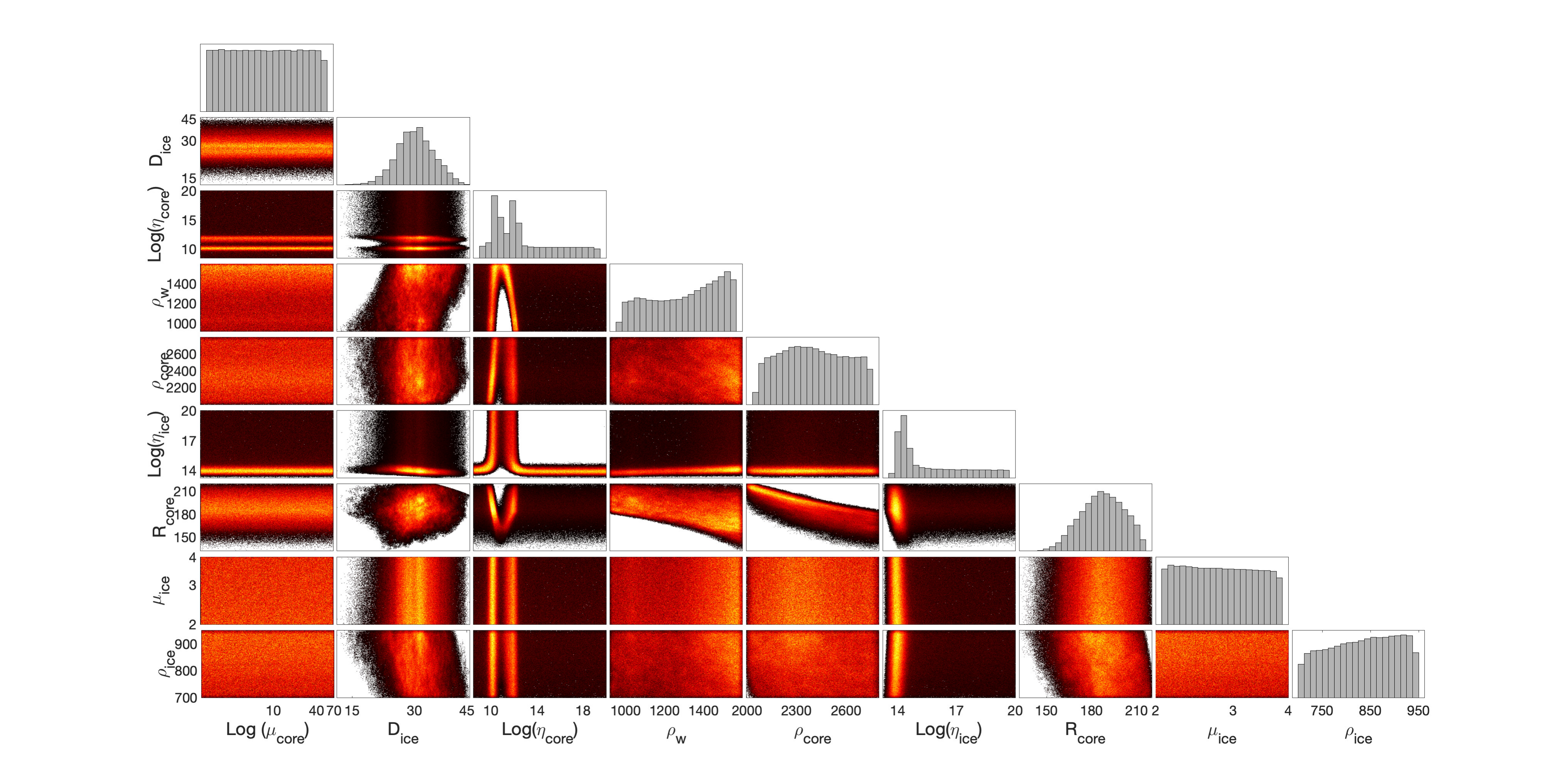}}
\vspace{-1.5cm}
\caption{\small{Exploration of the key geophysical interior model parameters corresponding to with Figure~\ref{fig:obsrvblsmassheatlibgeneral}}. Multiple peaks observed in the PDFs of the viscosities of ice shell and the core correspond to separate tidal dissipation regimes that can match the inferred global heat loss.  Shear moduli are presented in GPa, $\rm D_{shell}$ and $\rm R_{core}$ are in km, and other parameters are presented in their SI units. Ice shell thickness, and shear moduli are plotted in logarithmic axes. Black and yellow colors indicate the lowest and highest likelihoods. }
 \label{fig:mdlprmtesmassheatlibgeneral}
\end{figure}

\textcolor{black}{Figure~\ref{fig:mdlprmtesmassheatlibgeneral} shows corner plots of the interior parameters associated with the case considered in Figure~\ref{figure:heatviscs}.
Here, the sensitivity of the adopted set of measurements used in Figure~\ref{fig:mdlprmtesmassheatlibgeneral} to parameters such as the density and rigidity of the layers is small. In addition, in this case we have assumed a very broad range of viscosities of the ice shell and core. As a result,} multiple peaks are apparent in the PDFs of the viscosities of the core and the shell, which are associated with the multiple regions highlighted in Figure~\ref{figure:heatviscs}, consistent with the estimated heat loss. These distinct peaks are indicative of tidal dissipation dominated by viscous deformation versus anelastic deformation. In other words, these peaks correspond to viscosities that are lower than or higher than the viscosity associated with the maximum heat production rate.

\noindent
\textcolor{black}{Assuming the same set of measurements, but limiting the viscosity ranges of the shell and core to $10^{11}<\eta_{shell}<10^{15}$~ Pa.s and $10^{11}<\eta_{core}<10^{20}$~Pa.s, and separating the two sets of models based on the two dissipation scenarios, we obtain the interior parameters shown in Figures~\ref{fig:prmtrsmassheatlibshelldiss} and~\ref{fig:prmtrsmassheatlibCorediss}.}
\textcolor{black}{We compare the amplitude and phase lags of $h_2$ and $l_2$  in Figure \ref{fig:h2shellvscorediss} for the two tidal dissipation scenarios. The 2-$\sigma$ standard deviations are indicated in this figure.}

When using the inferred geodetic quantities by \citet{park_etal24} and separating the interior parameters for the two tidal dissipation scenarios, we obtain constraints on the interior parameters and the correlations between them, as shown in Figures~\ref{fig:mdlprmtrsinversionboth90shell} and \ref{fig:mdlprmtrsinversionboth90core}.

\renewcommand{\thefigure}{19a}

\vspace{-4cm}
\begin{figure}[ht]
    \rotatebox{270}{
 \includegraphics[width=1.5\textwidth]{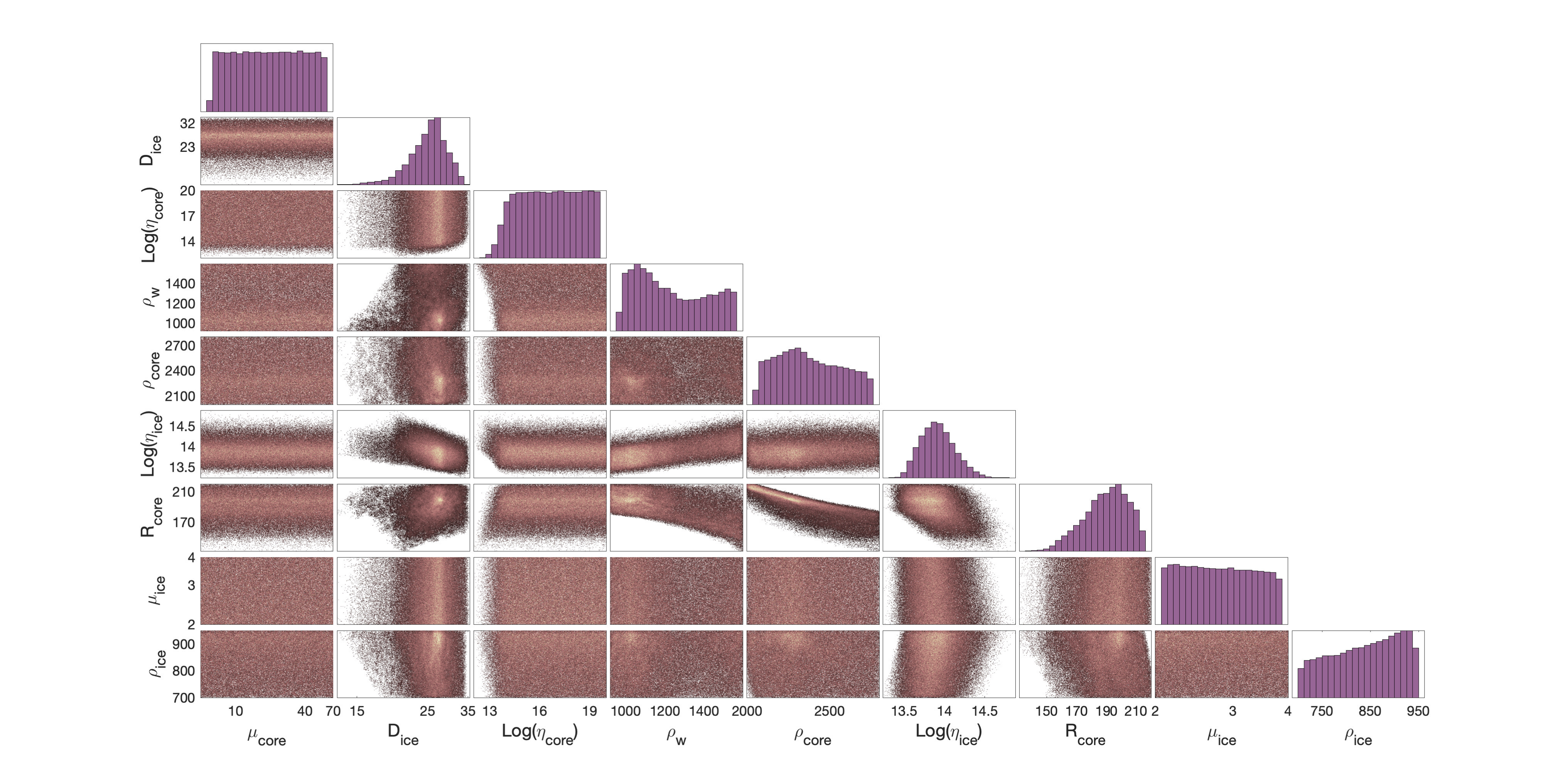}} 
 \vspace{-1.cm}
\caption{\textcolor{black}{Exploration} of the correlation between the model parameters associated with Figure~\ref{fig:observblessheatlibshellcorediss} for the case tidal dissipation is dominated in the shell. Black and yellow colors indicate lowest and highest likelihoods.  Units and plotting conventions are similar to Figure~\ref{fig:mdlprmtesmassheatlibgeneral}. }\label{fig:prmtrsmassheatlibshelldiss}
\end{figure}

\clearpage
\renewcommand{\thefigure}{19b}

\begin{figure}
    \rotatebox{270}{
\includegraphics[width=1.5\textwidth]{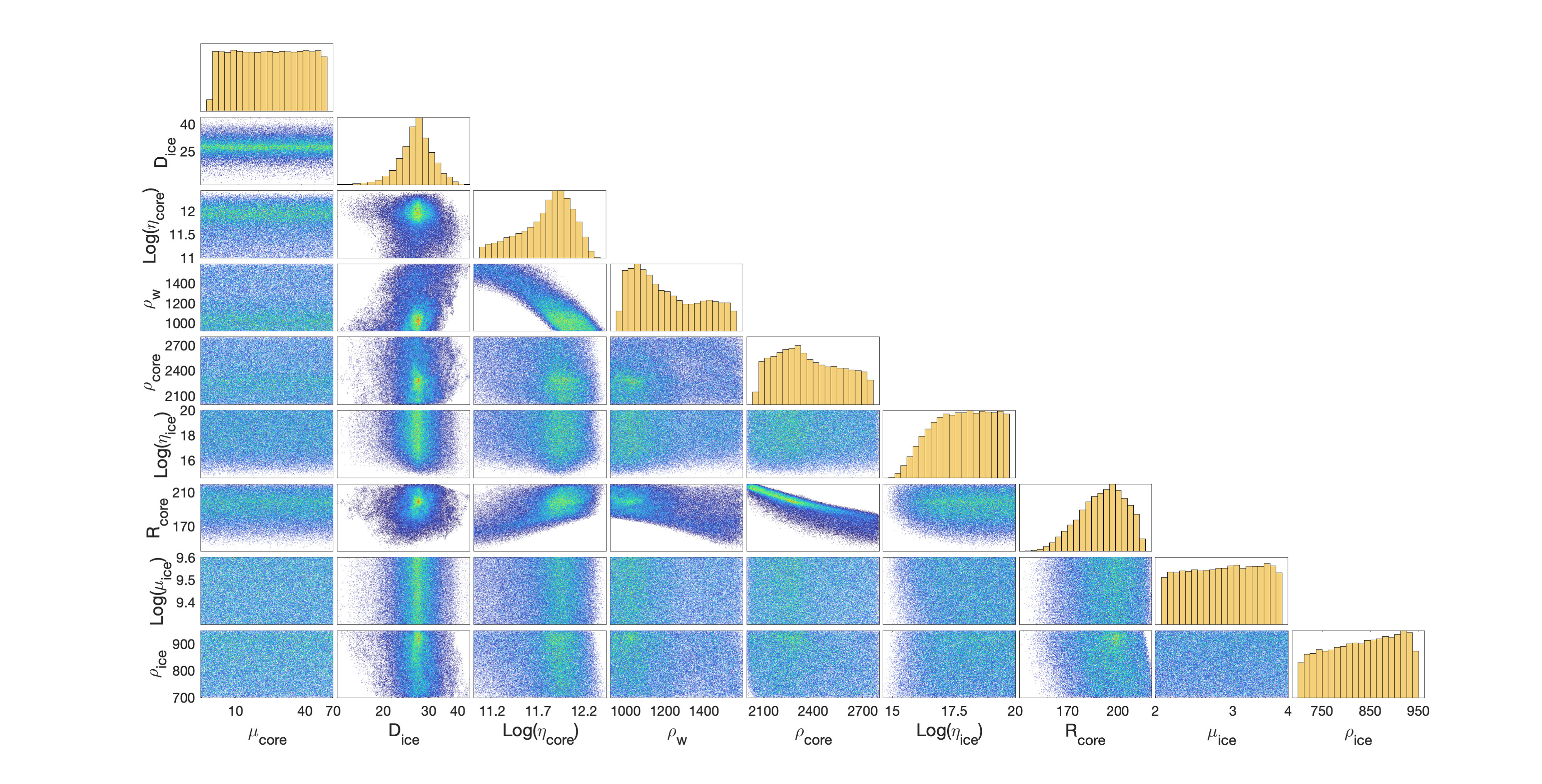}}
\vspace{-1.cm}
\caption{\textcolor{black}{Exploration} of the correlation between the model parameters associated with Figure~\ref{fig:observblessheatlibshellcorediss} for the case tidal dissipation is dominated in the shell. Units and plotting conventions are similar to Figure~\ref{fig:mdlprmtesmassheatlibgeneral}. Blue and red colors indicate the lowest and highest likelihoods.}\label{fig:prmtrsmassheatlibCorediss}
\end{figure}

\renewcommand{\thefigure}{\arabic{figure}}
\setcounter{figure}{19}

\begin{figure}[ht] 
\includegraphics[width=1.\textwidth]{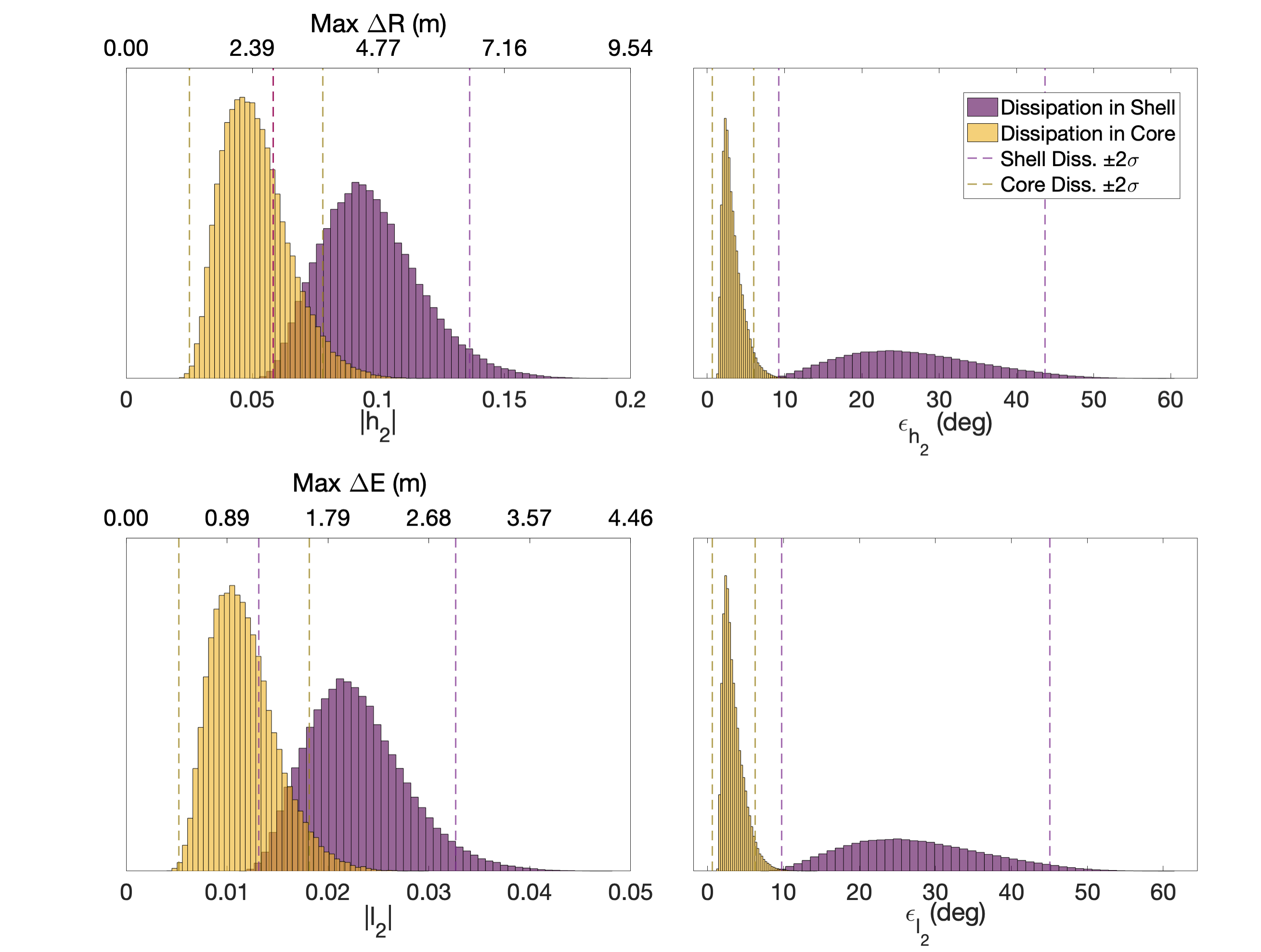}
\caption{\small{Vertical and horizontal displacement Love numbers ($|h_2|$, $|l_2|$) and phase lags ($\epsilon_{h_2}$, $\epsilon_{l_2}$) are presented for the core-dominated and shell-dominated tidal dissipation. The heat flux is assumed to be $\rm15~GW<\dot E<40~GW$, covering estimates in \citet{hemingway_etal18} and \citet{park_etal24}, and the direct measurements \citep{howett_etal11}. Libration is conservatively assumed to be between $\rm 0.082^\circ<\gamma<0.169^\circ$, covering the estimates in \citet{thomas_etal16}, \citet{nadezhdina_etal16}, and \citet{park_etal24}. The top axis in the plots for $h_2$ and $l_2$ indicate the associated maximum radial and maximum easterly displacements in the bottom axis, respectively.}}
\label{fig:h2shellvscorediss}
\end{figure}

\clearpage

\renewcommand{\thefigure}{21a}

\begin{figure}[ht]
\centering
   \rotatebox{270}{
\includegraphics[width=1.5\textwidth]{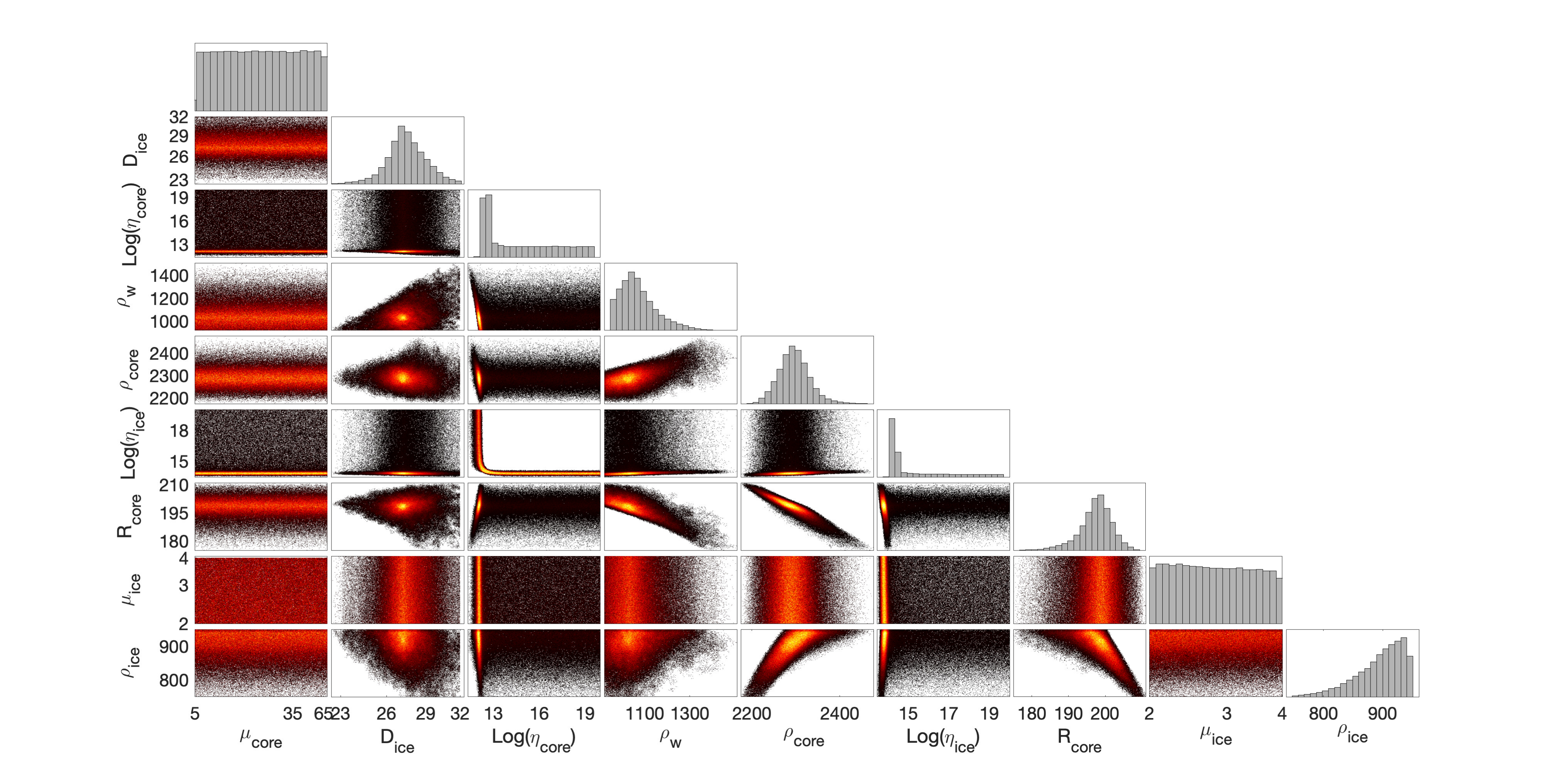}}
\vspace{-1cm}
\caption{\textcolor{black}{Exploration of the correlation between the model parameters associated with Figure~\ref{fig:Inversionobsrvblsboth}. The ranges of the plausible interior parameters are shown in Table~\ref{table:mdlprmtrscurrentdata}. Black and yellow colors indicate the lowest and highest likelihoods. Units and plotting conventions are similar to Figure~\ref{fig:mdlprmtesmassheatlibgeneral}.}}
\label{fig:Inversionmdlprmtrsboth}
\end{figure}

\clearpage

\renewcommand{\thefigure}{21b}

 \begin{figure}[ht] 
\rotatebox{270}{
\hspace{-2cm} \includegraphics[width=1.5\textwidth]{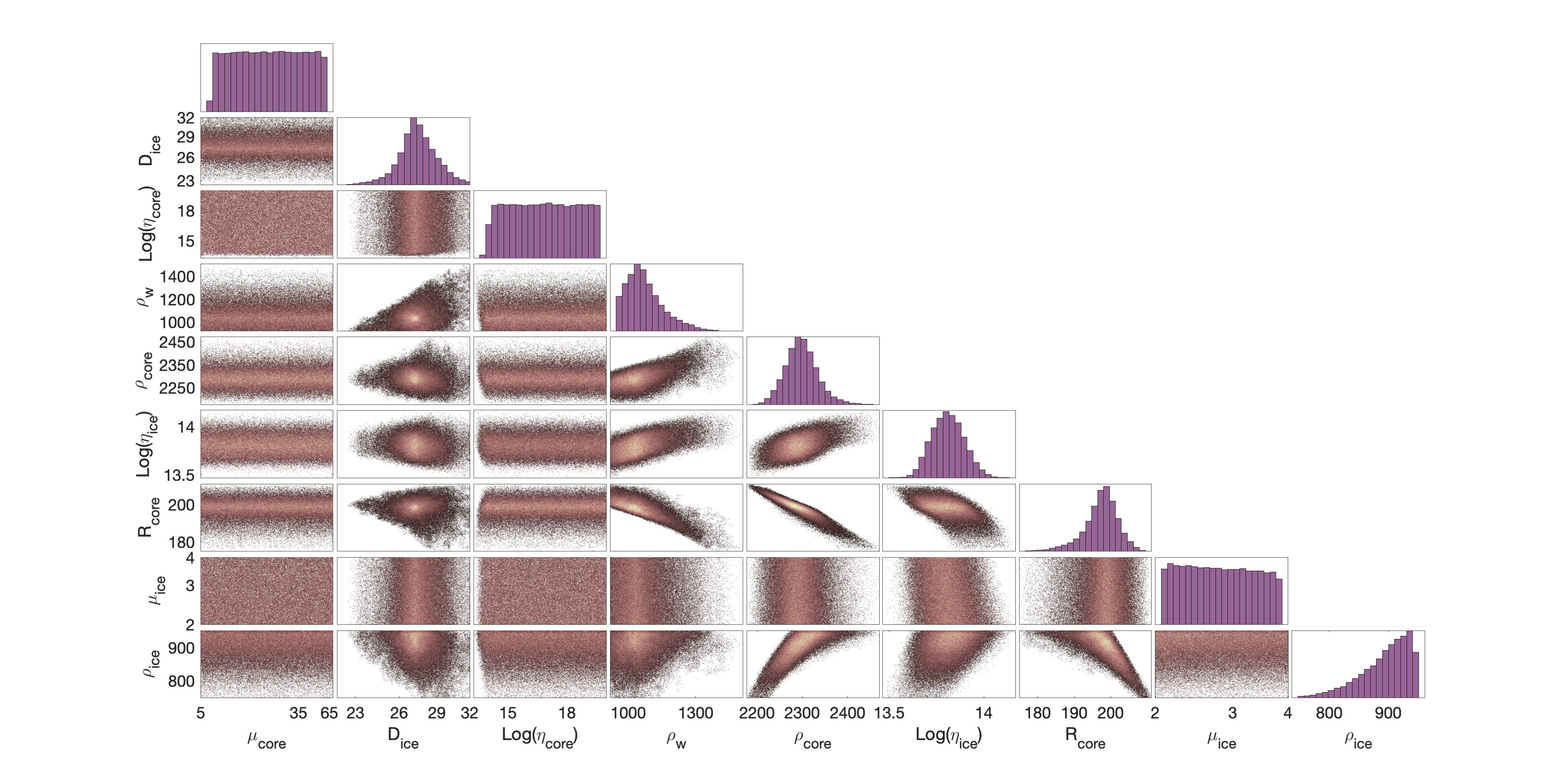}}
\caption{\textcolor{black}{Model} parameters for the case where at least 90\% of the tidal heating  is focused in ice shell associated with Figures~\ref{fig:Inversionobsrvblsboth} and \ref{fig:Inversionmdlprmtrsboth}.  The tidal heating in the shell is calculated using the formulation by \citep{beuthe19} (equation~\eqref{heatingwithbeuteapp}).  Darkest and brightest colors indicate lowest and highest likelihoods. Units and plotting conventions are similar to Figure~\ref{fig:mdlprmtesmassheatlibgeneral}.}\label{fig:mdlprmtrsinversionboth90shell}
\end{figure}

\renewcommand{\thefigure}{21c}

\clearpage

\vspace{5cm}
 \begin{figure}[ht] 
\rotatebox{270}{
\hspace{-2cm}\includegraphics[width=1.5\textwidth]{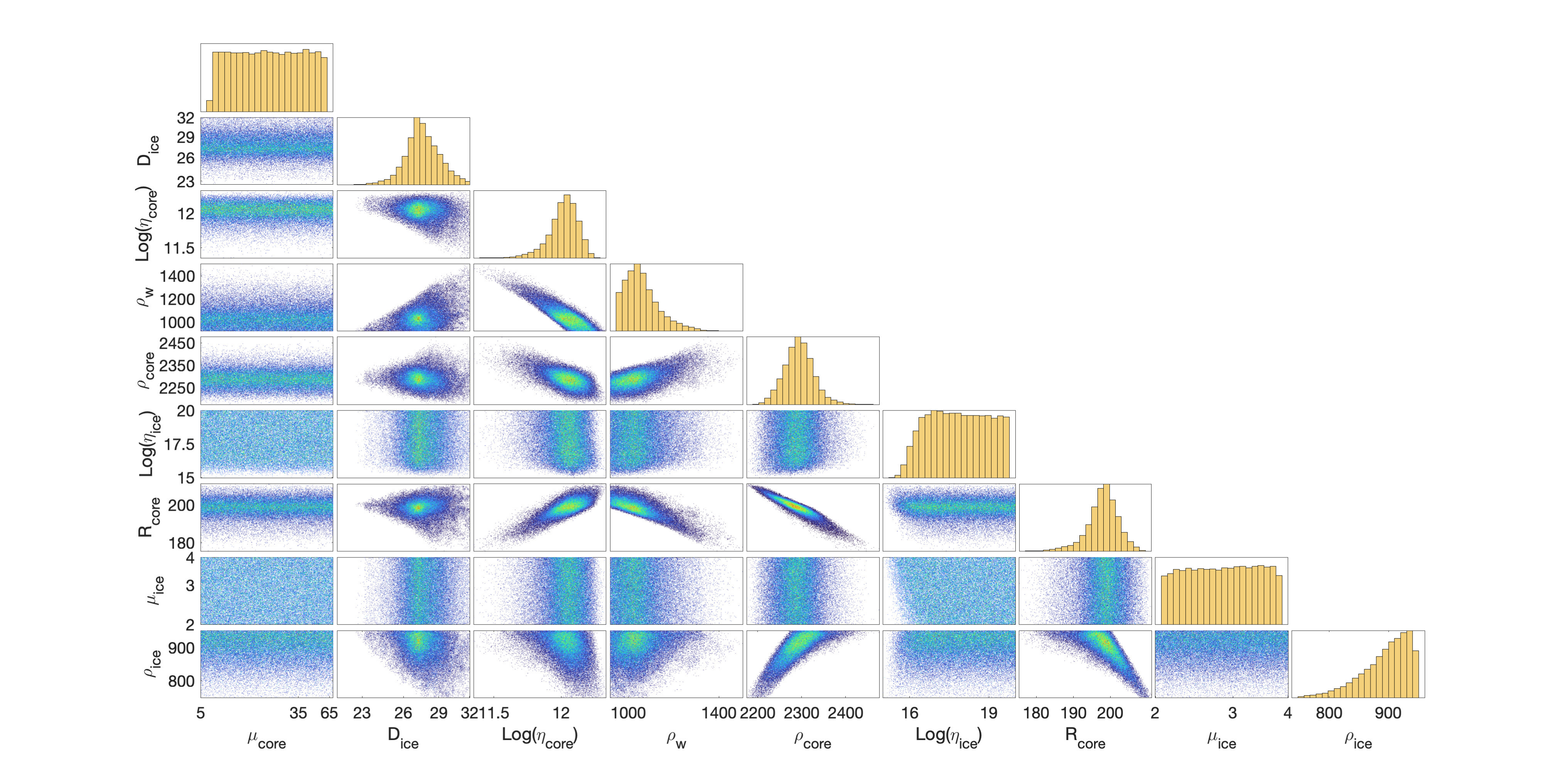}}\\
\caption{\textcolor{black}{Model} parameters for the case where at least 90\% of the tidal heating  is focused in the rocky core  associated with Figures~\ref{fig:Inversionobsrvblsboth} and \ref{fig:Inversionmdlprmtrsboth}. The tidal heating in the core is calculated by subtracting the heat in the shell from the total tidal heating computed based on equation~\eqref{tidalheatingEquation}. Units and plotting conventions are similar to Figure~\ref{fig:mdlprmtesmassheatlibgeneral}.}
 \label{fig:mdlprmtrsinversionboth90core}
\end{figure}

\renewcommand{\thefigure}{\arabic{figure}}
\setcounter{figure}{21}

\clearpage

\section{Effect of shell heterogeneous structure on the Love numbers}\label{sec:Heterogeneityapp}
Here, we assess the effect of lateral variations in ice shell thickness on the tidal response and its effect on the required measurements of $k_{2m}$ and $h_{2m}$. We use the shape model shown in Figure~\ref{fig:3dshellmap}. We build on the analysis of \citet{berne_etal23} and 
conduct simulations of tidal deformation for four models: 1. without lateral variations in ice shell thickness; 2. with lateral variations in ice shell thickness; 3. with a dissipative core and without lateral variations in ice shell thickness; and 4. with a dissipative core and lateral variations in ice shell thickness. For each model class, we vary the mean ice shell thickness across 15--50~km consistent with the findings in Section~\ref{sec:twoscenarios}. 
We assume a core of viscosity $10^{20}$~Pa.s for the case of a non-dissipative core, and $10^{12}$~Pa.s for the case of a dissipative core. All other parameters are chosen to be the same as those used in Figures~\ref{figure:heatviscs}~and~\ref{figure:tidalresponseviscs}. To compute $k_{2m}$ and $h_{2m}$, we use a numerical finite element model \citep{berne_etal23}.

\begin{figure}[htbp]  
    \centering
    \includegraphics[width=\linewidth]{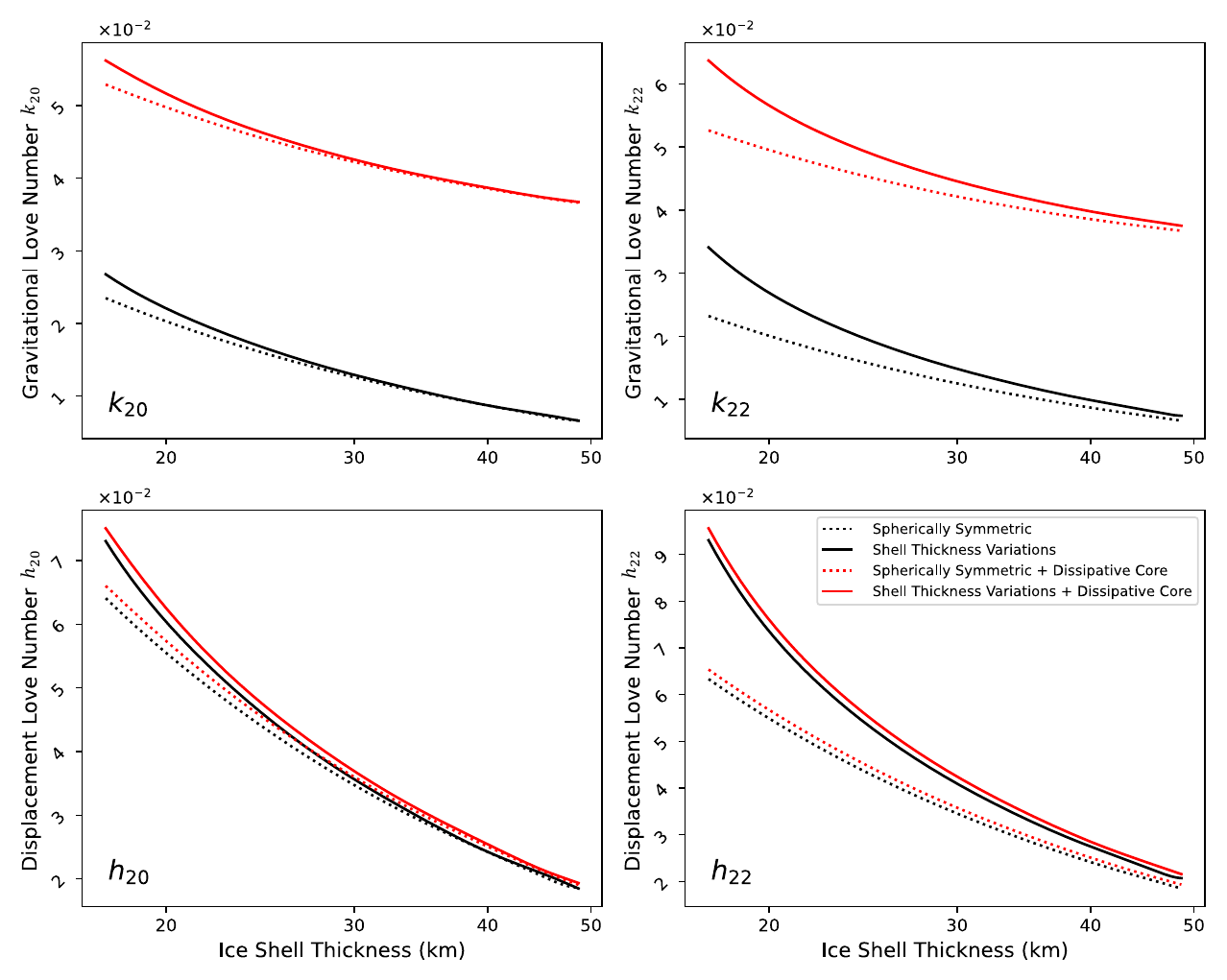}  
    \caption{Top: $k_{20}$ (left column) and $k_{22}$ (right column), against mean ice shell thickness for models that assume an ice shell of constant thickness (dotted line), an ice shell with lateral variations in thickness (solid line), or a dissipative core (dotted and solid red lines). Bottom: Similar to top row except showing displacement Love numbers $h_{20}$ and $h_{22}$ plotted against mean ice shell thickness. x-axes are plotted on a log-scale.}  
    \label{fig:alex_fig}  
\end{figure}

If the core is tidally active and is the primary location of the tidal heating, then its effect of $k_{2m}$ is very significant, regardless of the shell thickness. In that case $k_{2m}$ increases by a factor of (2--4). If the shell thickness is small, the effect of structural heterogeneity on both $k_{2m}$ and $h_{2m}$ is important. For a thin shell, $h_{20}$ can vary by a maximum of 10\% while $h_{22}$ can vary by approximately 50\%, as a result of the shell thickness variation. For thick shells, the effects of the tidal activity of the core and the structural heterogeneity of the shell on $h_{2m}$ are small because the tidal displacement is dominated by the shell rigidity. For a thin shell, the effects of the structural heterogeneity of the shell and the tidal activity of the core on $k_{2m}$ become superimposed. However, the effect of core activity is more significant. If the tidal activity is focused in the core, then its effect on the shell displacement ($h_{2m}$) through the additional gravitational perturbation can be more significant than the shell structural heterogeneity and overshadow it.

\end{appendix}

\clearpage

\bibliography{sample631}{}
\bibliographystyle{aasjournal}




\end{document}